\documentclass{aa}
\usepackage{graphicx}
\usepackage{txfonts}
\begin{document}
\title{Photometric study of the IC 65 group of galaxies
\thanks{Based on observations obtained at the Calar Alto
observatory, Almeria, Spain.}
\thanks{Figures 5, 7, and 9, Tables 2, 6 and 7 and Appendicies are only available in electronic form 
at http://www.aanda.org}
}
\author{ J. Vennik\inst{1} \and U. Hopp\inst{2}}

\offprints{J. Vennik, \email{vennik@aai.ee}}
\institute{Tartu Observatory, 61602 T\~oravere, Tartumaa, Estonia,\\
\email{vennik@aai.ee}
\and Universit\"atssternwarte M\"unchen, Scheiner Str. 1,
          D-81679 M\"unchen, Germany,\\ 
\email{hopp@usm.uni-muenchen.de}
}
\date{ Received ; accepted  }

\abstract
{A large fraction of the stellar mass is found to be located in groups of the size of the
Local Group.
Evolutionary status of poor groups is not yet clear and many 
groups could still be at an early dynamical stage or even still forming, especially
the groups containing spiral and irregular galaxies only.}
{We carry out a photometric study of a poor group of late-type galaxies around IC 65, with the aim:  
(a) to search for new dwarf members
and to measure their photometric characteristics; 
 (b) to search for possible effects of mutual interactions on the morphology and star-formation
characteristics of luminous and faint group members; (c) to evaluate the evolutionary status of this 
particular group.}
{We make use of our $BRI$ CCD observations,
 DPOSS blue and red frames, and the 2MASS $JHK$ frames.
 In addition, we use the \ion{H}{i} imaging data, the far-infrared and radio data from the literature. 
Search for dwarf galaxies is made using the SExtractor software. Detailed surface photometry 
is performed with the MIDAS package.
}
{Four LSB galaxies were classified as probable dwarf members of the group and the $BRI$ physical 
and model parameters were derived for the first time for all true and probable group members. 
Newly found dIrr galaxies around the IC 65 contain a number of \ion{H}{ii} regions, which show a range of 
ages and propagating star-formation. Mildly disturbed gaseous and/or stellar morphology is found 
in several group members. 
}
{Various structural, dynamical, and star-forming characteristics let us conclude that the IC 65 group is a typical 
poor assembly of late-type galaxies at an early stage of its dynamical evolution with some evidence of 
intragroup (tidal) interactions.} 


   \maketitle

\section {Introduction}

Most galaxies reside in groups of the size of the Local Group which typically consist of a
few bright members and dozens of dwarf galaxies (Tully \cite{tully87}). 
While rich clusters of galaxies reveal the sites of the highest concentration 
of luminous and dark matter, 
less massive aggregates of galaxies are distributed in 
less dense regions of the Universe and/or trace the cosmic filaments (Grebel \cite{grebel07}).
The physical nature of many compact as well as loose groups
has been confirmed through the detection of the X-ray emission of their hot
inter-galactic matter (Mulchaey \cite{mulchaey00}).  Zabludoff \& Mulchaey
(\cite{zabludoff98}) discuss the evidence that a large fraction of the loose
groups are still in at an early dynamical stage or even still forming, especially
those containing spiral and irregular galaxies only.
Plionis et al. (\cite{plionis04}) noted that groups are considerably more 
elongated than clusters, and they suggested that the poorest groups and/or subclumps 
in filaments may still be in the process of getting assembled through galaxy infall 
along {\bf a} filament.

The group environment can play an important role in the evolution of its members.
A variety of gravitational (tidal) and hydrodynamical mechanisms are at work in groups and 
clusters that can 
severely alter the galaxy properties by modifying their original morphology,
triggering star formation and/or nuclear activities 
(Mamon \cite{mamon07})
According to the hierarchical scenario for the structure formation 
today rich clusters could have been assembled from groups of galaxies. 
Therefore, groups of galaxies may 
represent sites for a \lq preprocessing\rq~ stage of the cluster galaxies, 
through some varieties of tidal interactions 
(e.g. merging of galaxies in slow collisions) otherwise ineffective in high velocity 
dispersion environments (Boselli \& Gavazzi \cite{boselli06}).
Galaxies of both low mass and low density are expected to reflect the environmental
influence on their evolution most prominently (Lake \& Moore \cite{lake99}). 

In rich clusters of galaxies, the impact of the environment on the galaxy
properties has already been well studied while the 
link between the environment and the galaxy evolution in poor groups is
still not fully understood and suffers from the shortage of observational
information.  Several recent studies have focused on the photometry of
dwarf galaxies in nearby groups (Bremnes et al. \cite{bremnes98}, \cite{bremnes99}, 
\cite{bremnes00}, Trentham et
al. \cite{trentham01}) and on the photometry of local field dwarfs (Barazza et al.
\cite{barazza01}, Parodi et al. \cite{parodi02}). These authors have shown
that the dwarf irregular galaxies which reside in a low density environment (e.g. in loose groups 
or in the field) have a statistically
lower scale length (and, consequently, higher central surface brightness) 
at a given luminosity than those galaxies residing in high density environments.
They argue that this could be an effect of a difference in the star-forming histories, 
in that the higher SB dwarfs 
in low density environments are also found to be bluer, or the photometric difference between the 
field/group and cluster dwarfs could primarily be a structural difference in that the larger scale 
length of cluster dwarfs could plausibly be an effect of frequent tidal encounters (harassment) in 
dense environments (Parodi et al. \cite{parodi02}). 

This paper is the first of a series addressed to the investigation to the galaxy properties 
in a sample of about ten poor/loose groups of galaxies which are located in low density environments, 
i.e. of reasonably isolated groups. The groups will be studied on the available sky surveys (DPOSS, SDSS) 
and by means of our own CCD imaging in at least three optical bands. The studied groups are selected 
in the redshift range of 1000 $\la cz \la$ 4000 km~s$^{-1}$ in order to map them optically with 
reasonable number of pointings. The study also aims to contribute to the detection and photometric 
charcterization of the faint galaxy population within the group area whose membership will be determined 
by follow up redshift surveys (e.g. Hopp et al. \cite{hopp07}). Some preliminary results of this ongoing 
project are presented in Vennik \& Tago (\cite{vt07}) and in Vennik \& Hopp (\cite{vennik07}).

Here, in the first paper of a series we present a search for new members and also a detailed study of the members
and new candidates in the area of the IC~65 group of galaxies.  This poor 
grouping of late type galaxies around the luminous spiral IC~65 is located
unfavourably close to the zone of strong Galactic extinction. Consequently, it has received little
attention in previous studies, and very limited optical
information exists only for the bright galaxies of this group. However, the \ion{H}{i} aperture
synthesis studies of van Moorsel (\cite{moorsel83}) provide detailed \ion{H}{i}
distribution maps, which reveal many irregularities in the \ion{H}{i} distribution
of the bright members.  The maps give hints also for additional dwarf galaxies undetected so far. 
This motivated us to carry out our own
dedicated optical observations which we supplement by an analysis of the
available near- and far-infrared and radio data as well as of the published \ion{H}{i} imaging and
kinematics.

The main purposes of the paper are (1) to search for new (dwarf) galaxies in the area of this 
particular group and to 
discuss their membership probabilities; (2) to obtain a homogeneous
photometric database and to derive structural characteristics of the bright
(certain) and faint (possible) group members; (3) to search for possible
effects of mutual interactions on the optical and \ion{H}{i} morphology and
star formation characteristics of luminous and faint group members; (4) to
use the new photometry and available kinematical data to analyse the dynamics 
and the evolutionary state of the group. 

The selection and classification of new group members are described in Section 2,
CCD observations and data reduction are presented in Section 3, data analysis
galaxy-by-galaxy is given in Section 4, and the dynamical and structural characteristics of the
group are discussed in Section 5. The results are summarised in Section 6.
Throughout this paper a Hubble parameter $H_0$ = 75~km~s$^{-1}$Mpc$^{-1}$ is
assumed. All magnitudes are given in the Vega-magnitude system. 

\section {The IC 65 group of galaxies} 

Probably the first study of this group dates back to van Moorsel 
(\cite{moorsel83}, hereafter vM83) who
carried out 21 cm neutral hydrogen observations of a sample of selected double
galaxies.  He included the pair of IC~65 with UGC~622 in his sample. In that
field two additional systems containing \ion{H}{i} were detected, namely UGC~608
and an unclassified \lq\lq edge-on\rq\rq galaxy. The latter is partly hidden by the bright
$7^{\mathrm th}$ $B$-magnitude blue (spectral type B8) Galactic star HD 5764 (BD+47
272). This late type \lq\lq edge-on\rq\rq galaxy was later catalogued as PGC/LEDA 138291.
\begin{figure}
\resizebox{\hsize}{!}{\includegraphics{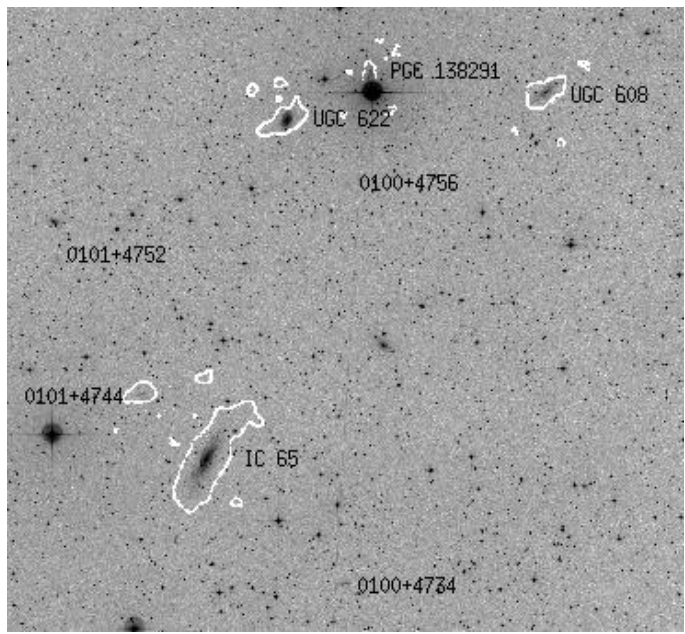}}
\caption{ DPOSS blue image of the IC 65 (LGG 16) group of galaxies: four bright group members 
are labelled with their catalogue numbers; four new probable dwarf galaxies are identified with 
their coordinates. The \ion{H}{i} contours (white)  have been obtained from vM83 and 
correspond to the 2.5~-~3 $\sigma$ noise level (2.3 - 8.6 $\times 10^{20}$ atoms~cm$^{-2}$). 
The size of the image is $60' \times 60'$. The north is at the top and the east is to the left.}
\label{LGG16_HI.fig}
\end{figure}
This dense association of four galaxies has not been included in earlier
catalogues of nearby groups of galaxies because of 
its location near to the Zone of Avoidance ($b \simeq -15^{\circ}$), and
because most of its members are relatively faint ($B >$ 14.5 mag) 
except the luminous principal galaxy IC~65.
Garcia (\cite{garcia93}) assigned
IC 65, UGC 608 and UGC 622 to his Lyon Group of Galaxies (LGG) 16. Later on he
showed that this particular group of three galaxies fulfils the selection criteria
for Hickson compact groups (Garcia \cite{garcia95}). 
The \ion{H}{i} observations of vM83 show that the principal galaxy IC 65
has at least one \ion{H}{i}-rich LSB anonymous dwarf companion, which is barely
visible on POSS plates.  This encourages us to look for further dwarf
companion candidates in the area of the group.

\subsection*{Search for additional members}

We carried out a systematic search for new dwarf members of the IC 65 group utilizing
the Digitized Second Palomar Observatory Sky 
Survey\footnote{http://astro.ncsa.uiuc.edu/catalogs/dposs/} (DPOSS).  
The blue (type IIIaJ) and red (IIIaF)
emulsions of POSS II films have a small grain size with a resolving power
about 250 lines/mm (Reid et al. \cite{reid91}), which leads to highly uniform
sky background. Consequently, the long-exposure Schmidt plates that reach a
stellar limiting magnitude of $B_{\rm J,lim} \sim$ 23 mag, are especially
suited for detecting low-surface-brightness (LSB) features (Binggeli et al.
\cite{binggeli85}).  We extracted a 60$\times$60 arcmin$^2$ field centered on
the position of LGG 16 with RA(2000) = $01^{\mathrm h} 00^{\mathrm m} 30\fs1$ 
and DEC(2000) =
$+47^{\circ}48\arcmin 38\arcsec$ .  Because of its low Galactic latitude this field suffers heavy
($A_B \simeq$ 0.66 mag, $A_R \simeq$ 0.4 mag, Schlegel et al.
(\cite{schlegel98})), and non-uniform Galactic extinction. Also, the area is
crowded by a large number of Galactic stars. Both factors make the
detection of the LSB features more difficult.

\begin{figure}
\hspace{2mm}
\resizebox{0.35\textwidth}{!}{\includegraphics{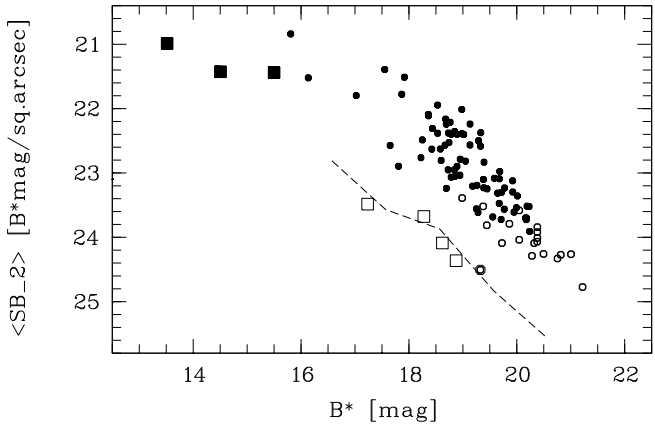}}
\\
\caption{Classification of galaxies in the field of the IC
  65 group. I. The mean surface brightness within $2''$ aperture
$<SB\_2>$ as a function of the total $B^*$-magnitude. The dashed line
shows the predicted position of dwarf galaxies according to
Ferguson \& Binggeli (1994, Fig.~3) (see the text for details).
Coding: {\it filled squares} - certain group members;
{\it open squares} -
large LSB galaxies -- probable new dwarf members of the group;
{\it open circles} - other LSB galaxies in the field;
{\it filled circles} - HSB galaxies in the field.
}
\label{MB_SB2.fig}
\end{figure}

\begin{figure}
\resizebox{0.36\textwidth}{!}{\includegraphics{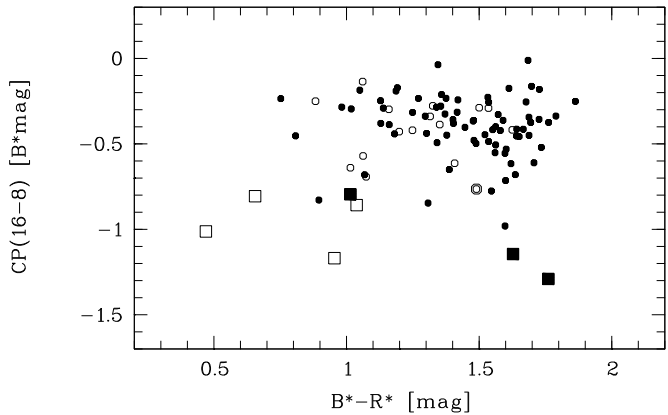}}
\\
\caption{Classification of the galaxies in the field of the IC
  65 group. II. Light concentration parameter, as defined in the text,
 versus central $B^*-R^*$ colour index within an 2$''$
  aperture. Coded as in Fig~\ref{MB_SB2.fig}.}
\label{BR_OCP.fig}
\end{figure}

The detection of galaxies depends on their apparent surface brightness ($SB$)
and on their apparent diameters. Galaxies with extremely low $SB$ 
disappear into the night sky background, while small galaxies can be hardly
distinquished from the stars.
A preliminary visual inspection of the selected area 
revealed four new large LSB dwarf galaxy candidates, which are
distributed within $13\arcmin$ ($\sim$ 146 kpc at the group distance)
around of the barycentre of the LGG 16. Based on their irregular morphology (LSB
irregular contours, no pronounced central light concentration but allowing for
luminous knots distributed in periphery), apparently blue colour,  
and relatively large diameters, we classified these
LSB galaxies for probable new dwarf member candidates of the IC 65
group of galaxies.
Eventually we registered a number of smaller LSB objects visible
both on blue and red frames.  The poor scale of the POSS II ( 67 $''$/mm) does
not permit to classify them confidently as dwarf galaxies, 
but we registered them as candidates for further investigation with
higher resolution.

\begin{table*}
\caption{Basic data of the observed galaxies}
\label{Basic.tab}
\centering
\begin{tabular}{llllcrrlrll}
\hline\hline
 Galaxy     & RA[2000]  & Dec[2000] & Type     & $D \times d$& $V_{\odot}$ & $V_\mathrm{rot}^\mathrm{m
ax}$ & $B_\mathrm{T}$ & $K_\mathrm{T}$ & $M_B^{b,i}$ & $M_K^{b,i}$ \\
 &&&&& \multicolumn{2}{c}{[km~s$^{-1}$]} & [mag] & \multicolumn{2}{c}{[mag]} & [mag] \\
\hline
 (1)        &     (2)      &    (3)      &  (4)     &   (5)       &  (6)   &     (7)   &  (8)  &
  (9) & (10) & (11) \\
\hline
\\
 UGC 608    & 00$^h$59$^m$02\fs 3 & 48$^o$01$'$02\farcs 3 & SABdm & 2\farcm 0~~0\farcm 9 &
 2755 & 104 & 15.42 & 12.2 & -18.67 & -20.8 \\
 PGC 138291 & 01~~00~~00.7 & ~48~~02~~14 & S(dm)    & 1.3:~~0.3: & 2598 & 62 & 17.2: &
 & -17.5: \\
 A0100+4756 & 01~~00~~08.0 & ~47~~56~~05 & dIrr     & 0.6:~~0.4: &   &      & 19.1: && -14.5
: \\
 A0100+4734 & 01~~00~~10.4 & ~47~~34~~05 & dIrr     & 0.8:~~0.3: &  &      & 17.55 && -16.1
: \\
 UGC 622    & 01~~00~~28.1 & ~47~~59~~42 & Scd:     & 1.1~~~0.7  & 2714 & 154 & 14.42 & 10.25
& -19.50 & -22.82 \\
 IC 65      & 01~~00~~55.5 & ~47~~40~~54 & SAB(s)bc & 4.4~~~1.2  & 2614 & 168 & 13.33 & 9.51 &
 -21.19 & -23.60 \\
 A0101+4744 & 01~~01~~16.1 & ~47~~44~~33 & dIrr     & 0.7:~~0.2: & 2760 & 65: & 18.20 & &
-15.4 \\
 A0101+4752 & 01~~01~~44.4 & ~47~~52~~06 & dIrr     & 0.4:~~0.3: &  &   & 18.7: && -14.9
: \\
\hline
\end{tabular}
\end{table*}

To quantify the selection criteria for new possible group
members, we attempted to linearize and calibrate the used DPOSS frames. 
For the density-to-intensity transformation 
we applied the average 
characteristic curves of representative IIIaJ and IIIaF plates as
given in Reid et al. (\cite{reid91}).  
The photometric zero points 
were determined comparing the $SB$
profiles of the galaxies UGC 608 and UGC 622 from the linearized DPOSS images 
with those determined in
exactly the same way from the CCD images (see Section 3).
We mark the calibrated DPOSS blue and red 
magnitudes as $B^*$ and $R^*$, respectively.
Detailed photometric calibration of the DPOSS has been published recently by
Gal et al. (\cite{gal04}), however, the IC 65 group area has not been
covered with their CCD calibration pointings, as evident in their Fig.~3.

The detection of objects has been performed on the linearized and calibrated blue
DPOSS image with the SExtractor software (Bertin \& Arnouts,
\cite{bertin96}). We conducted experiments to optimize between the
detection of as many LSB galaxies as possible and
minimizing the impact of spurious detections on the resulting catalogue
of candidate galaxies. 
Finally, all objects with a fixed $SB$ threshold 25.5 $B^*$mag~arcsec$^{-2}$ ($\sim 1.4 rms$ 
above the sky background level) and with the minimum consecutive area of 15 pixels were detected. 
The catalogue still contains some spurious detections, among them
satellite trails and emulsion flaws. Most of those can easily be 
identified and removed from the catalogue by comparing the blue and red
image. Therefore, we aligned and rescaled the 
red image to the blue one. SExtractor was now used in its
double-image mode, taking the detection from the blue frame and the
photometry from the red one.  This procedure allows to
remove image artefacts and also yields pixel-to-pixel colour maps for
securely detected objects. In total, 12275 objects were selected within
a 3600 arcmin$^2$ area.  
The accuracy of the photometric data
obtained this way from linearized DPOSS frames is 
sufficient to put the final object selection and classification to a 
more quantitative basis.

The final galaxy selection in the SExtracted catalogue was made in
several steps. First, we attempted to separate non-stellar objects
applying three selection criteria: the stellarity index 
$<$ 0.8, 
the image major diameter $> 4\arcsec$ (considering the stellar FWHM $\simeq 3\arcsec$), and
the mean surface brightnesses $<SB>$
$>$ 23 $B^*$mag~arcsec$^{-2}$, within 25.5 $B^*$mag~arcsec$^{-2}$ isophote.
This reduced the original catalogue by almost two orders of magnitude
as expected for the low Galactic latitude of the field.  A total of
348 nonstellar objects remained. 
In a crowded sky such as the one of the IC 65 group, and on the poor
photographic material, the automatic selection is 
insufficient. We need to take the morphologies into
account. Therefore, all 348 extended objects were carefully inspected
by eye on both the blue and red DPOSS frames and finally classified
according to their appearance on both frames.
In effect, a final list of 105 galaxies with diameters greater 
than 4\arcsec was established.
The remaining 243 objects were
classified as plate flaws (42), satellite track or stellar spike
fragments (97), or partly overlapping faint stellar images (104). 

To disentangle the group dwarf members from the background
field galaxies, we applied the following general considerations. 
First, we consider that according to Binggeli
(\cite{binggeli94}) both the dwarf ellipticals and dwarf irregulars
follow the common absolute magnitude -- central $SB$
correlation (but see also Carrasco et al. (\cite{carrasco06}) on this issue). 
In Fig.~\ref{MB_SB2.fig} the central $SB$ measured by
SExtractor within a 2$''$ aperture is plotted versus
the apparent total $B^*$ magnitude.  The dashed line shows the 
correlation as derived by Ferguson \& Binggeli (\cite{ferguson94},
Fig. 3), transformed to the distance of the IC 65 group, and corrected for
the Galactic absorption. 
Evidently, five LSB
galaxies of our catalogue 
closely fit this empirical relation.

A second consideration for separating group and field galaxies 
uses the lower light concentration in dwarf galaxies when compared 
to the HSB background galaxies. As suggested by Trentham et al. (\cite{trentham01}),   
we can define the light concentration parameter ($CP$) 
as a difference between various aperture magnitudes 
measured with SExtractor software. After attempting with different
aperture combinations we decided to use the concentration parameter 
as the difference of integrated $B^*$ magnitudes 
within $16''$ and $8''$ apertures, respectively:
\begin{equation} 
CP(16-8) = B^*(16^{\prime\prime})-B^*(8'').  \protect\label{CP}
\end{equation} 
The $CP$ is more negative for galaxies of larger scale length,  
i.e. of lower $SB$, for a given apparent magnitude and $CP$ is close to
zero for stars. 
At the distance of the IC 65 group of galaxies 
the chosen $CP$ characterizes light distribution on linear scales 
between about 0.75 kpc and 1.5 kpc. 
In Fig.~\ref{BR_OCP.fig} $CP(16-8)$ is plotted versus 
$B^*-R^*$ colour index measured again within a 2$''$ aperture. The
luminous group members (filled squares) show low concentrations $-1.3
< CP < -0.8$, as expected. Four of the five LSB dwarf galaxy
candidates selected in Fig.~\ref{MB_SB2.fig}, classify
into the category of low concentration galaxies. Furthermore, these
four galaxies show the bluest central colours of the sample ($0.45 <
B^*-R^* <$ 1.05), indicative of being star-forming dwarf irregular
galaxies. For one of these four LSB galaxies vM83 estimated 
a heliocentric radial velocity $V_{\odot}$ = 2760 km~s$^{-1}$ in \ion{H}{i} line, which 
is in accord with the mean velocity of the four bright group members 
$<V_{\odot}> = 2670 \pm 76$ km~s$^{-1}$, obtained from the NED\footnote{http://nedwww.ipac.caltech.edu}. 
These arguments strengthen our opion that all first four LSB
galaxies are indeed new dwarf companions of the IC~65 group. We rate
them as confidence class 1.  
The fifth dwarf galaxy candidate, selected in
Fig.~\ref{MB_SB2.fig} (and marked with a double circle 
in Figs.~\ref{MB_SB2.fig},~\ref{BR_OCP.fig}) 
appears significantly redder ($B^*-R^* \simeq$
1.5). It is also more distant from the group centre compared to other
four candidates and therefore it was assigned a lower membership probability.
Furthermore, there are 17 galaxies with LSB morphologies, but
all of them with isophotal magnitudes fainter than 19.0 $B^*$mag. 
At these fainter magnitudes and/or smaller diameters we actually lose 
the ability to distinguish group members from background galaxies on $SB$
grounds. Therefore, we classify these 17 LSB galaxies, with caution, as
possible background galaxies (rated 2). 
The majority of the selected galaxies 
are HSB and/or red galaxies and can therefore be classified with confidence as  
probable field galaxies (rated 3), located in the
background of the IC 65 group.

{\it To summarize, we have now a list of five certain group members, confirmed by redshift,
supplemented by three new LSB probable dwarf member
candidates.}  Table~\ref{Basic.tab} contains basic data for the
group galaxies: 
(1) name of the galaxy, anonymous galaxies start with an \lq\lq A\rq\rq, (2, 3) their 2000.0 epoch RA and DEC, 
(4) morphological type either obtained from the RC3\footnote{Third Reference Catalogue of Bright Galaxies 
(de Vaucouleurs et al.~\cite{rc3})} or our classification, (5) major and 
minor diameters, (6) heliocentric radial velocity from the 
NED; for A0101+4744 the velocity 
is obtained from vM83. (7) maximal \ion{H}{i} rotational velocity, taken from vM83, 
(8) total $B$ and $K_s$ magnitudes from our photometry, (9, 10) absolute $B$ and $K_s$ magnitudes 
calculated for the distance modulus $m-M$ = 32.93 and corrected for the Galactic and internal 
(following Karachentsev et al.(\cite{karachentsev04})) absorption. 

\section {Observations and data reduction}

\subsection {Optical data}

Broad band $B$ (Johnson) and $R, I$ (Cousins) frames were obtained
during two observing runs in 1995 and 1999 with the Calar Alto 1.23 m
telescope. Its CCD camera is equipped with the 1024 $\times$ 1024
pixel chip that yields a field of view of $ 8.7' \times 8.7'$ and a
spatial scale of 0.51 $''$/pixel.  The seeing varied between 1.5$''$
and 2.5$''$ (FWHM).  More observational details are summarized in
Table~\ref{Obs}.

\begin{table}
\caption{\normalsize Log of the direct imaging with the Calar Alto (CA)
1.23 m telescope}
\label{Obs}
\begin{tabular}{clclc}
\hline\hline 
Date & Detector & Band & Exposure & Sky  \\
&&& [sec] & [mag/$\Box''$] \\
\hline
\\
01-02.09.95   & Tek\#6     & $B$ & 1800           & 21.5 \\
              &            & $R$ & 1200           & 20.7 \\
16-20.01.99 & Tek\#7c-12   & $B$ & 300, 600       & 22.3 \\
              &            & $R$ & 100, 150, 600  & 21.5 \\
              &            & $I$ & 150, 600       & 19.7 \\
\hline
\end{tabular}
\end{table}

The observed frames were reduced by using the ESO MIDAS\footnote{MIDAS is developed 
and maintained by the European Southern Observatory} software package. 
The raw CCD frames were bias subtracted and flat-fielded using twilight
sky flats. An additional defringing procedure was applied to the $I$
band images. Cosmic ray hits were removed using the FILTER/COSMIC task.
Then, images of the same filter were registered and co-added.
 
Unfortunately, the observing session in 1995 was under non-photometric
weather conditions. Two nights in January 1999 (the 18th and 19th)
were of photometric quality and the data obtained during these nights
were calibrated by means of 138 measurements of standard stars in the
star cluster NGC 7790 (Christian et al. \cite{christian85}).
Deep images 
obtained under non-photometric conditions have been calibrated by
means of comparing total fluxes of a sample of unsaturated stellar
images on both photometric and non-photometric frames.

\subsection{Near-infrared data}

In addition to our optical observations in $B, R$ and $I$
passbands, we extracted calibrated, full resolution 
$J, H$ and $K$ frames with 1.0$'' \times 1.0''$ sampling from the 
2MASS\footnote{http://www.ipac.caltech.edu/2mass/} archive
(Jarret et al. \cite{jarret00}).
As the NIR domain is largely dominated by stellar 
radiation and much less affected by extinction than the optical bands,
they are sensitive to population changes and can uncover structural
properties not visible in the optical bands either due to
extinction or gaseous emission or both. Therefore, we included
the archival NIR survey data in an attempt to further explore 
morphological components of galaxies and to enable a
multi-wavelength comparison of their structural parameters over a wide
wavelength range from almost 400 to 2200 nm. 

The surface photometry performed on both the optical and NIR images,  
the relevant error analysis and profile fitting are described in Appendix~A 
and are only available in electronic form. 

\begin{table*}
\caption {Model-free photometric parameters of the observed galaxies. 
{\it Col.~1:} name of the galaxy. {\it
Cols.~2,3:} equivalent radii in arcseconds of the effective and 25$^{th}~
B$ mag~arcsec$^{-2}$ isophotes, respectively. {\it Cols.~4,5:} central $SB$
and $SB$ at the effective radius (i.e. $\mu^{\rm eff} = \mu(r^{\rm eff}$)),
respectively.  {\it Col.~6:} total $B$ magnitude within $r_{25}$. {\it
Col.~7:} asymptotic $B$ magnitude. {\it Cols.~8-10:} total $B-R$,
$R-I$ and $B-J$ colour indices, respectively. {\it Col.~11:} light concentration
index $c_{31} = r(3/4L_{\rm T})/r(1/4L_{\rm T})$, defined by de Vaucouleurs (\cite{dev77}). 
{\it Cols.~12, 13:} galaxy
mean minor-to-major axis ratio ($b/a$), and position angle ($P.A.$),
determined as an average between the 24 and 25.5 $B$mag~arcsec$^{-2}$
isophotes. {\it Col.~14:} average {\it decentering degree (decen.)} in
the optical passbands, calculated as the displacement
of the centre of the external isophotes near the Holmberg radius
(at 26.5 $B$mag~arcsec$^{-2}$) with respect to the centre
of the innermost ellipse as defined in Marquez \& Moles
(\cite{marquez99}).}
\label{Model-free.tab}
\centering
\begin{tabular}{lccccccccccccc}
\hline\hline
 Galaxy &$r^\mathrm{eff}_B$ & $r_{25,B}$ & $\mu^0_B$ & $\mu^\mathrm{eff}_B$ & $B_{25}$ & $B_T$ & $B-R$ & $R-I$
 & $B-J$ &
$c31$ & $b/a$ & $P.A.$ & decen.\\
& \multicolumn{2}{c}{[arcsec]} & \multicolumn{2}{c}{[mag~arcsec$^{-2}$]} & [mag] & [mag] & [mag] & [mag] & [mag] &&& [$^\circ$] &\\
\hline
 (1) & (2) & (3) & (4) & (5) & (6) & (7) & (8) & (9) & (10) & (11) & (12) & (13) & (14)\\
\hline
\\
 UGC 608   & 11.1 & 26  & 20.8 & 23.11 & 15.08 & 14.75 & 0.74  & 0.33 & 1.60 &  3.50 & 0.42 & 128 & 0.8:
\\
PGC 138291 & 7.7  & 13: & 21.9 & 23.24  & 16.7: & 16.5: & 1.0: & 0.2: && 2.1  & 0.14 & 165 \\
A 0100+4756 & 6.3  & 4:  & 23.9:& 25.0  & 20.3: & 18.5  & 0.76:  & 0.2: && 2.15 & 0.6: & 40: \\
A 0100+4734 & 9.1  & 8.4 & 21.7 & 24.4  & 17.65 & 16.85 & 0.74   & 0.25 && 2.56 & 0.4  & 108 \\
 UGC 622   & 14.4 & 34.6& 20.4 & 21.86 & 13.86 & 13.76 & 1.28  & 0.61 & 2.81 & 1.90 & 0.65 & 159 & 6.3
\\
 IC 65     & 23.5 & 54.4& 19.7 & 22.10 & 12.80 & 12.69 & 1.07  & 0.56 & 2.47 & 2.81 & 0.27 & 153 & 8.2
\\
A 0101+4744 & 6.8  & 7.8 & 22.15 & 24.15  & 18.07 & 17.55 & 0.80  & 0.1: && 2.36 & 0.33 & 100 \\
A 0101+4752 & 5.4  & 6.0 & 23.05 & 24.15  & 18.63 & 18.00 & 0.65   & 0.2: && 2.14 & 0.75 & 17: \\
\hline
MCG+8-3-3 & 14.2& 24.7& 20.5:  & 23.32 & 14.89 & 14.60 & 1.51  &      && 3.52 & 0.65 & 49  \\
MCG+8-3-6 & 8.8 & 18.3& 20.4: & 22.55  & 15.27 & 15.13 & 0.97  & 0.52 && 2.60 & 0.66 & 72 \\
\hline
\end{tabular}
\end{table*}
\begin{table*}
\caption {Exponential model parameters of the observed galaxies. {\it Col.~1:}
name of the galaxy. {\it Cols.~2-4:} exponential model central $SB$ in $B, R$
and $I$ band, respectively, corrected for the Galactic
absorption. {\it Cols.~5-7:} exponential model scale length in $B, R$ and
$I$ band, respectively. {\it Cols.~8-10:} difference between the total light
emitted by the model exponential disk and the measured
asymptotic total light ($\Delta m = m^{\rm exp} - m_{\rm T}$) in $B, R$ and $I$
band, respectively.}
\label{Model.tab}
\centering
\begin{tabular}{lccccccccc}
\hline\hline
 Galaxy & $\mu_{0,B}^\mathrm{exp}$ & $\mu_{0,R}^\mathrm{exp}$ & $\mu_{0,I}^\mathrm{exp}$ & $h_B$ & $h_R$ & $h_I$ & $\Delta
m_B$ & $\Delta m_R$ & $\Delta m_I$ \\
 & \multicolumn{3}{c}{[mag~arcsec$^{-2}$]} & \multicolumn{3}{c}{[arcsec]} & \multicolumn{3}{c}{[mag]} \\ 
\hline
 (1) & (2) & (3) & (4) & (5) & (6) & (7) & (8) & (9) & (10) \\
\hline
\\
 UGC 608        & 21.93 & 20.98 & 20.90 & 10.0 & 9.20 & 11.0 &  0.19 &  0.03 & -0.17 \\
PGC 138291      & 22.04 & 21.19 &       & 6.50 & 6.90 &      & -0.56 & -0.49 &       \\
A 0100+4756     & 23.56 & 22.80 & 22.51 & 5.50 & 5.60 & 4.40 & -0.60 & -0.64 & -0.31 \\
A 0100+4734     & 22.81 & 22.17 & 21.89 & 6.50 & 7.10 & 7.30 & -0.11 & -0.22 & -0.30 \\
 UGC 622        & 18.84 & 17.89 & 17.50 & 5.40 & 5.90 & 6.30 & -0.58 & -0.45 & -0.36 \\
 IC 65          & 20.53 & 19.61 & 19.31 & 14.6 & 14.7 & 15.5 &  0.02 &  0.13 &  0.26 \\
A 0101+4744     & 22.25 & 21.49 & 21.50 & 3.80 & 3.70 & 4.00 & -0.19 & -0.14 & -0.22 \\
A 0101+4752     & 22.65 & 22.20 & 22.41 & 3.80 & 4.20 & 6.40 & -0.24 & -0.41 & -0.90 \\
\hline
MCG+8-3-3       & 21.84 & 20.37 &       & 10.6 & 10.7 &      &  0.12 &  0.14 &       \\
MCG+8-3-6       & 21.45 & 20.10 & 19.21 & 6.90 & 5.70 & 5.10 &  0.13 &  0.17 &  0.04 \\
\hline
\end{tabular}
\end{table*}

\subsection{Main results of surface photometry}

We discuss the optical morphology (and \ion{H}{i} imaging data, if
available) of each galaxy in Section 4. For
each galaxy we present a grey-scale optical image. 
For IC 65 and UGC 622 useful composite NIR images are given, too.  
We further apply contours of the
Laplacian filtered image (as described in Appendix A.1) to reveal specific morphological
features in luminous parts of the galaxies. We also show 
grey-scale colour index images of the IC 65 and UGC 622  
for which the S/N allowed to construct reliable colour maps.  
These maps are especially useful for visualizing changes in stellar
population and dust distribution. 
Finally, we show $SB$ and colour profiles (see Appendix A.2) of each galaxy. For those bright
galaxies with a regular appearance, we add (only in the electronic version)   
the radial profiles of the parameters of fitted ellipses  
($b/a, P.A.$ and decentering). 

The observed (model-free) photometric parameters, as measured from
radial profiles, are given in Table~\ref{Model-free.tab}. 
All magnitudes are corrected for the Galactic extinction 
obtained from the NED.  Uncertain values are marked with a colon. 
The results of the profile fitting are summarized in Table~\ref{Model.tab}. 
Tables~\ref{Model-free.tab} and~\ref{Model.tab}
also list the photometric parameters of two luminous early type
galaxies MCG+8-3-3 and MCG+8-3-6 in the area of the group but located in its background 
(Hopp et al. \cite{hopp07}).

\section {Description of individual galaxies} 
\subsection {IC 65} 
\begin{figure*}
\resizebox{0.31\textwidth}{!}{\includegraphics{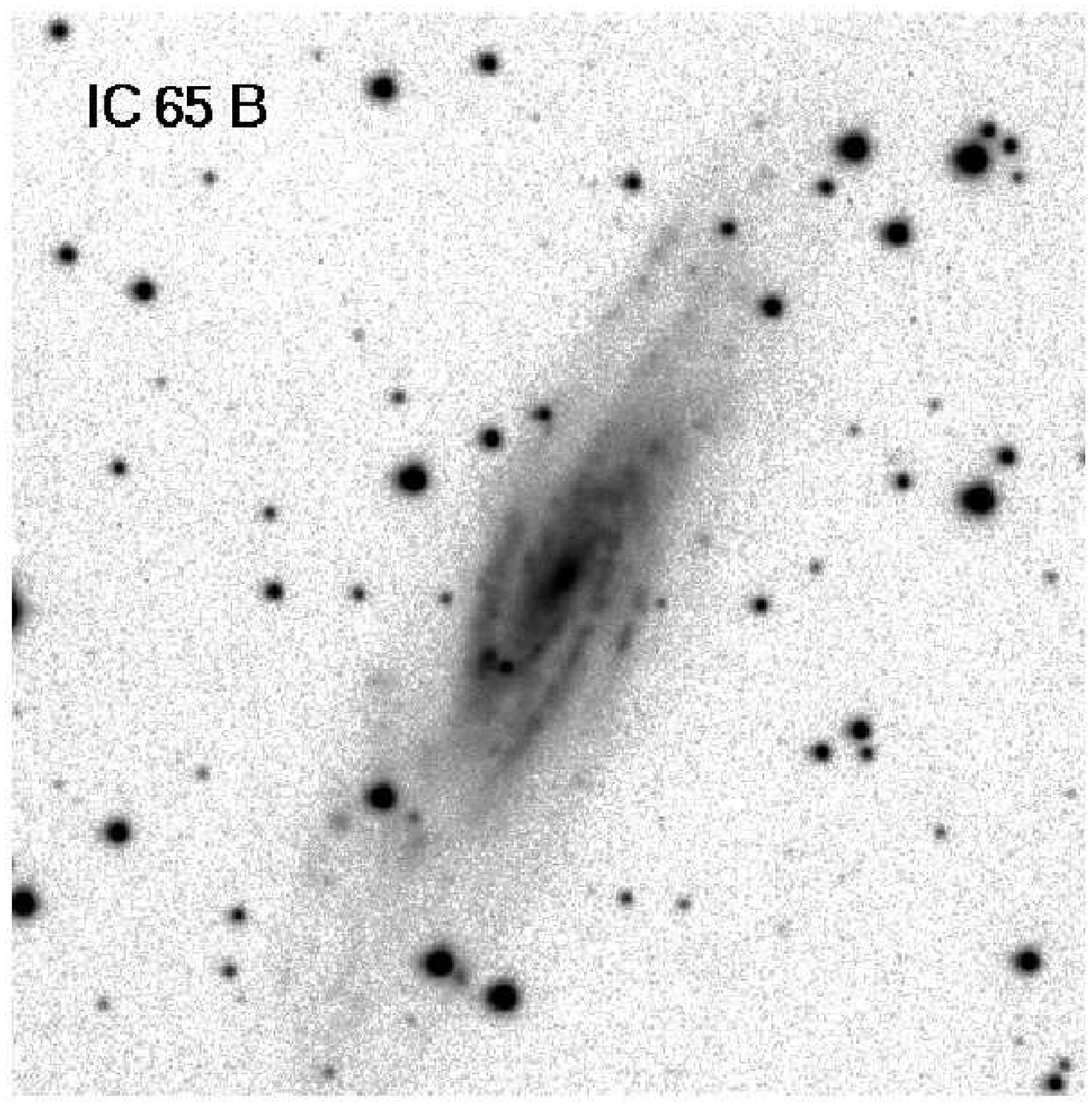}}
\resizebox{0.305\textwidth}{!}{\includegraphics{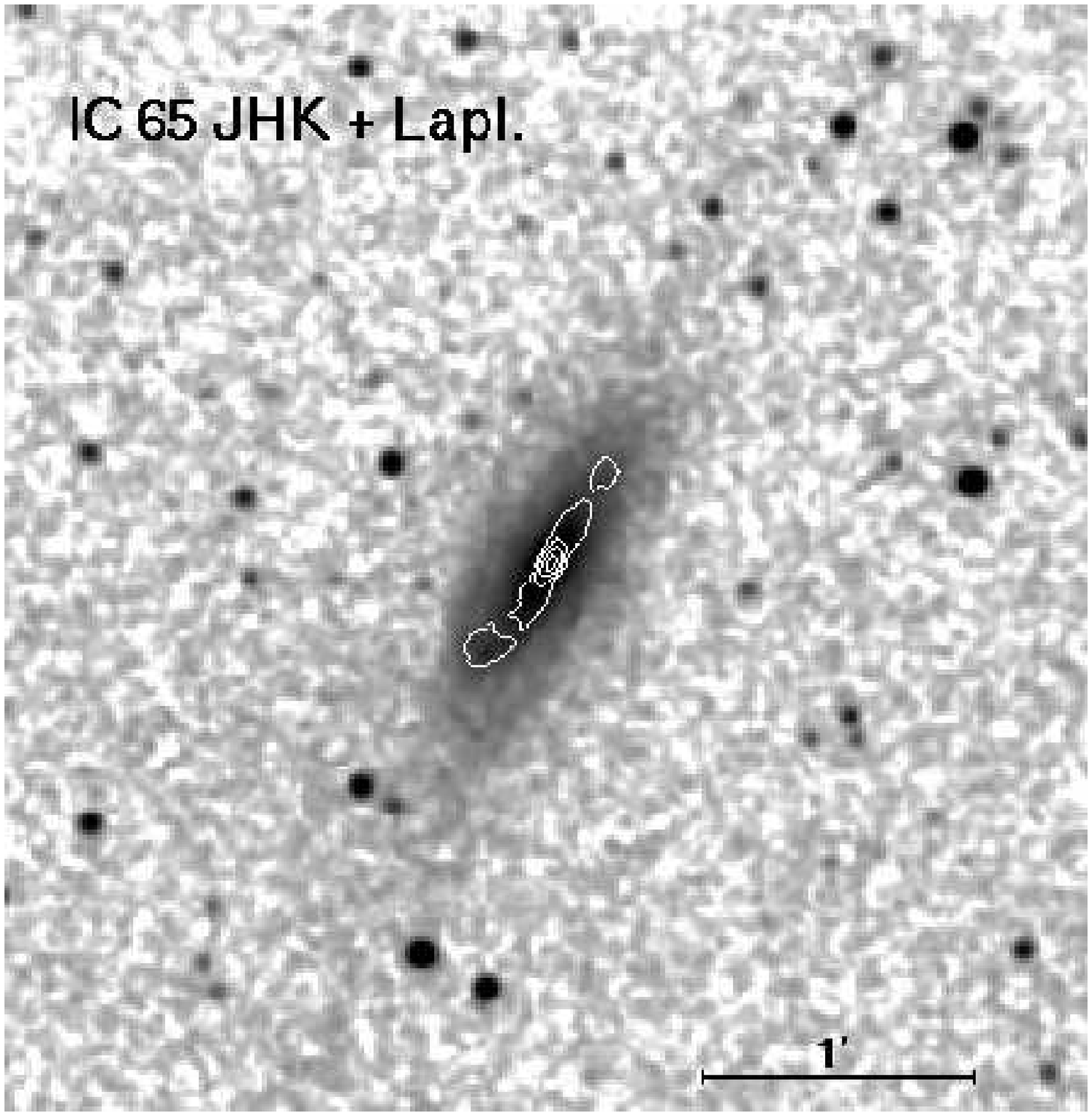}}
\resizebox{0.31\textwidth}{!}{\includegraphics{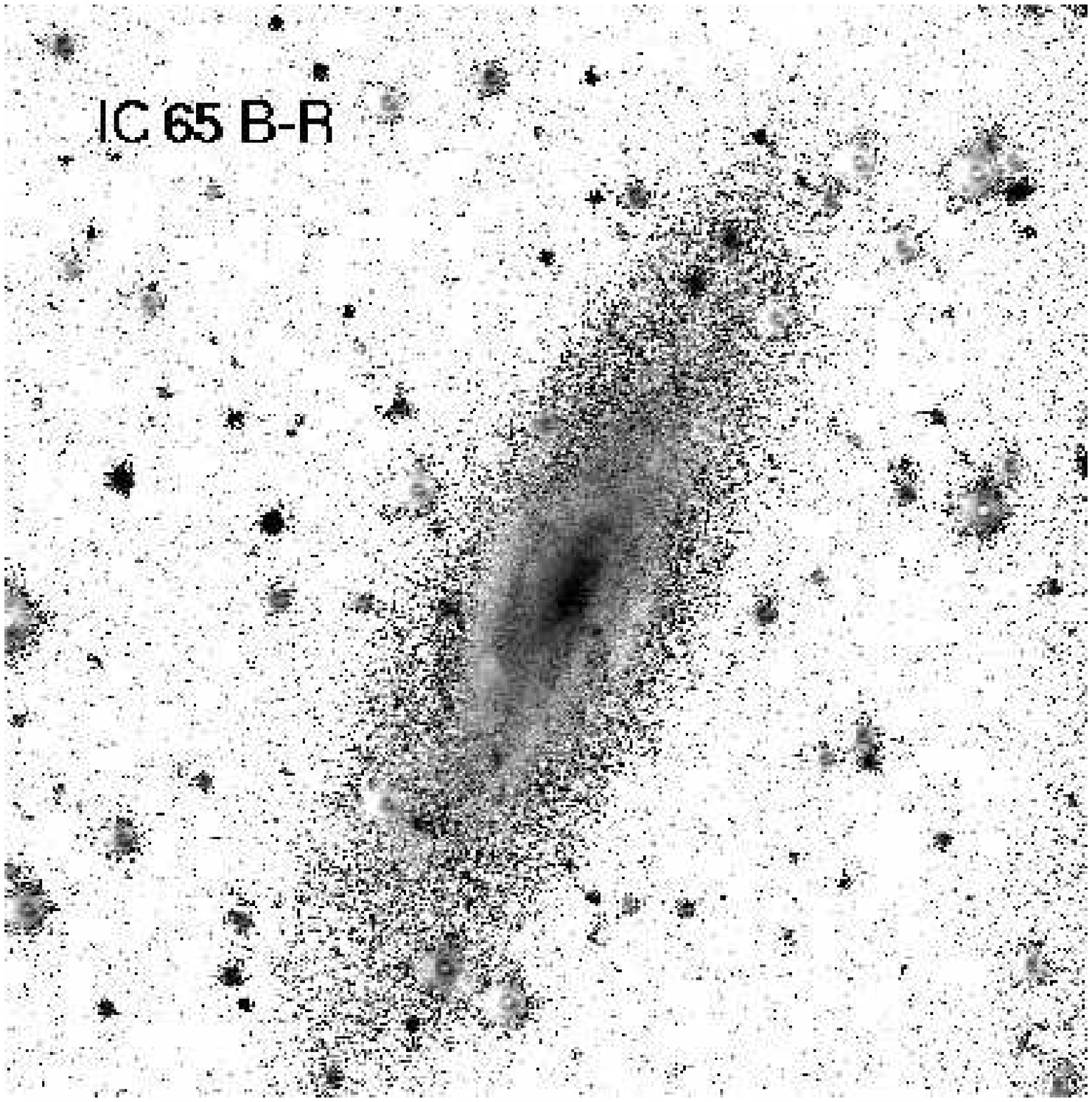}}
\\\\\\
\resizebox{0.36\textwidth}{!}{\includegraphics[width=45cm]{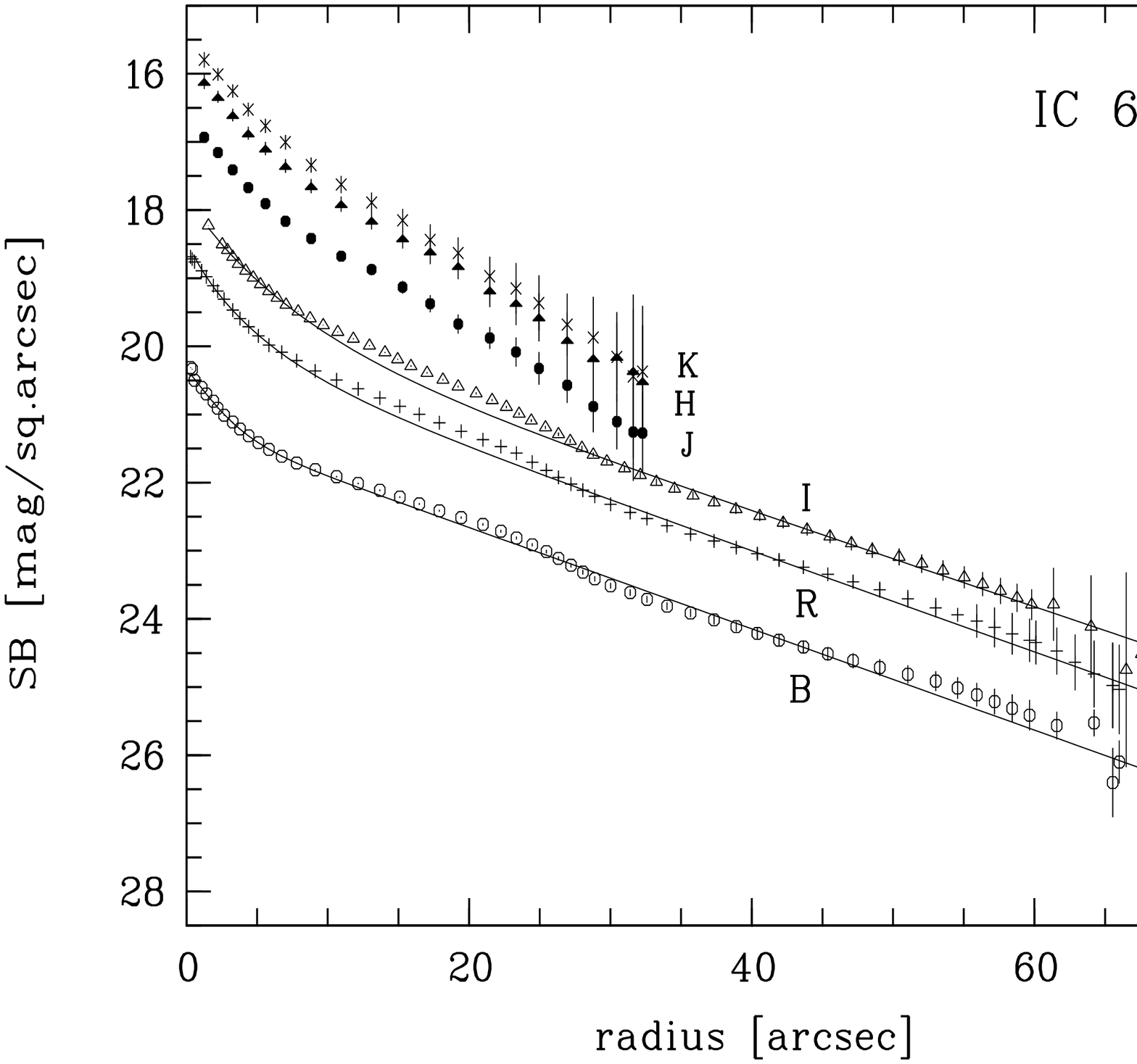}}
\hspace{25mm}
\resizebox{0.36\textwidth}{!}{\includegraphics[width=45cm]{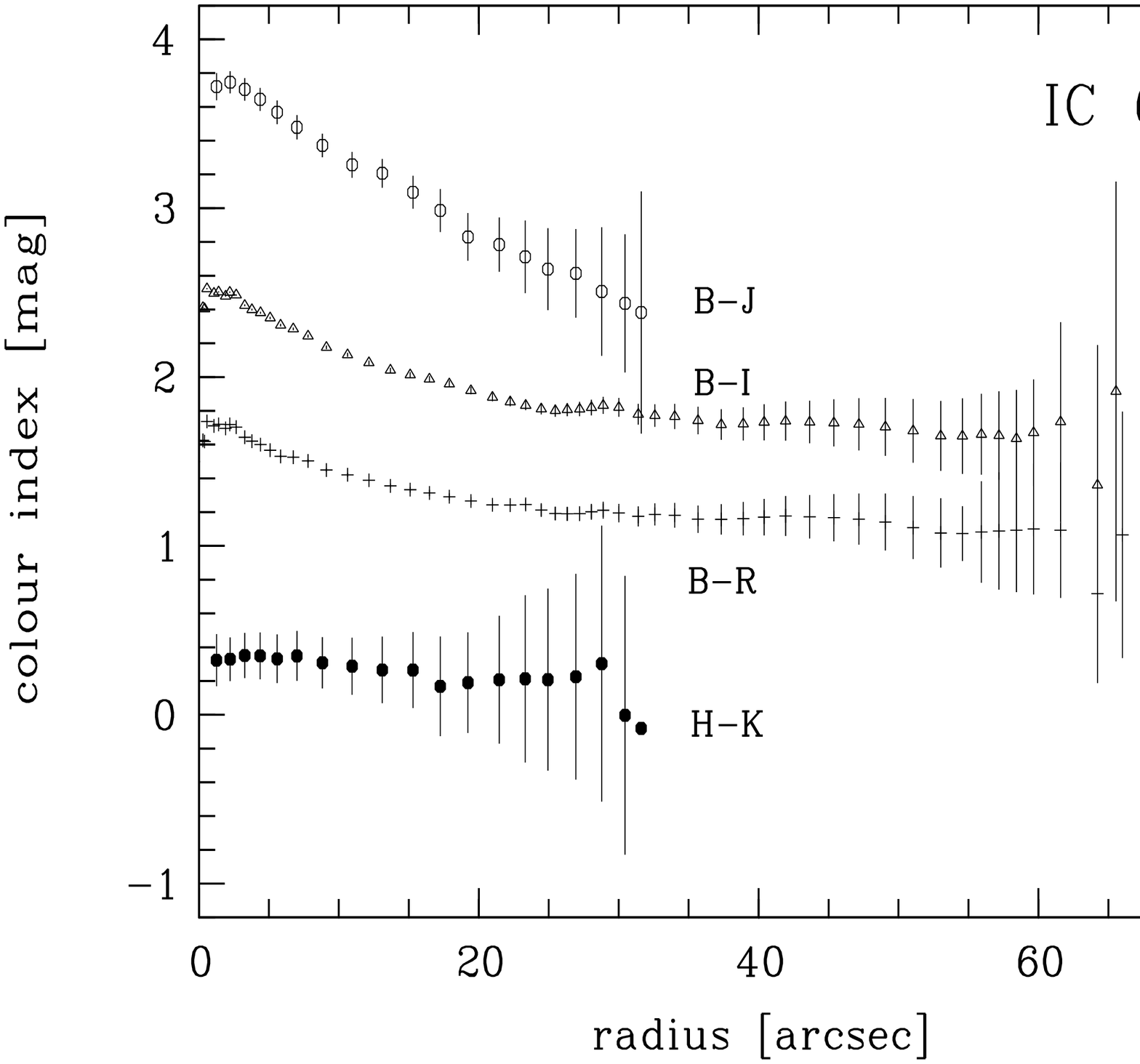}}
\caption{Light distribution in IC 65 I. {\bf 2D-frames.}
{\bf Top left:} The $B$ band CCD image;
{\bf top center:} the $H, J$ and $K$ composite image
              with $K$-band Laplacian-contours which delineate
              the nucleus, the bar, and a possible inner ring;
{\bf top right:} the $B-R$ colour index image in the range 0.7~-~1.8 mag.
              The dark shade is red, the light shade is blue.
The frames have a field size of about $4'\times4'$; the north is at he top and
the east is to the left.
{\bf 1D radial profiles.}
{\bf Bottom left:} Surface brightness ($SB$) profiles in $B, R, I, J$ and $K$.
The continuous lines represent the sum of the 2D-fits of the exponential disk and
bulge components, as described in the text.
{\bf Bottom right:} Colour index ($CI$) profiles, from bottom to top: $H-K,
B-R, B-I$ and $B-J$.
The observed $SB$ and $CI$ profiles are not corrected for Galactic absorption.
The radii are equivalent radii (i.e. $radius = \sqrt{ab})$)}
\label{IC65pics.fig}
\end{figure*}

\begin{figure*}
\resizebox{0.36\textwidth}{!}{\includegraphics{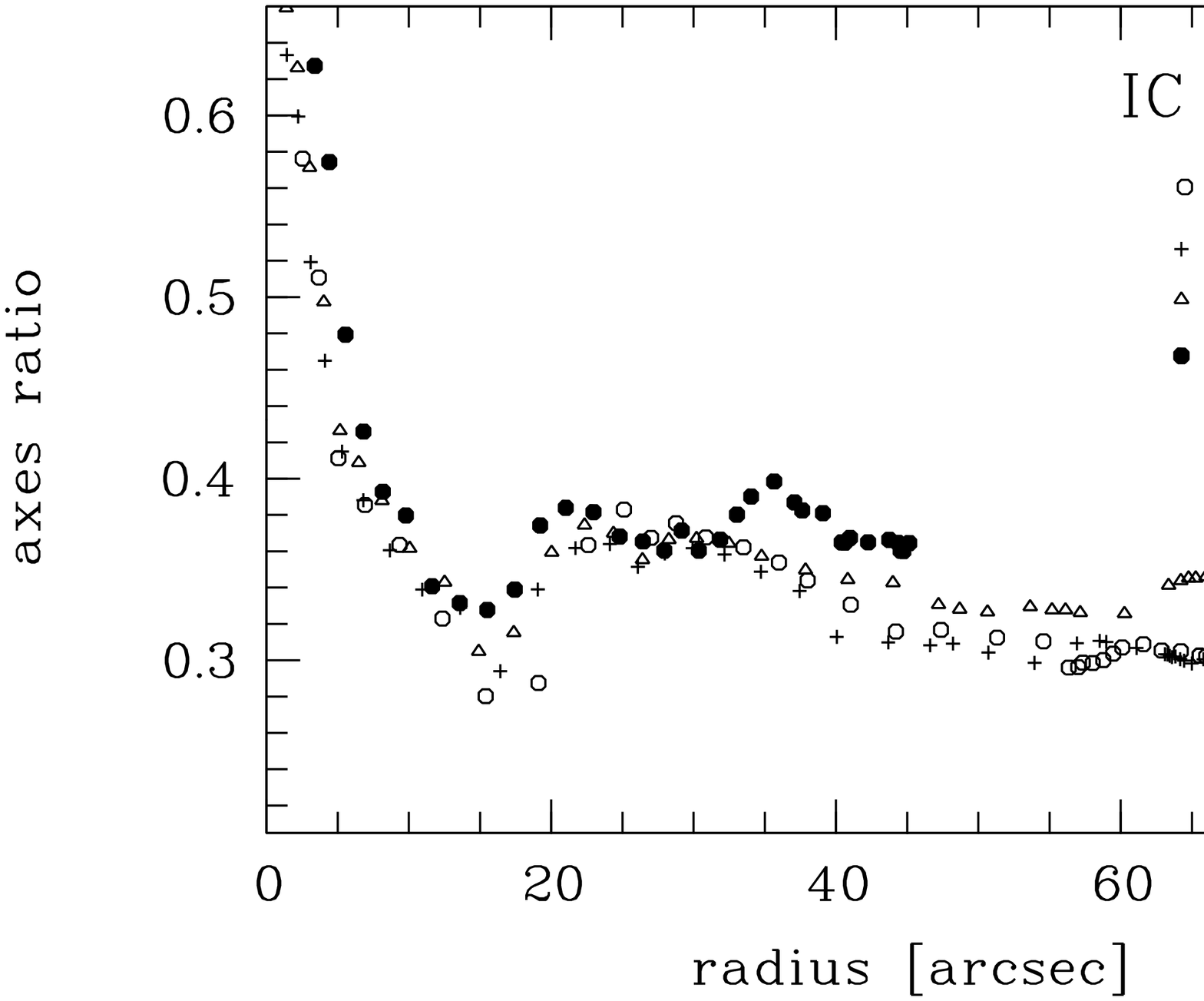}}
\hspace{25mm}
\resizebox{0.36\textwidth}{!}{\includegraphics{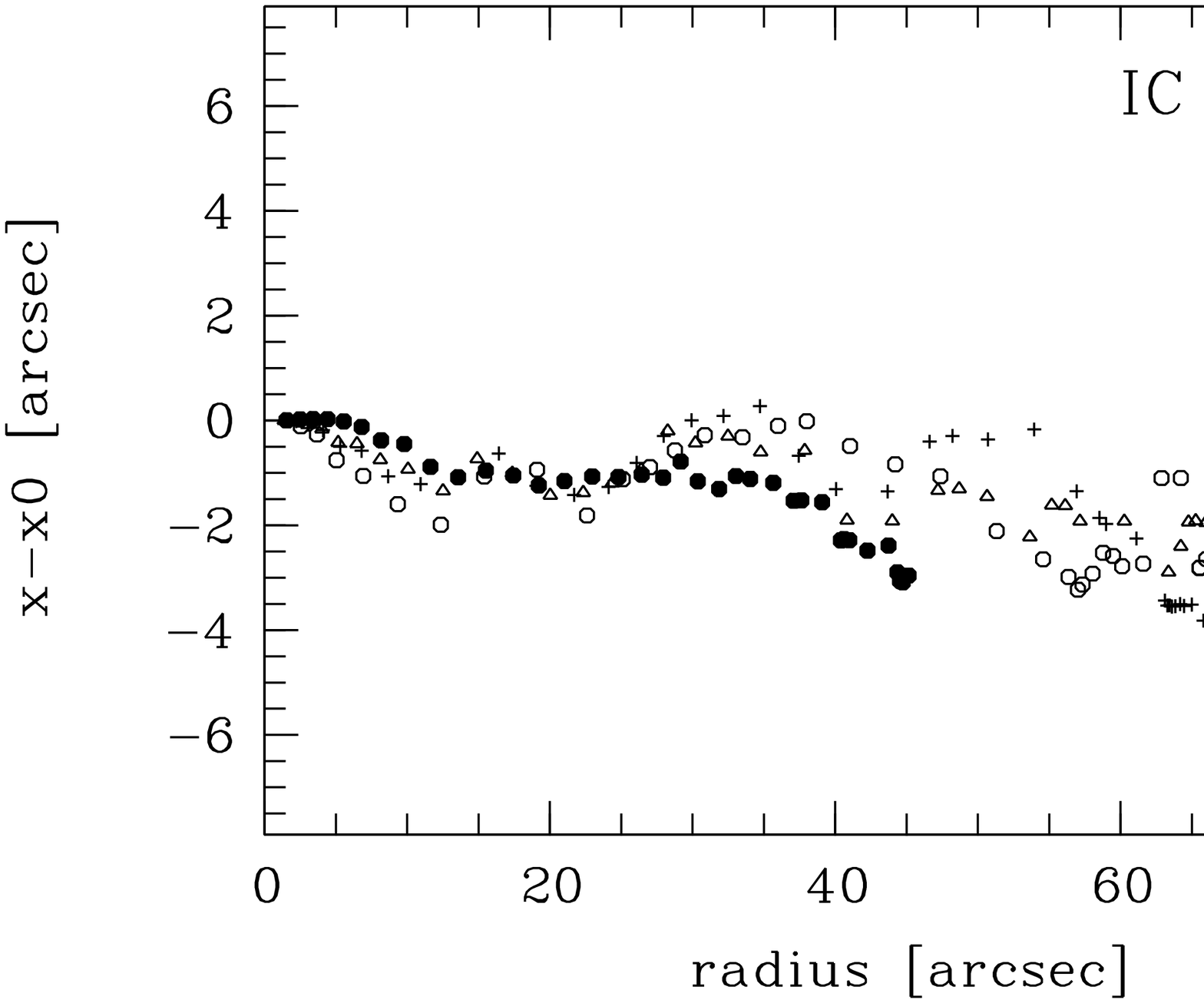}}
\\\\
\resizebox{0.36\textwidth}{!}{\includegraphics{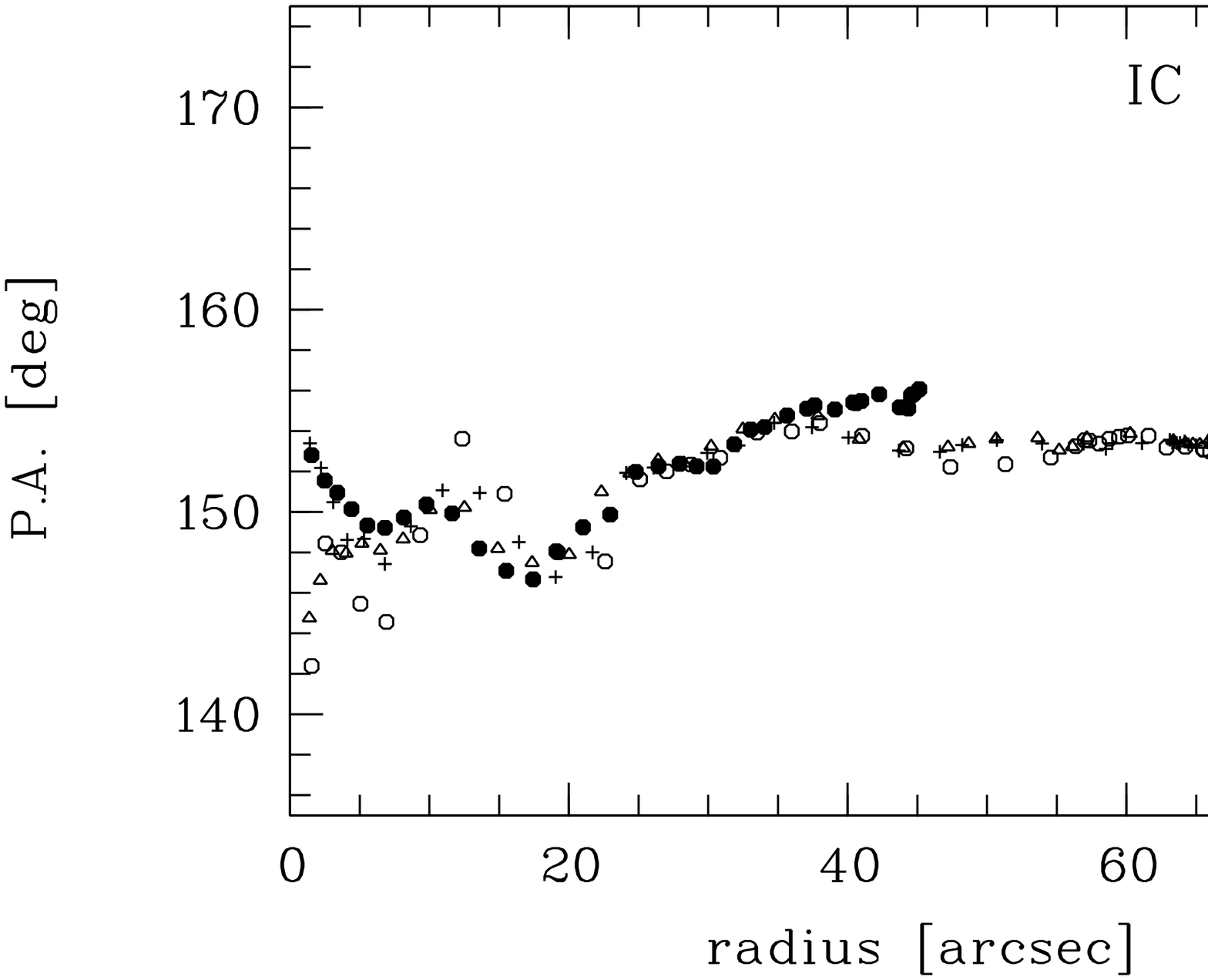}}
\hspace{25mm}
\resizebox{0.36\textwidth}{!}{\includegraphics{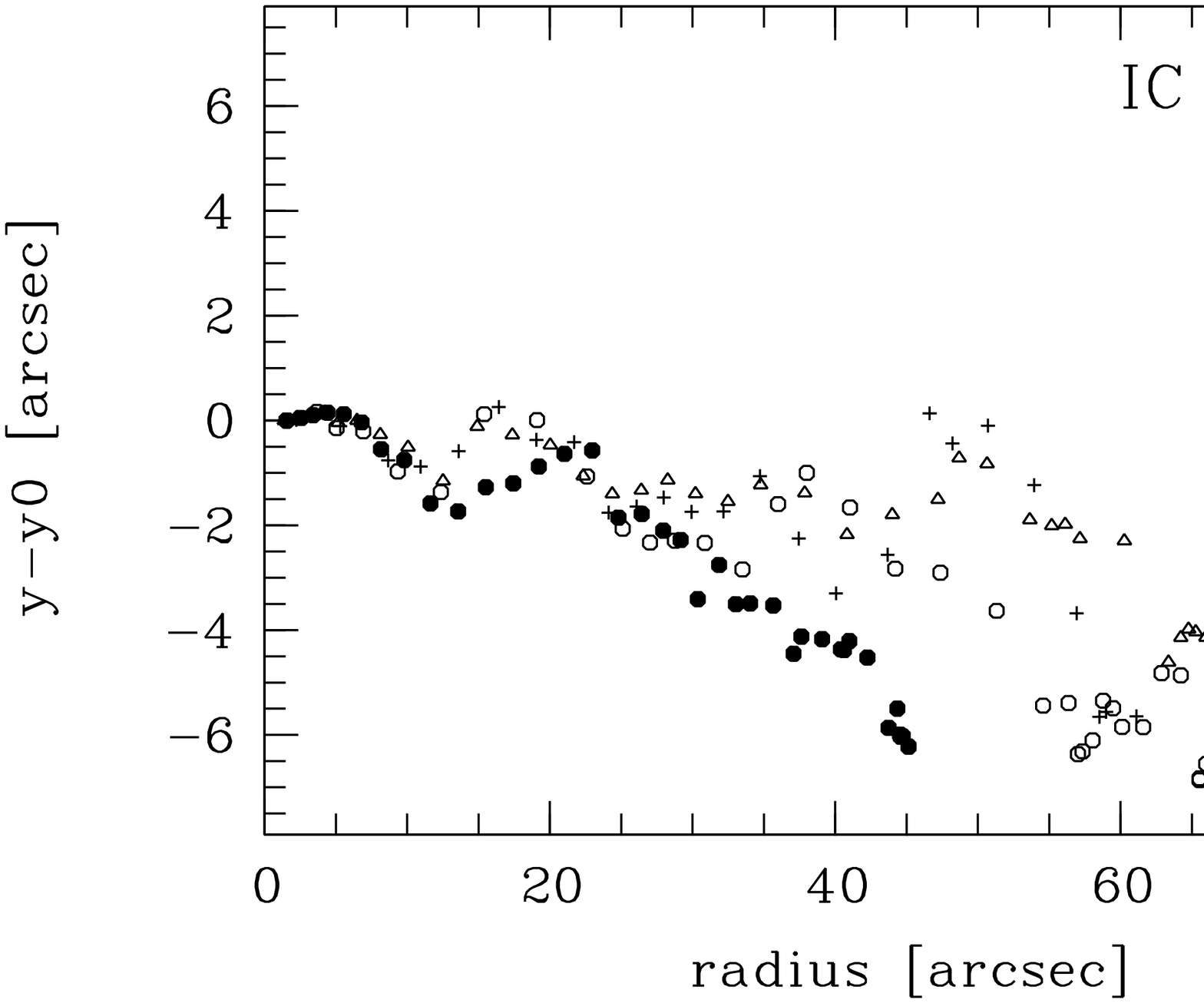}}
\\
\caption{Light distribution in IC 65 II. Parameters of the fitted free ellipses: 
the axes ratio ($b/a$, {\bf top left}),
position angle from the north through the east ($P.A.$, {\bf bottom left}), displacement
of the centre of fitted ellipses in x (x-x0, {\bf top right}) and in y
(y-y0, {\bf bottom right}). The radii are equivalent radii (i.e. $radius = \sqrt{ab})$).
}
\label{IC65pics.fig2}
\end{figure*}

IC 65 is by far the brightest and most massive, i.e. the principal galaxy in this group.
Its optical and NIR images as well as various light distribution characteristics are shown in 
Figs.~\ref{IC65pics.fig} and \ref{IC65pics.fig2}. 
The $B$ band image depicts an oval central bulge and/or a short bar
with the long symmetric double-armed spiral pattern starting at its ends.
The spiral arms consist of a number of bright \ion{H}{ii} regions  
giving them a filamentary appearance.
The short-exposure $JHK$
composite image has a smoother intensity distribution compared to
optical images. This reflects the dominance of an older stellar
populations and the reduced effect of dust in the NIR images.
The contours of the Laplacian filtered image (see Appendix A.1) 
delineate the possible bar in this NIR image.
A small nucleus is located in the middle of the bar. The two round
features beyond the ends of the bar may
be either the starting points of the spiral arms or intersections of an
inner tilted ring with the plane of the sky. 
The presence of a bar is further evidenced by the typical behaviour of 
fitted ellipses (Fig.~\ref{IC65pics.fig2}): nearly constant $P.A. \simeq 148^{\circ}$ and systematically 
decreasing axes ratio within $15''$ (equivalent radius of the bar).
The total length of the bar is $\simeq 32''$ ($\sim 6$ kpc), and the axes
ratio $b/a \simeq$ 0.25.

The optical image of IC 65 appears slightly asymmetric with an extension to the 
south-east. This asymmetry can be quantified by means of the decentering 
degree calculated from the drift of the centre of the  
fitted free ellipses (Marquez \& Moles 1999). For the IC 65 the decentering amounts 
to $\sim$ 8.2\% in the radius range of $60'' - 70''$.
In contrast, the \ion{H}{i} isophotes appear regular within the optical limits of the galaxy
(vM83, Fig.~27), but show an extension to the north-west at the
lowest measured \ion{H}{i} column density level of 2.3 $\times 10^{20}$ atoms~cm$^{-2}$,
well outside the optical image. The total extent of neutral hydrogen exceeds nearly 
twice the optical image of the IC 65.  

The $B-R$ colour image shows blue spiral arms with
typical colours in the range of 0.7 - 0.95 and  
a red ($B-R \simeq$ 1.55) central bulge/bar region.  
The area of red colours extends outside the bulge/bar region towards the SW 
which probably indicates an
enhanced amount of dust on the line of sight in that area. 

The optical $SB$ profiles allow to distinguish the following structural components: 
a small bulge and/or nucleus within $\sim 10''$, a bar with round features beyond its end 
within $\sim 25''$, and an extended nearly exponential disk with imprints of 
the spiral-arm-pattern superimposed to the underlying disk. 
The comparison of our data with
the $B$ profile, published by Blackman \& Moorsel (\cite{blackman84})
shows good accordance within 40$''$, which is the extent of their
photographic $SB$ profile.  Our new $I$-band photometry agrees  
with the results in Haynes et al. (\cite{haynes99}).
We performed the pilot
modeling with only two major components: the bulge and the disk,
neglecting the bar and the spiral arm components. The resulting
bulge+disk models in $B$ and $R$ are shown in Fig.~\ref{IC65pics.fig} by 
contionuous lines. The optical colour-index profiles show 
a strong negative colour gradient 
within $r \leq 25''$,  
followed by a nearly flat disk region. 
The central red colours 
 $(B-R)_0 \simeq 1.5, (B-I)_0 \simeq 2.2, (B-J)_0 \simeq 3.3, (H-K)_0 \simeq 0.3$ 
nearly correspond to 5 - 8 Gyr old stellar populations with sub-solar abundances,  
however large negative colour gradients, particularly in the NIR-optical colours 
indicate the presence of an amount of dust in that area.
The mean disk colours of $(B-R) \simeq 0.95, (B-I) \simeq 1.4, (B-J) \simeq$ 2.1, 
and $(H-K) \simeq 0.2$ nearly correspond to stellar populations, which are several 
Gyrs younger than the nuclear ones. 
Here, and in the following, we compare the observed colours to those predicted by 
the evolutionary models of Bruzual \& Charlot (\cite {bruzual03}) to derive  a guess 
of the age and evolutionary status of the stellar populations of the galaxies.
\begin{figure*}
\resizebox{0.31\textwidth}{!}{\includegraphics{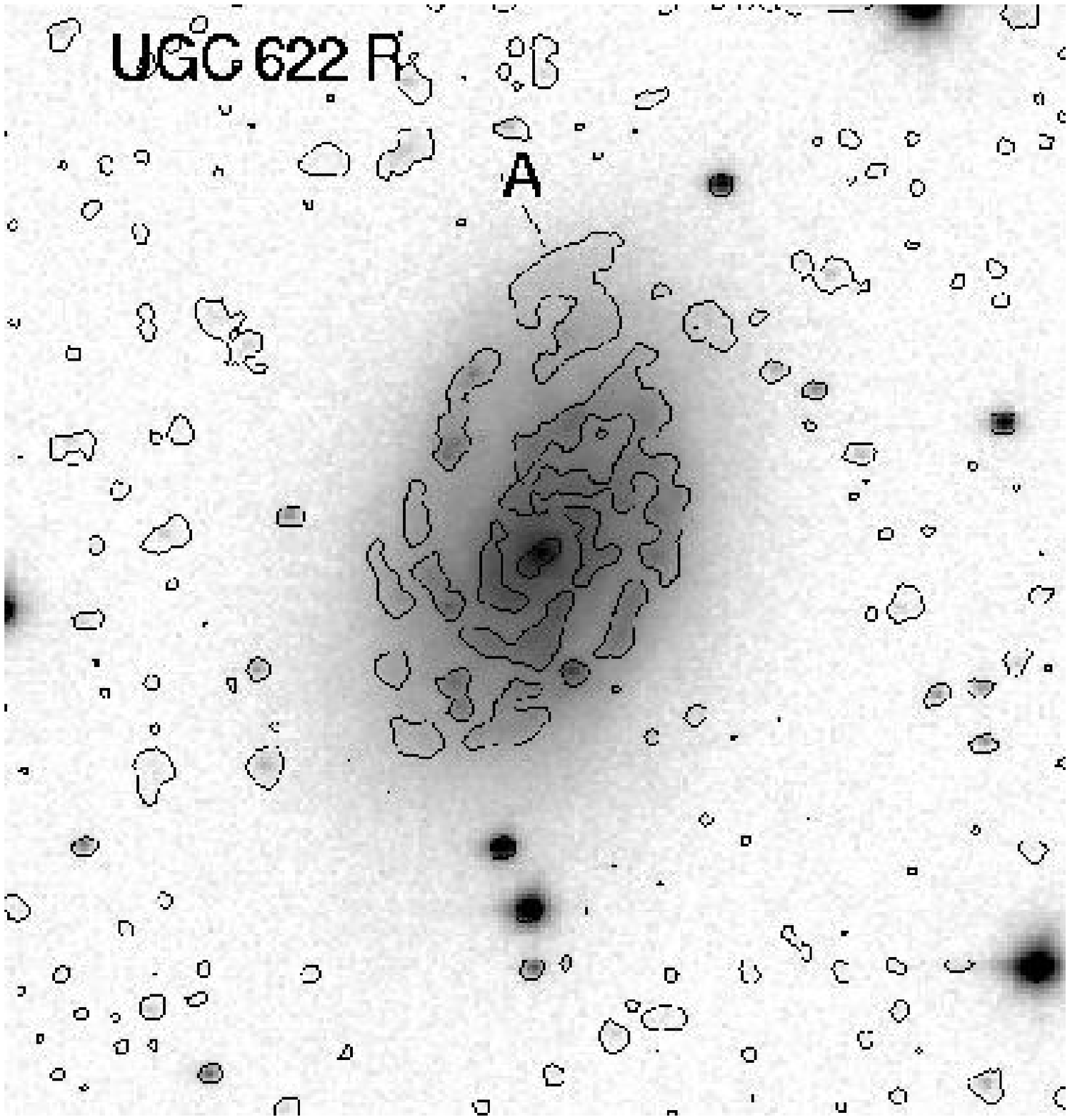}}
\resizebox{0.305\textwidth}{!}{\includegraphics{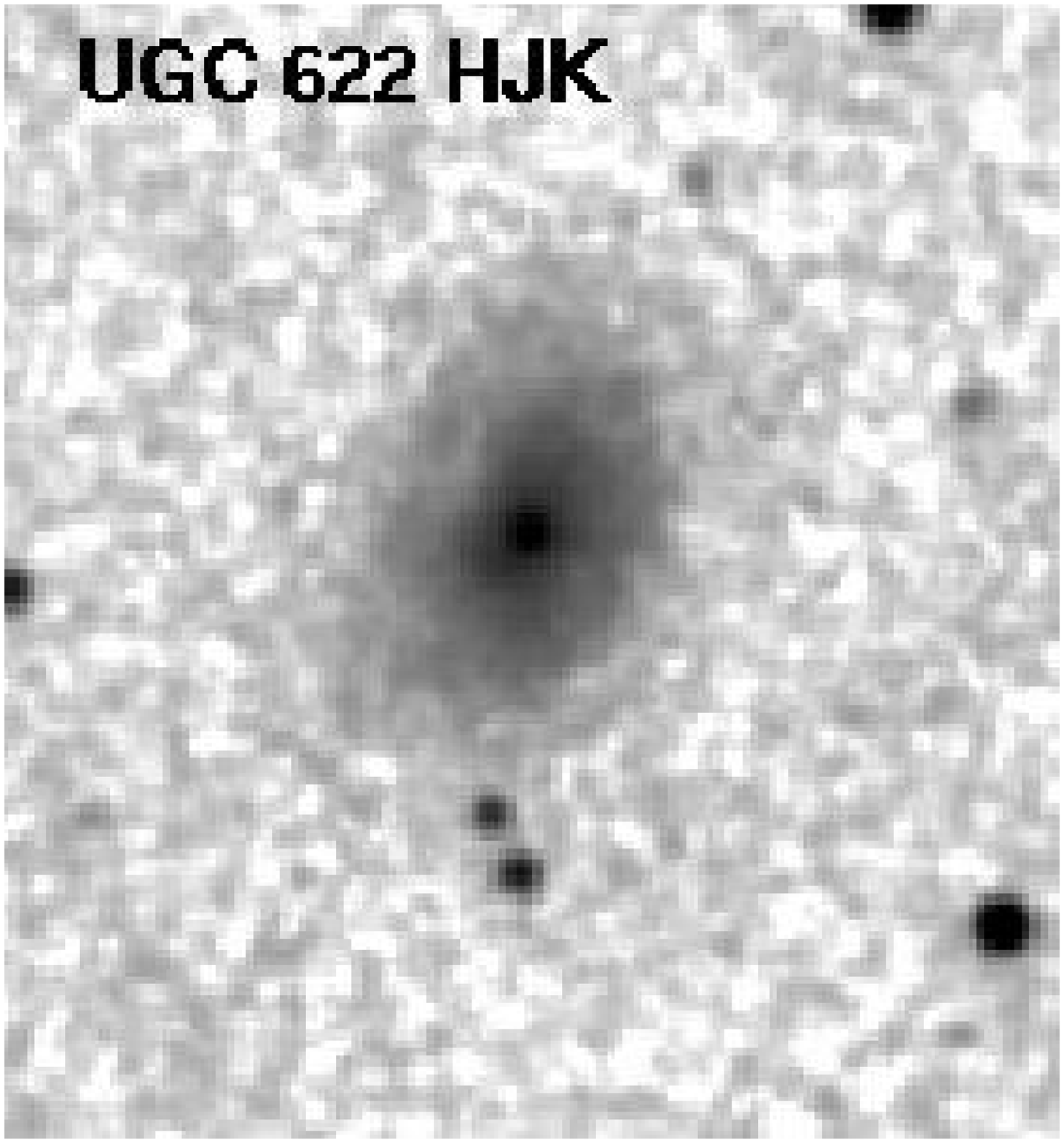}}
\resizebox{0.31\textwidth}{!}{\includegraphics{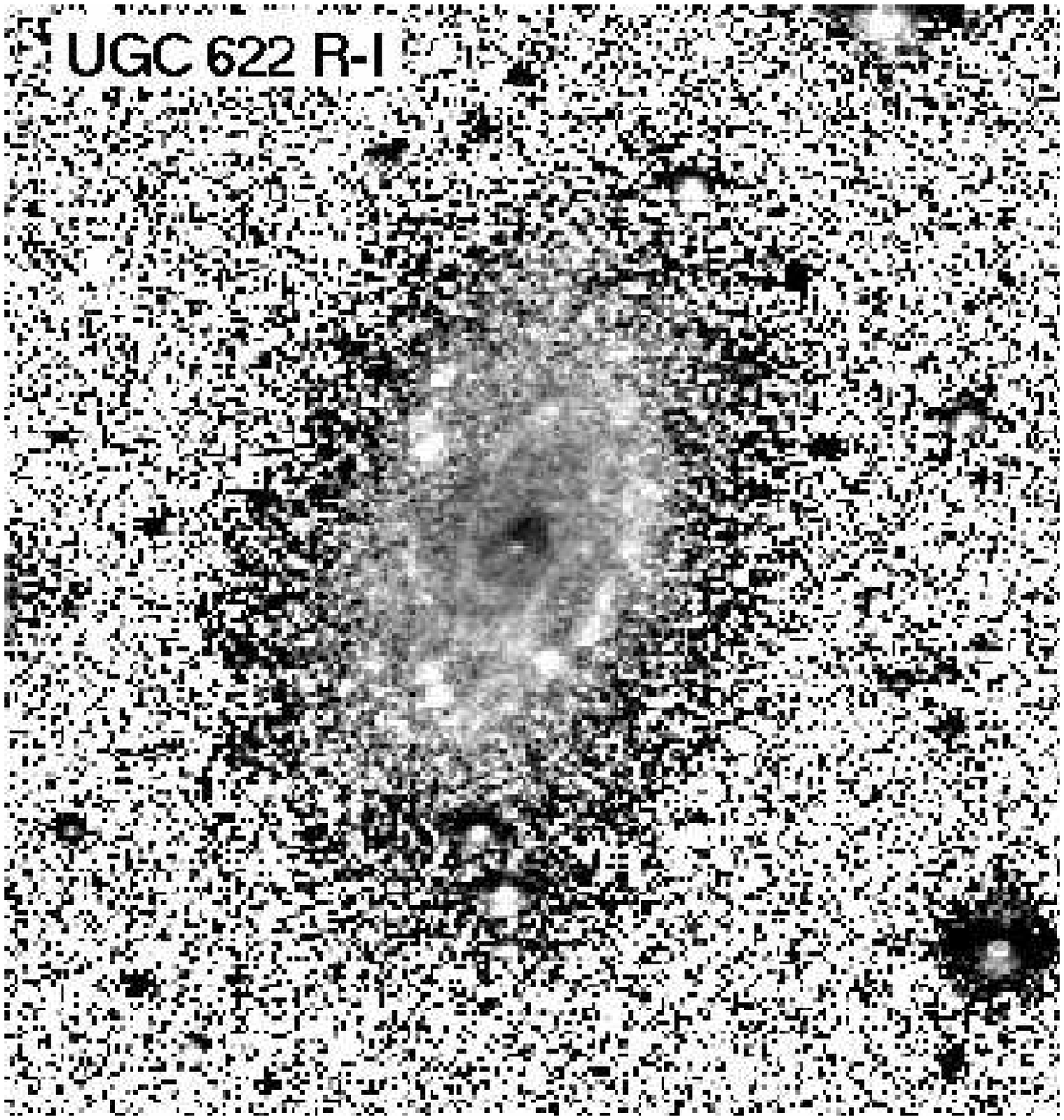}}
\resizebox{0.5\textwidth}{!}{\includegraphics[angle=270]{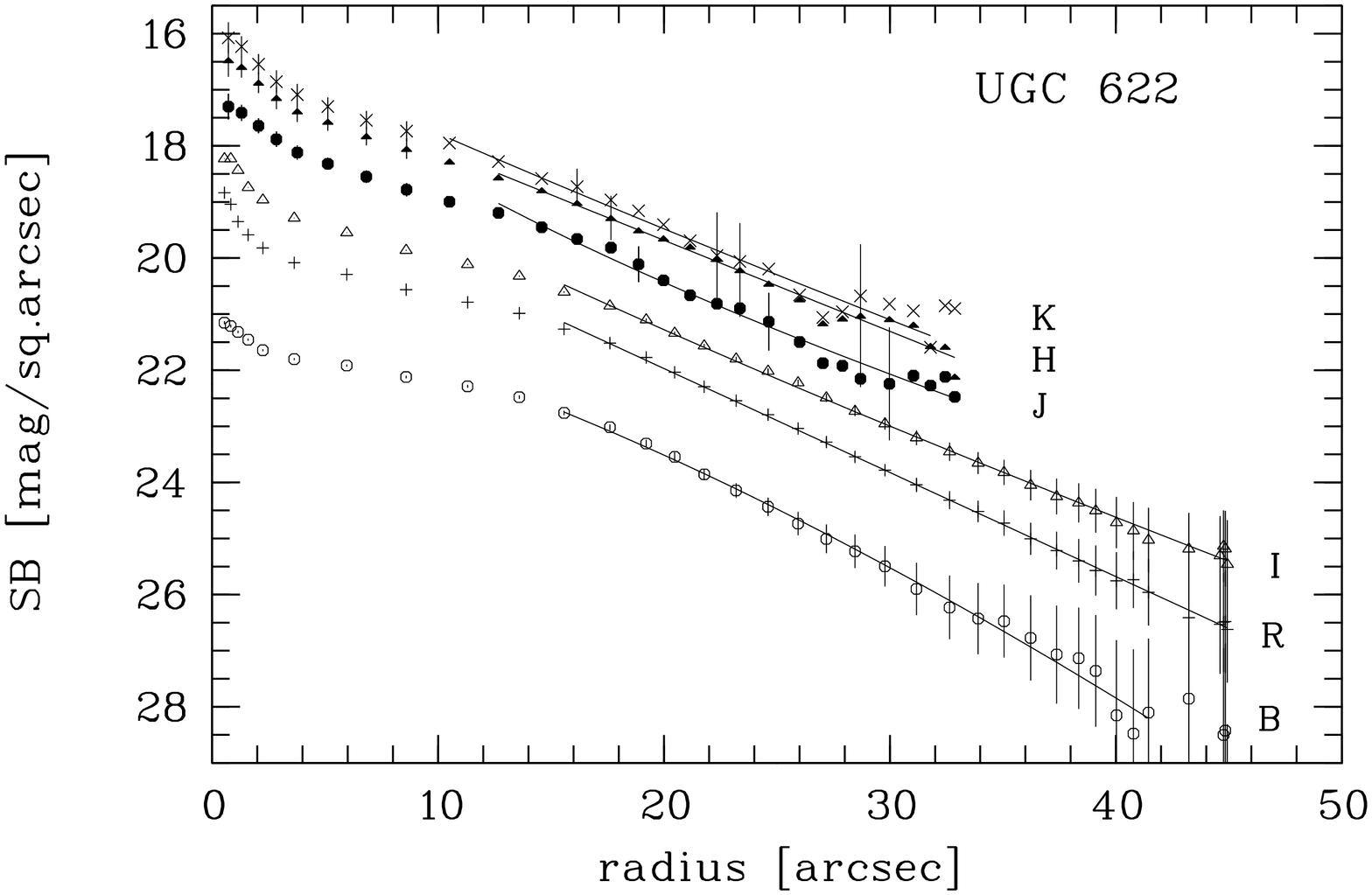}}
\resizebox{0.5\textwidth}{!}{\includegraphics[angle=270]{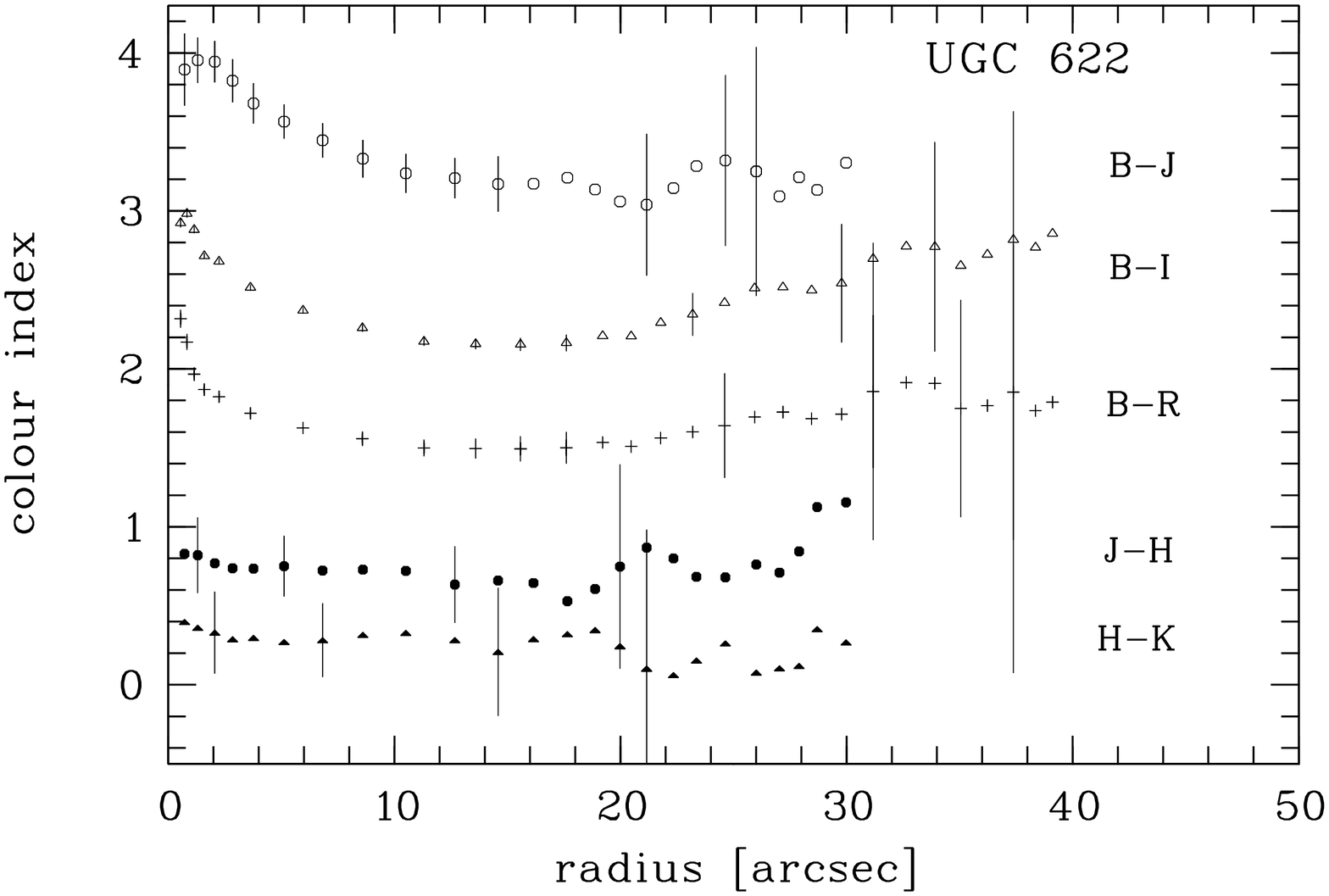}}
\caption{Light distribution in UGC 622 I. {\bf 2D-frames.} 
{\bf Top left:} the $R$ band CCD image with
             Laplacian contours delineating spiral arms,
             which consist of a number of star-forming
             \ion{H}{ii} regions; 
  {\bf top centre:} the composite $J, H,$ and $K$ band image; 
  {\bf top right:} the $R-I$ colour index image in the range
             0.5 - 1.0 mag. The dark shade is red,
             the light shade is blue.
All frames have a field size of nearly $2'\times 2'$. The north is top and the east is to the left.
{\bf 1D radial profiles.} Coded as in Fig.~\ref {IC65pics.fig}.
}
\label{UGC622pics.fig}
\end{figure*}
\begin{figure*}
\resizebox{0.36\textwidth}{!}{\includegraphics{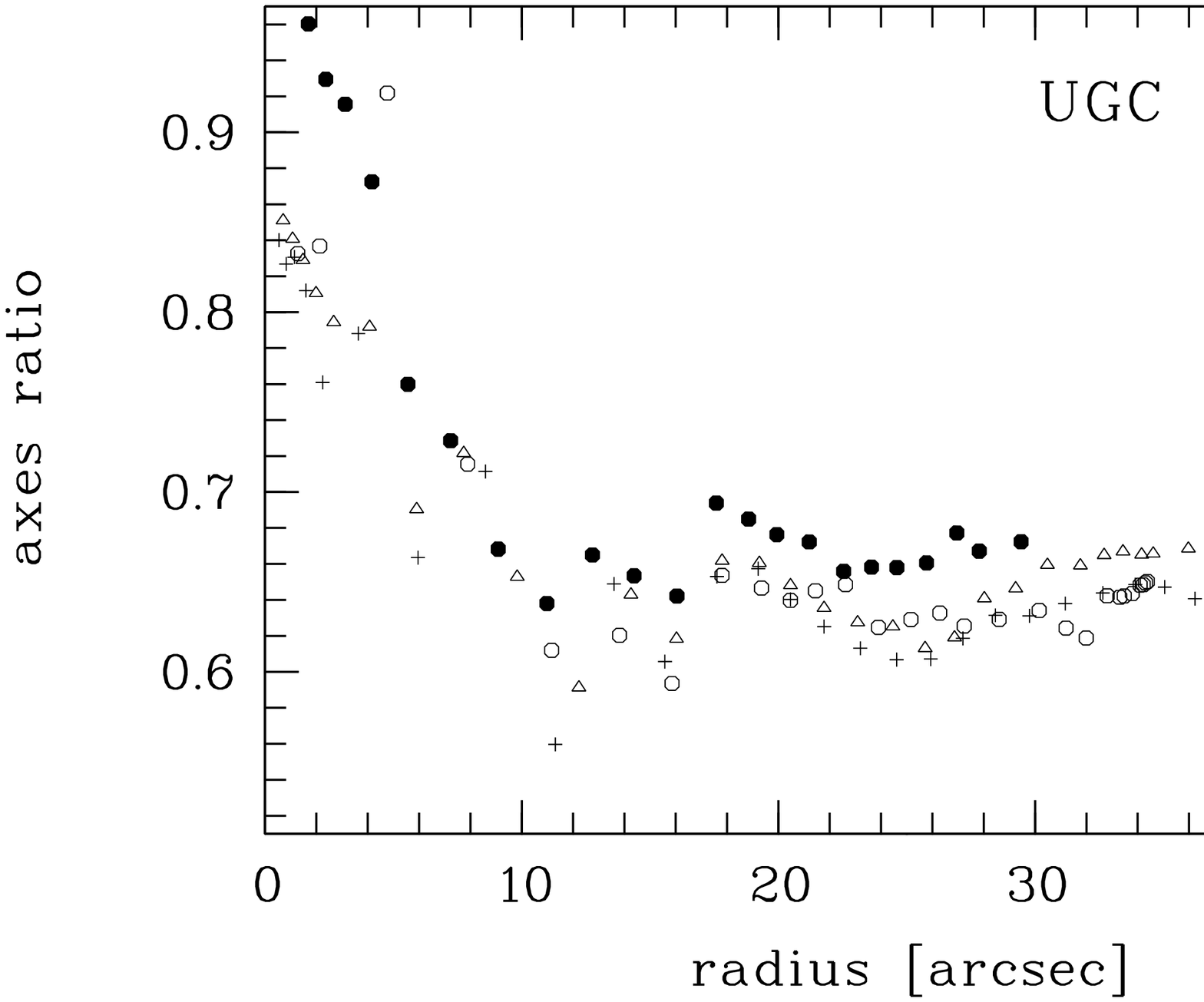}}
\hspace{25mm}
\resizebox{0.36\textwidth}{!}{\includegraphics{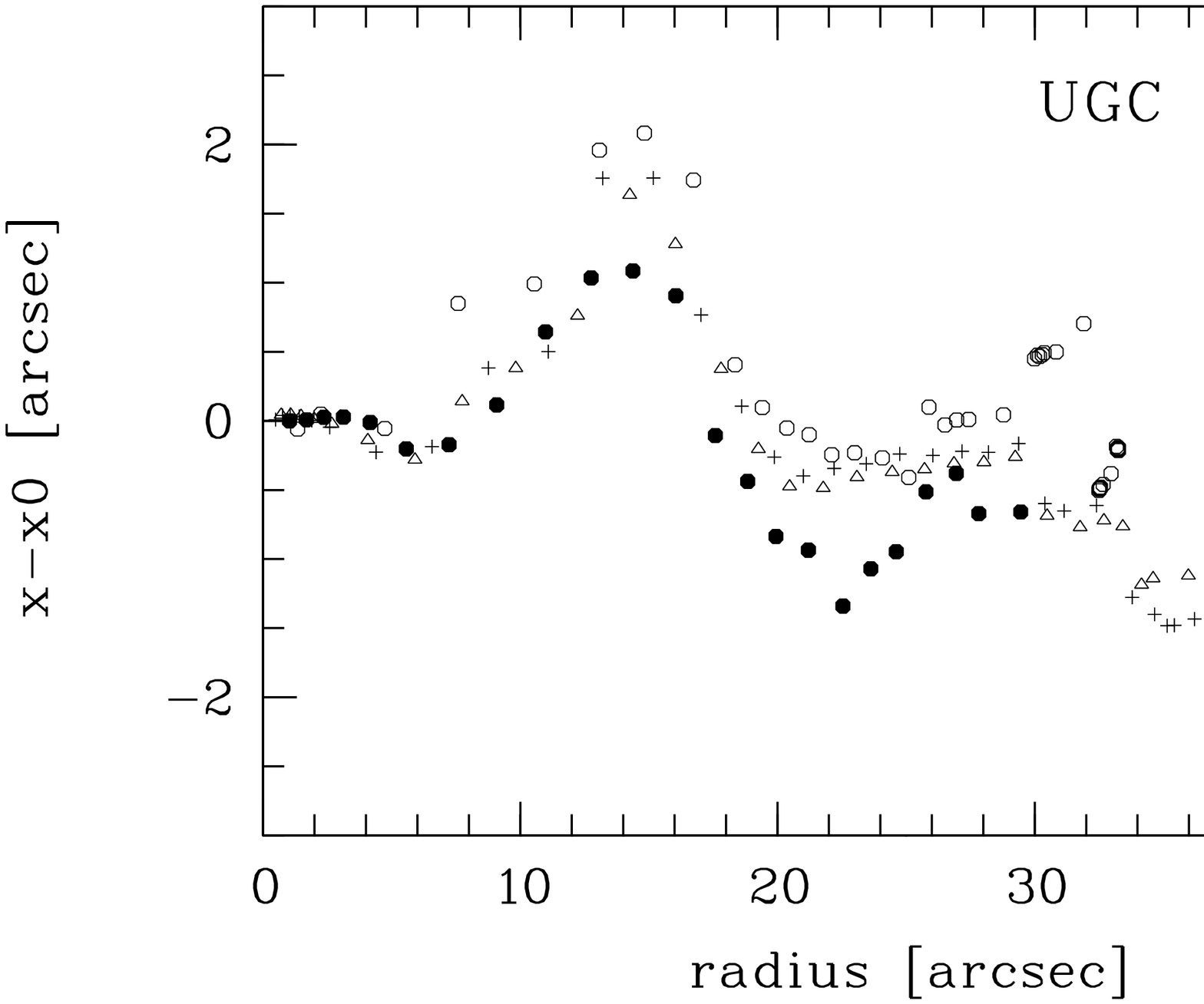}}
\\\\
\resizebox{0.36\textwidth}{!}{\includegraphics{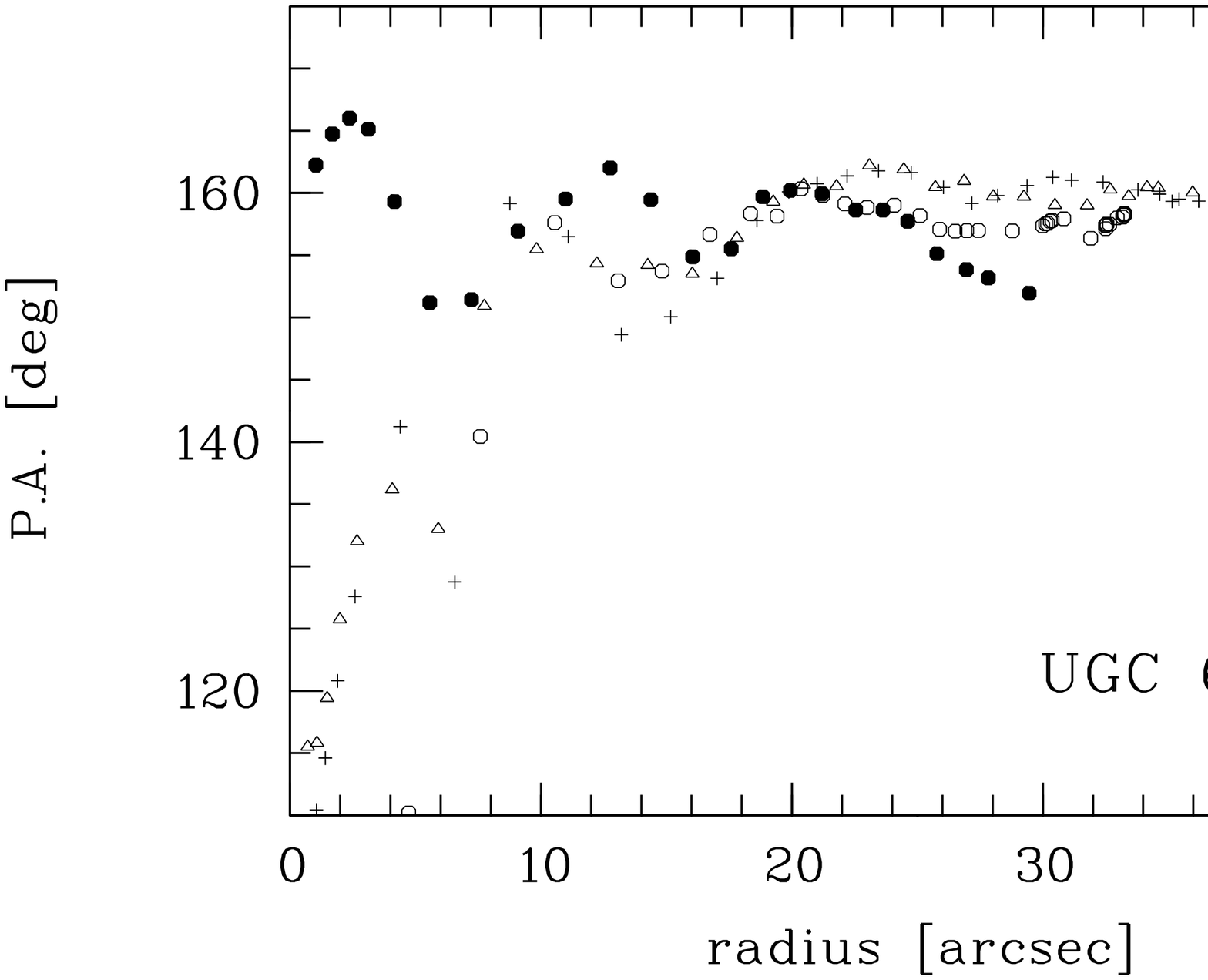}}
\hspace{25mm}
\resizebox{0.36\textwidth}{!}{\includegraphics{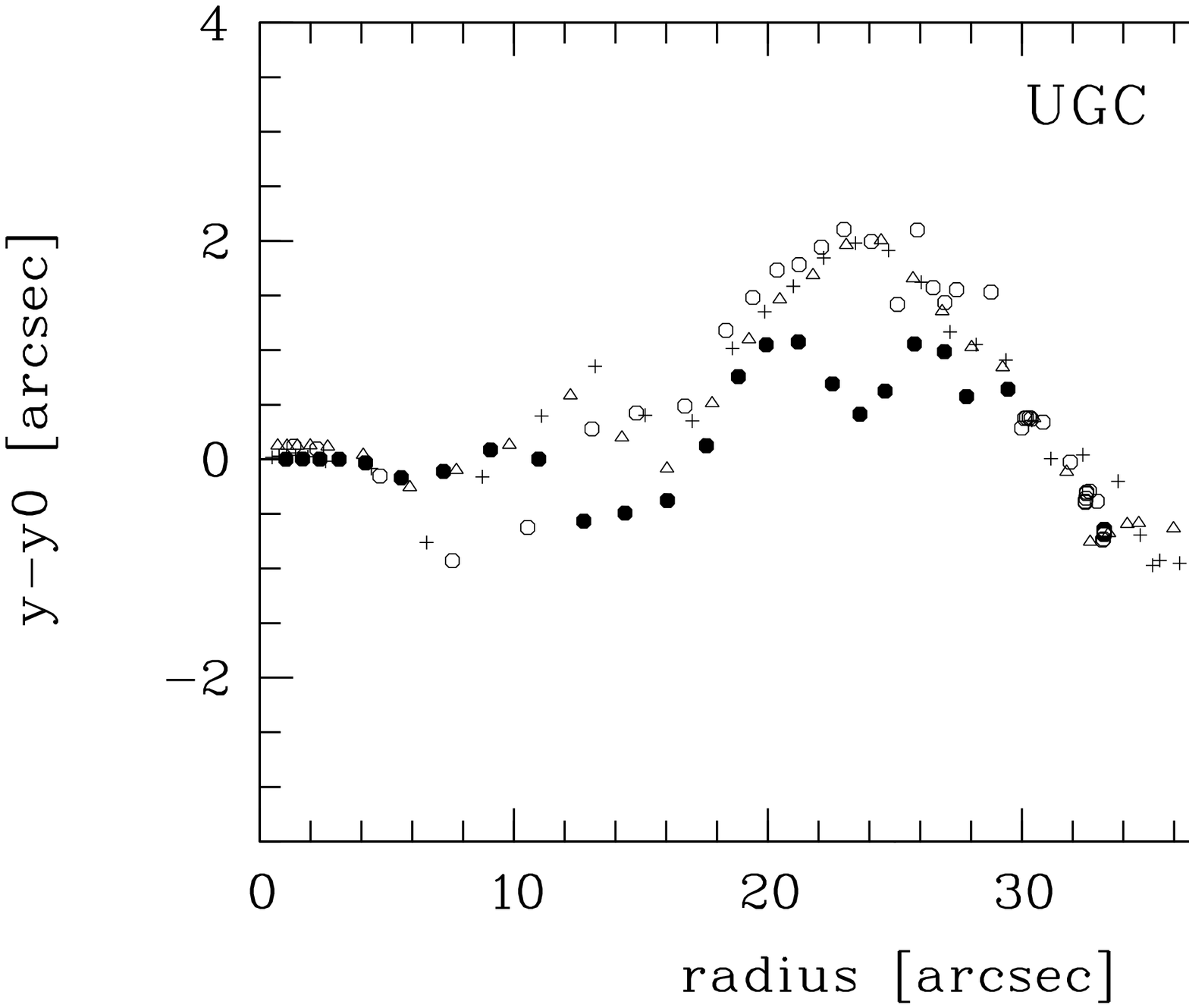}}
\\
\caption{Light distribution in UGC 622 II.
Parameters of the fitted free ellipses. Coded as in Fig. \ref{IC65pics.fig2}.
}
\label{UGC622pics.fig2}
\end{figure*}
\subsection {UGC 622} 
The results of the surface photometry of the UGC 622 are summarized in  
Figs. \ref{UGC622pics.fig}, \ref{UGC622pics.fig2}.
This late-type galaxy has 
an almost stellar nucleus and/or bulge 
and faint flocculent spiral arms. Numerous star-forming regions, distributed 
along the arms are outlined with the help of the Laplacian contours.   
The NIR composite image shows a prominent 
nucleus and an almost featureless smooth disk. 
The red colour of the circumnuclear area ($B-R \simeq 2.05, R-I \simeq 0.75$) may be 
partly attributed 
to dust and its asymmetry may be caused by inclination. 
Almost all knots and blobs
distributed along the spiral arms are blue with colours in the range of 0.30 $< R-I <$ 0.50 and 
1.10 $< B-R <$ 1.22. 
The colours of the underlying disk ($<B-R> \simeq 1.25, <R-I> \simeq 0.6$) 
refer to a relatively young (3 - 5 Gyr) stellar population 
with roughly 0.4Z$_{\sun}$ (Bruzual \& Charlot \cite{bruzual03}).

UGC 622 shows a classical type II light profile (Freeman \cite{freeman70}), where  
the central light cusp is followed by two exponential sections
with different gradients. 
The inner exponential in the range of $3'' \leq r \leq 15''$ has a scale length
of $h_B \simeq 12.1''$ in $B$ band.  The outer ($r \geq 15''$)
exponential has a steeper slope with scale lengths of $h_B \simeq 5.4''$
in $B$ band and $h_K = 6.6''$ in $K$ band.  The optical and
NIR light profiles show similar behaviour, but the difference between
the inner and outer exponential disk gradients becomes less pronounced
at longer wavelengths - an effect of
dust extinction in optical passbands.  
The $V$-band $SB$ profile, determined by Heraudeau et
al. (\cite{heraudeau96}), appears very similar to our optical $SB$
profiles.
The optical colour index profiles 
show a red (dusty) nuclear region, uniform colours within the inner disk,
along with some evidence for the colour gradient in the outer disk. 
%
The conspicuous twisting and shifting of the fitted free ellipses (Fig.~\ref{UGC622pics.fig2})  
could be explained as an effect of the patchy spiral arms.  
 
VM83 noted asymmetries
in the distribution of atomic hydrogen of the UGC 622.  The \ion{H}{i} map
(his Fig. 28 and our Fig.~1) shows an extension towards SE ($P.A. \sim 145^o$), which
differs from the orientation of the stellar disk ($P.A. =
160^o$). 
Weak \ion{H}{i} emission has been detected 
at the distance of 1.1$'$ and 2$'$
to the north and at 2.5$'$ to the NE of UGC 622.  
The northernmost \ion{H}{i} detection nearly coincides with an optical LSB feature, which  
has been registered by running SExtractor on the DPOSS blue frame.
Other \ion{H}{i} detections, classified as \lq\lq barely significant\rq\rq~ by vM83 
have not been confirmed later neither by the WSRT wide-field \ion{H}{i} survey of 
van Braun (\cite{braun03}) nor could they have been identified in our optical search. 
\subsection {UGC 608}
\begin{figure*}
\includegraphics[height=37mm]{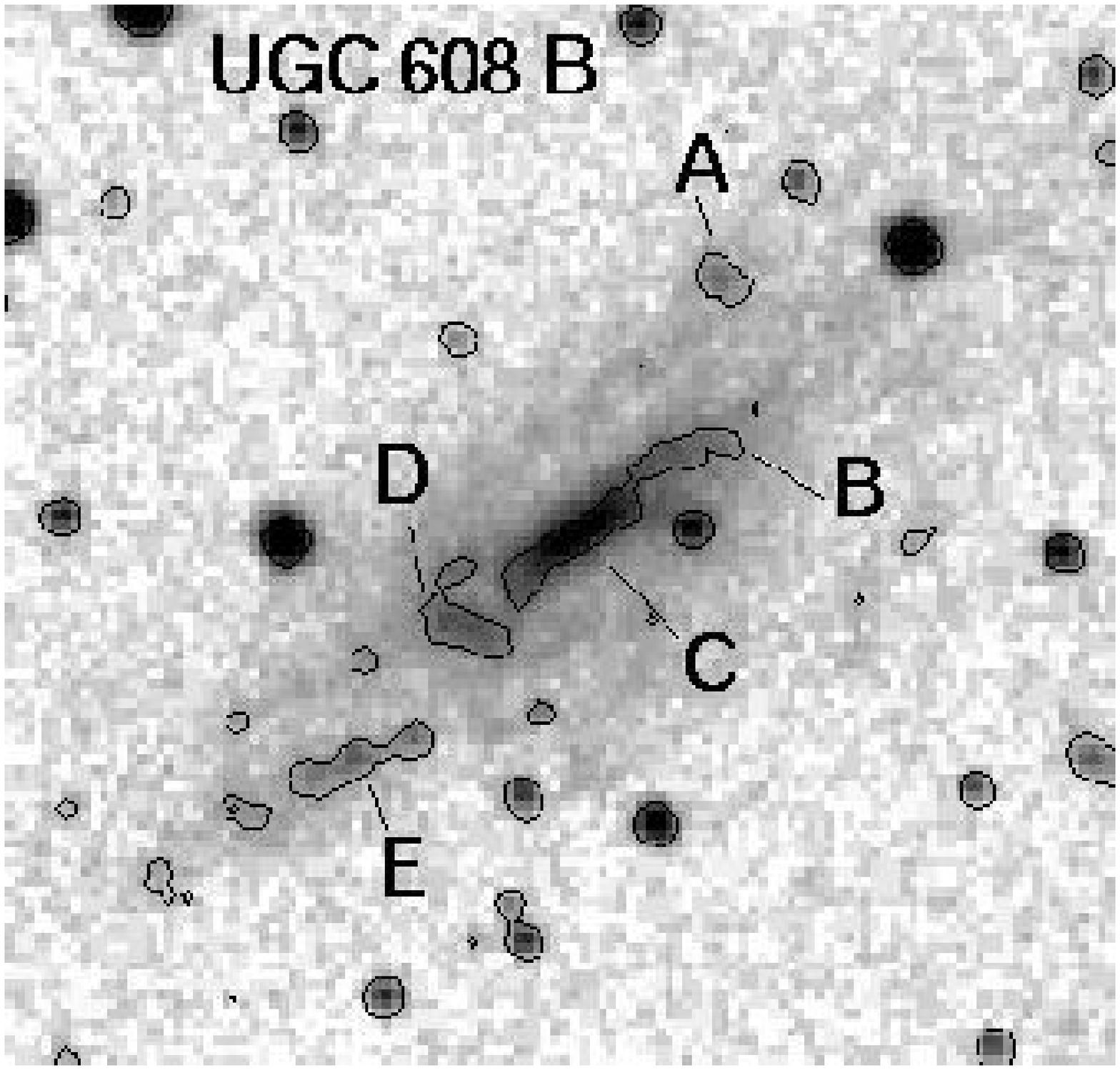}
\hspace{1mm}
\includegraphics[height=38mm]{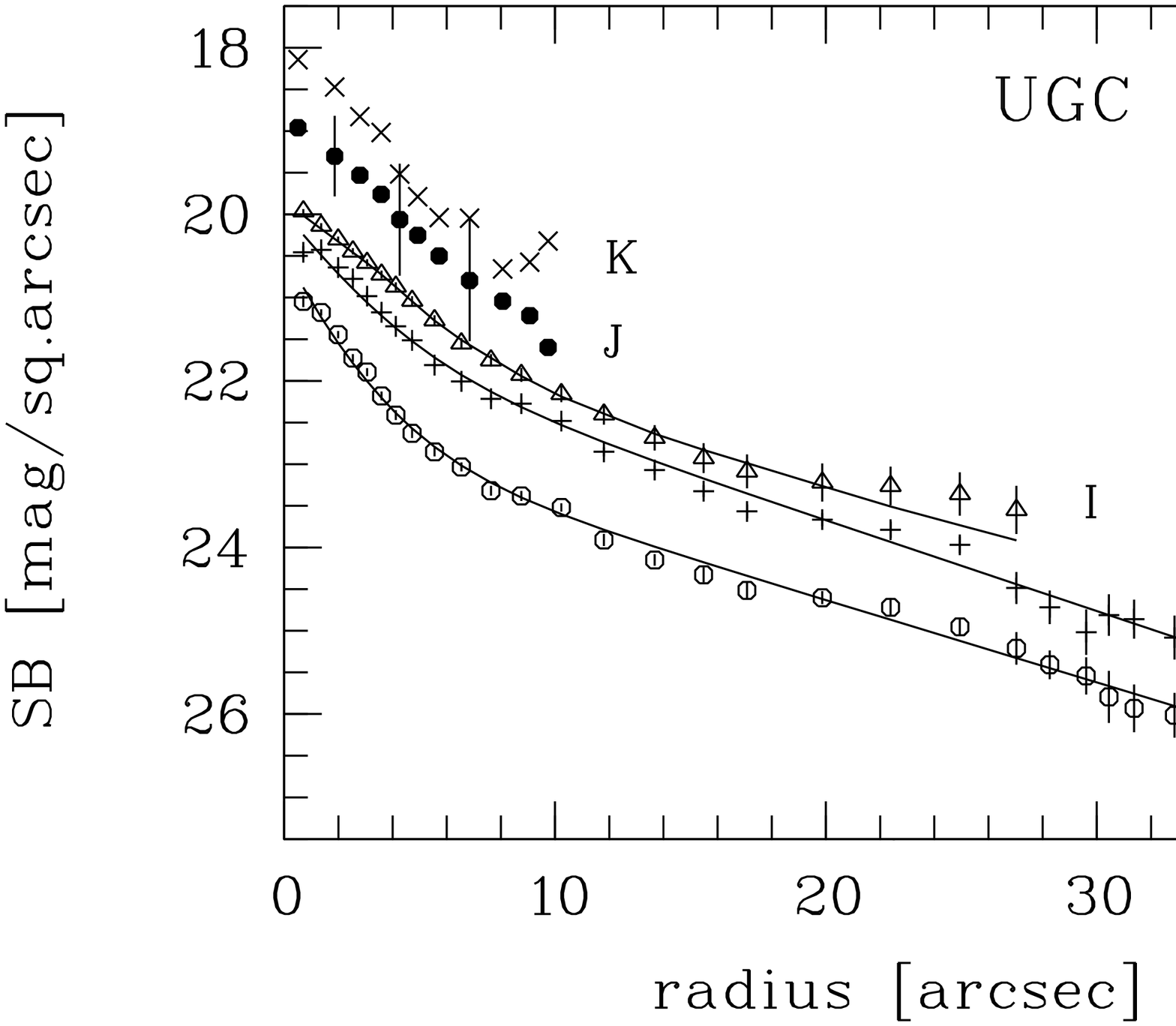}
\hspace{17mm}
\includegraphics[height=40mm]{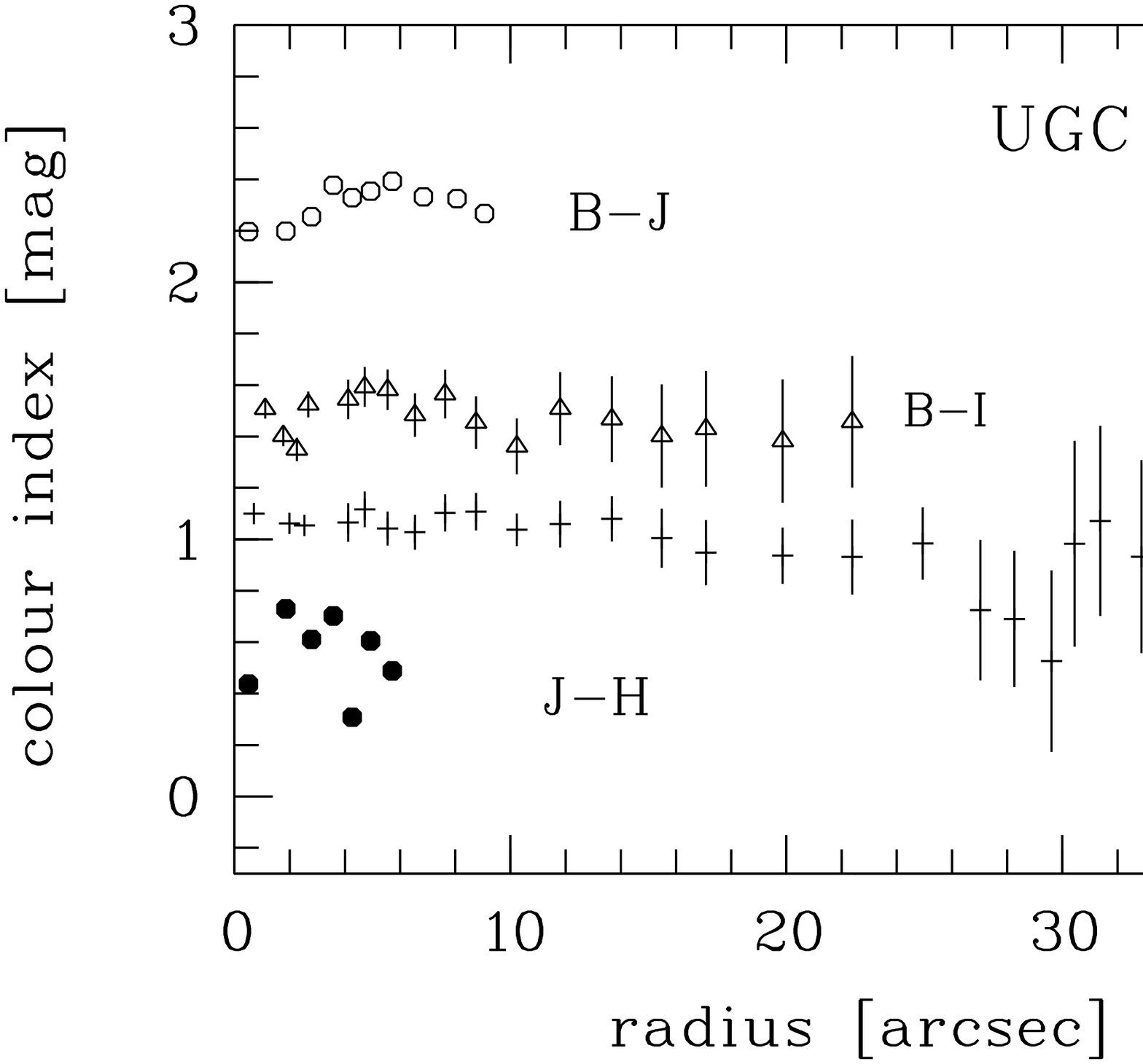}
\caption{Light distribution in UGC 608 I. The blue image (2$' \times 2'$) obtained from DPOSS
with Laplacian contours superimposed and brightest knots labelled A through E ({\bf left}); 
 the $SB$ profiles ({\bf centre}), and the
colour-index profiles ({\bf right}). 
}
\label{UGC608pics.fig}
\end{figure*}

This is a late type SBdm spiral galaxy
with a short luminous bar ($2a \simeq 3.2$ kpc, $b/a$ = 0.25, $P.A. \simeq 120^{\circ}$)
embedded into a LSB disk containing faint and knotty spiral arms
(Fig. \ref{UGC608pics.fig}).
The most prominent, generally blue knots and blobs are delineated with the help of
Laplacian contours and labelled from A through E.
The $B-R$ colour index image is noisy, and
reliable individual colour estimates were obtained only for the
brightest features.  The luminous
bar (C) is slightly redder ($0.85 \leq (B-R)_{\rm bar} \leq$
1.0) than the surrounding LSB disk area, which is
remarkably blue ($(B-R)_{\rm disk} \simeq$ 0.70). The optical colours $B-R$ and $B-I$
show marginal bluening towards the periphery.
Comparing the colours with the stellar population models  of Bruzual \& Charlot
(\cite {bruzual03}) we can conclude that both the centre and the underlying stellar disk
of the UGC 608
are populated with young (a few Gyrs) and metal-poor ($Z \simeq$ 0.001)
stellar populations
 which is quite typical of late type spirals.
The fuzzy red ($B-R \simeq$ 2.2) non-stellar object A is
probably a distant galaxy, seen through the disk of the UGC 608.
\begin{figure*}
\resizebox{0.36\textwidth}{!}{\includegraphics{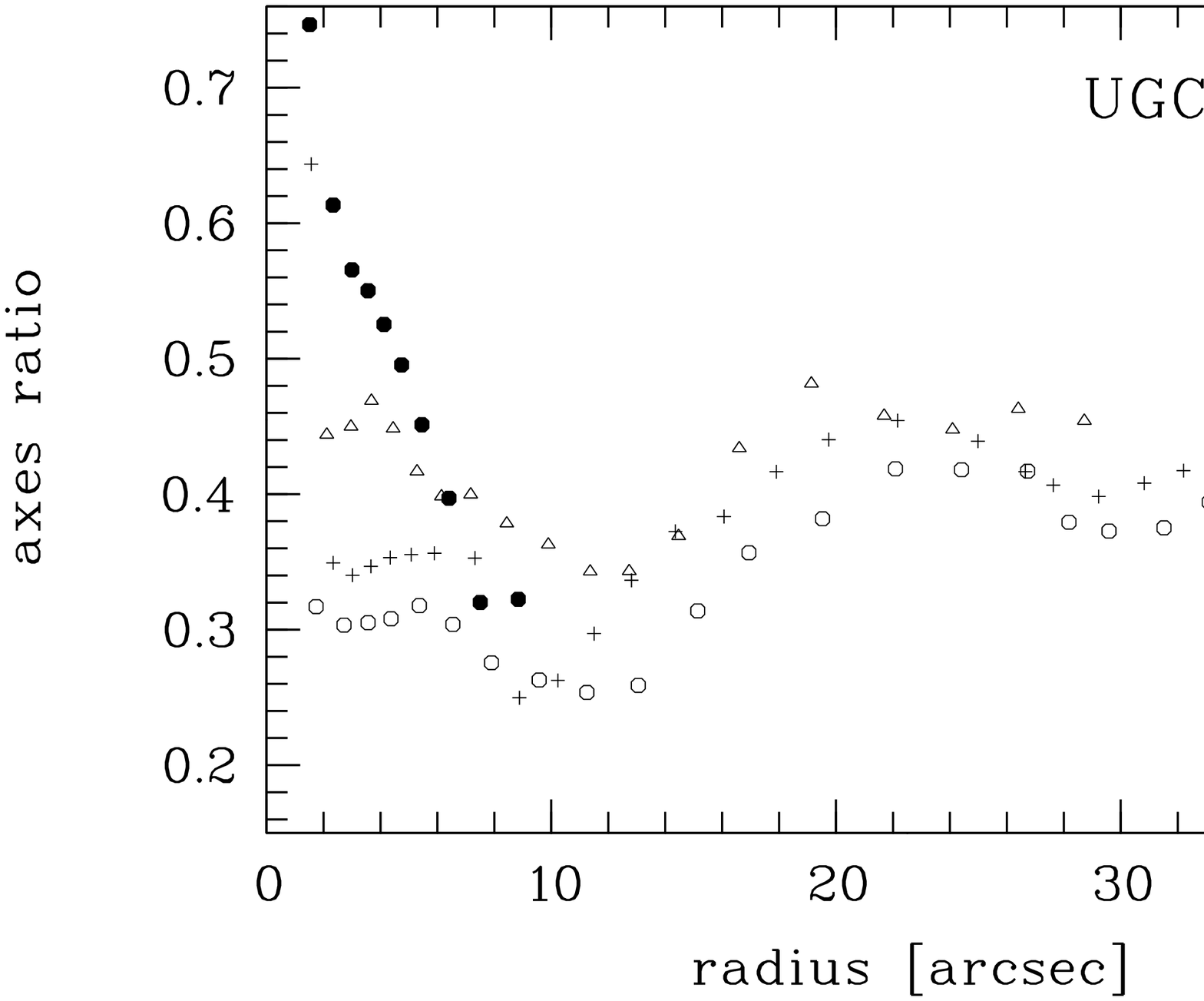}}
\hspace{25mm}
\resizebox{0.36\textwidth}{!}{\includegraphics{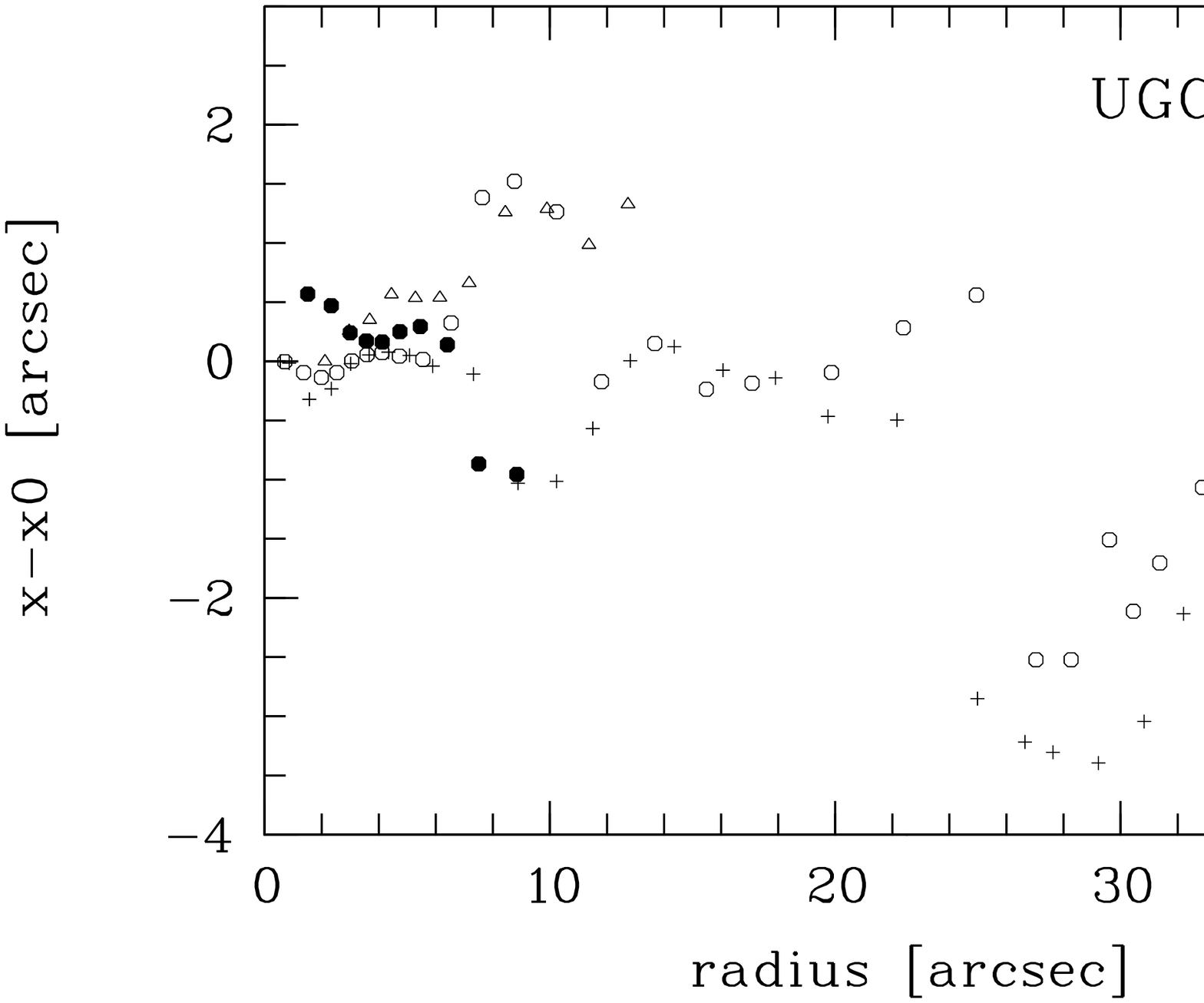}}
\\\\
\resizebox{0.36\textwidth}{!}{\includegraphics{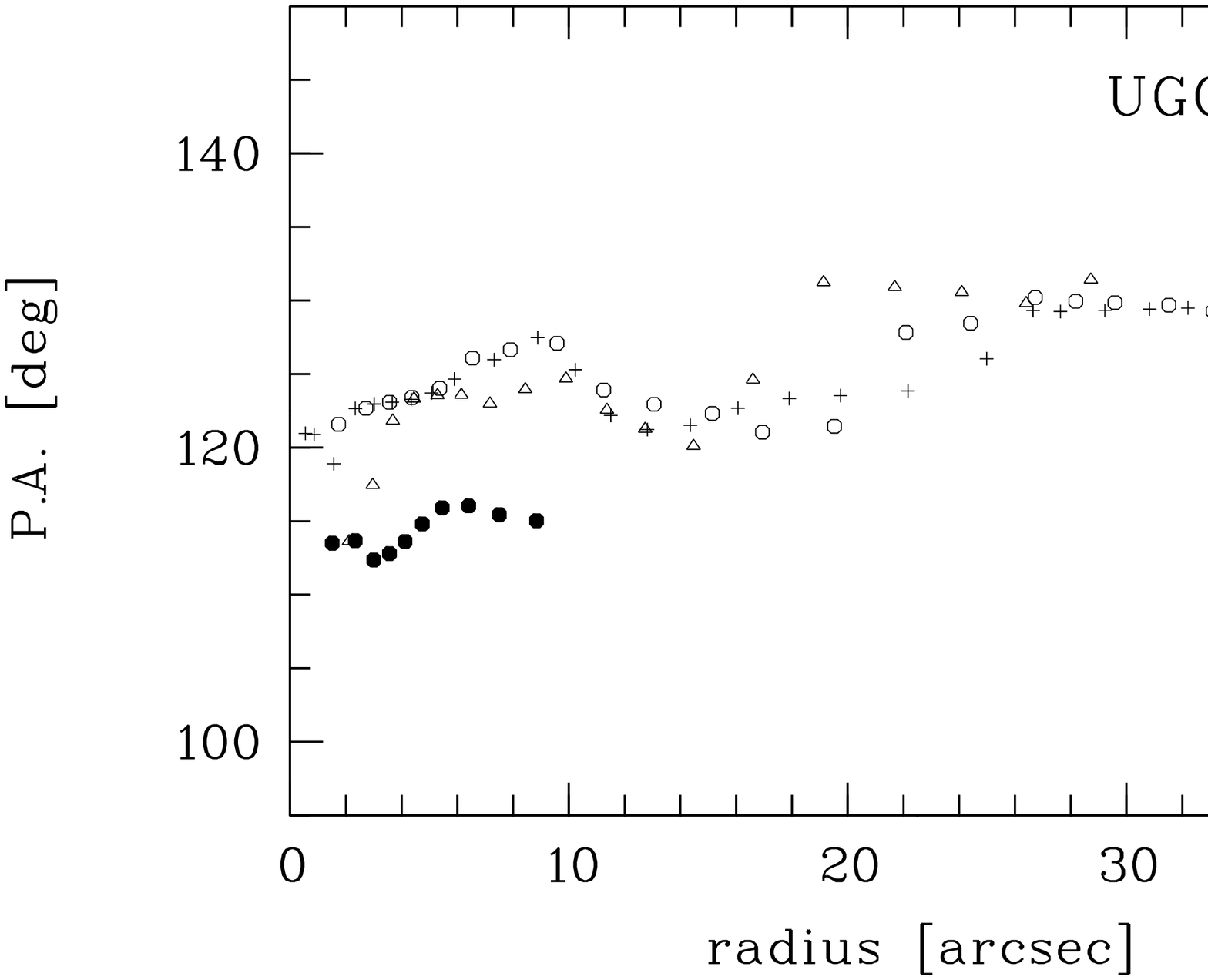}}
\hspace{25mm}
\resizebox{0.36\textwidth}{!}{\includegraphics{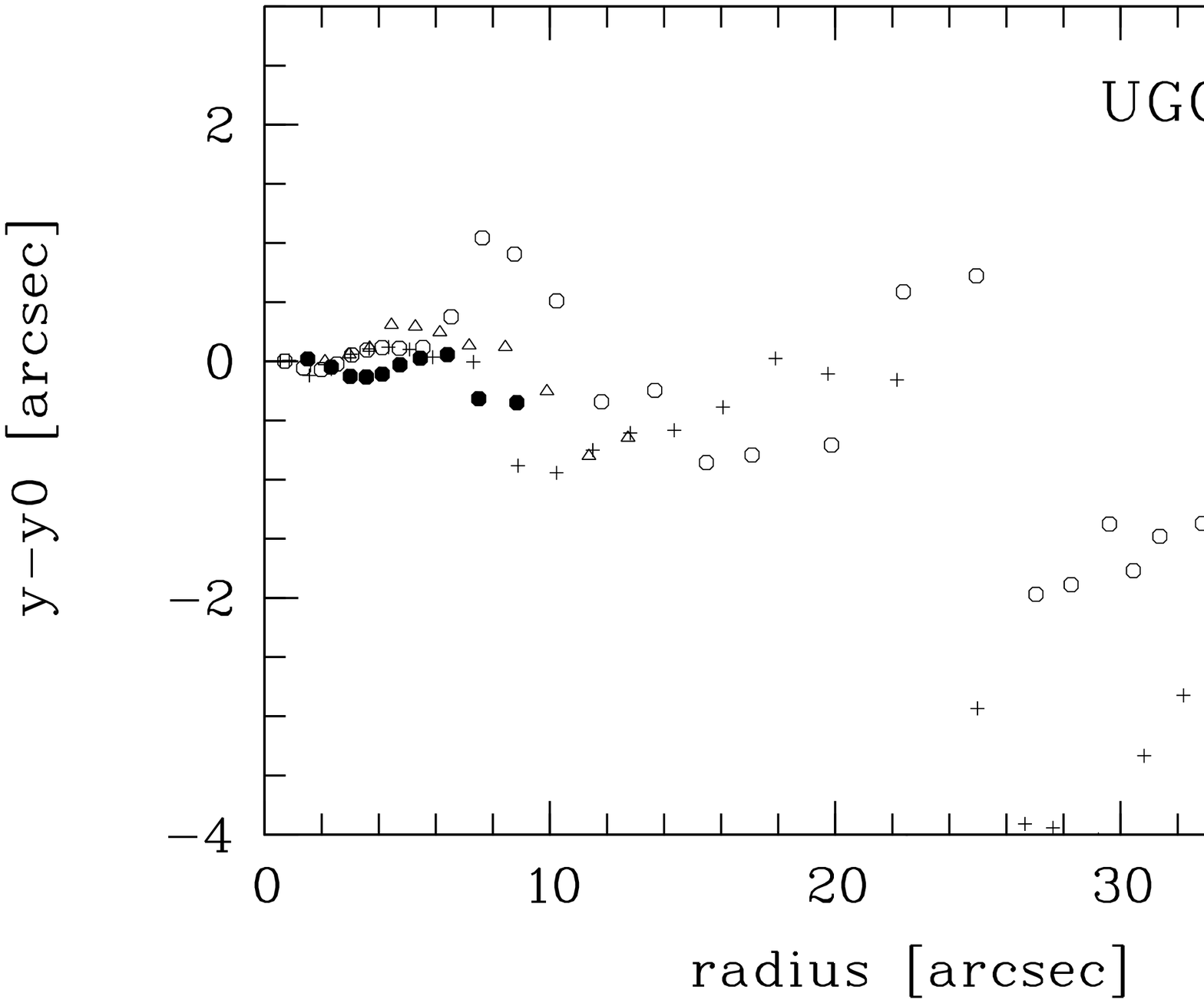}}
\\
\caption{Light distribution in UGC 608 II.
Parameters of the fitted free ellipses. Coded as in Fig. \ref{IC65pics.fig2}.
}
\label{UGC608pics.fig2}
\end{figure*}
The $B$ and $R$ band $SB$ profiles show
a dominating exponential disk component.
Our $I$ band image is
rather noisy and the corresponding $SB$ profile is less reliable.
Outside the bar region $r \geq 7''$ the free fitting ellipses trace
the faint spiral pattern, and the center of the successively fitted ellipses
slightly oscillates around its starting value, mostly because of
spiral arms and bright knots (Fig. \ref{UGC608pics.fig2}).
The \ion{H}{i} map of UGC 608 (vM83, Fig. 30) shows
a number of local maxima distributed in a generally regular gaseous disk.
Several of these gaseous features coincide
with optical star-forming knots (e.g. the triple knot E). The \ion{H}{i} halo is
$\sim 30'$ (32 kpc) in its largest extent, which is by a factor of about 2.4 larger than 
the optical diameter of the galaxy at the 25.0 $B$mag~arcsec$^{-2}$ level. 
%
\subsection {PGC 138291} 
\begin{figure*}
\includegraphics[height=45mm]{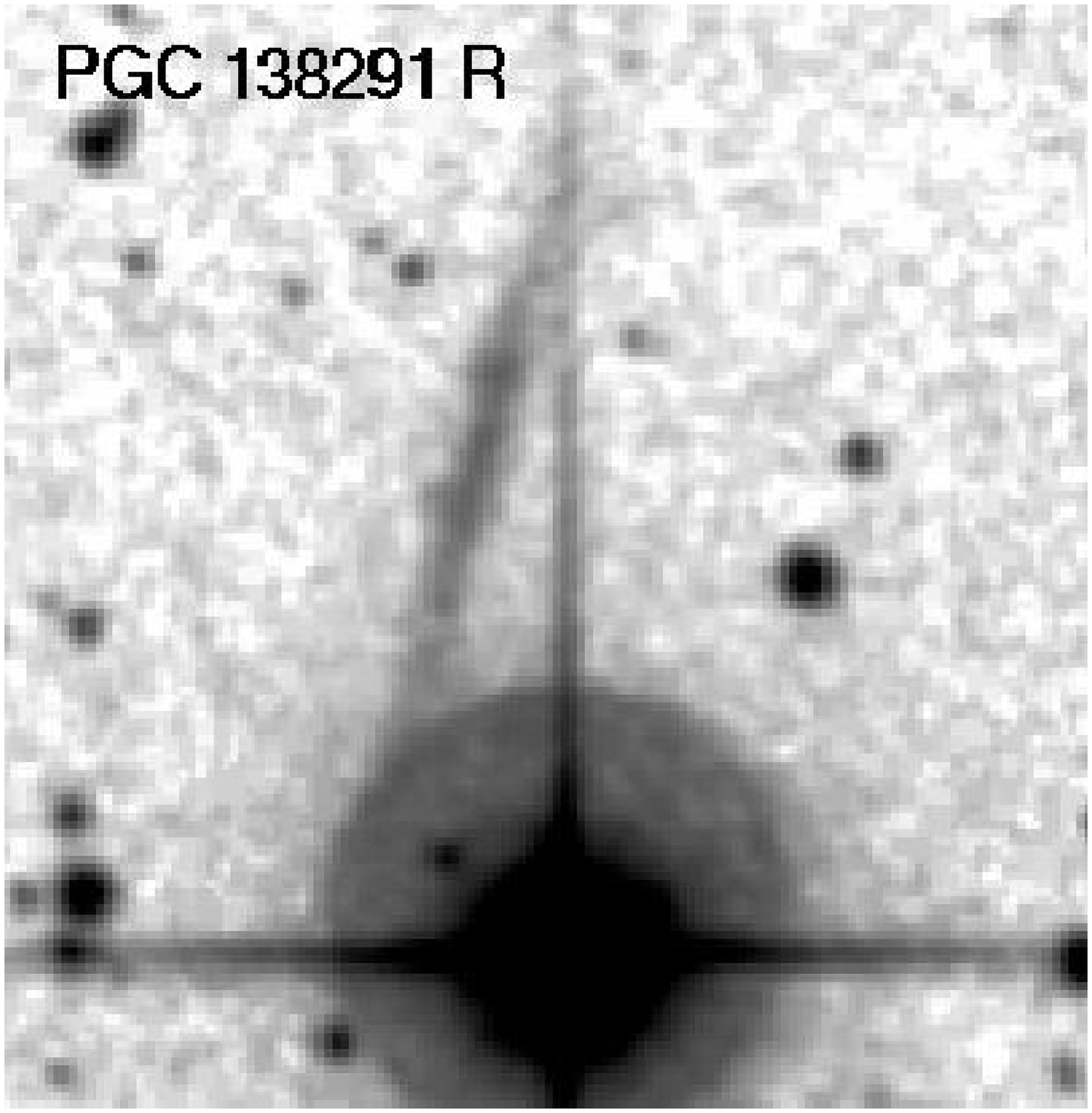}
\hspace{6mm}
\includegraphics[height=45mm,width=48mm]{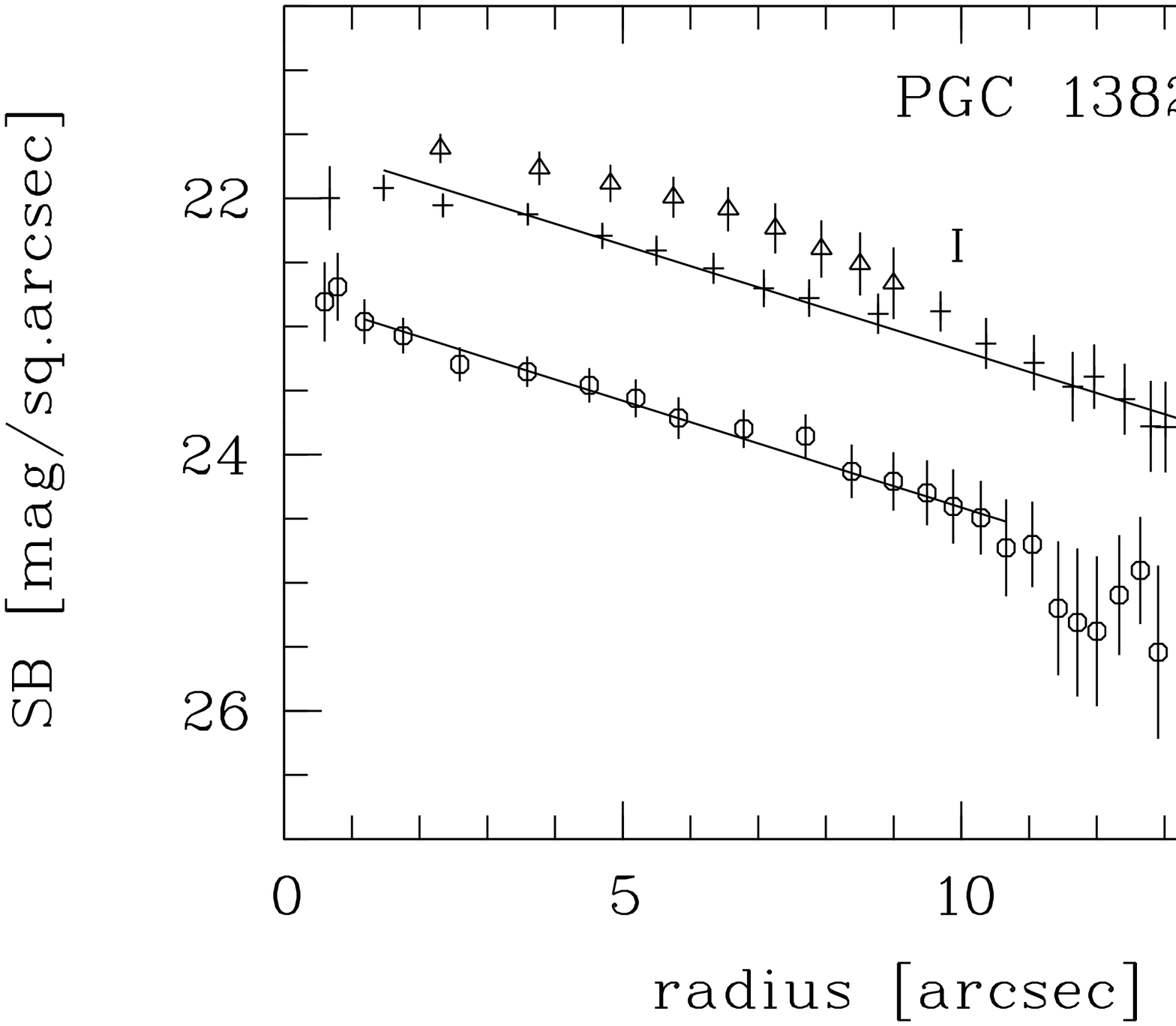}
\hspace{20mm}
\includegraphics[height=45mm,width=42mm]{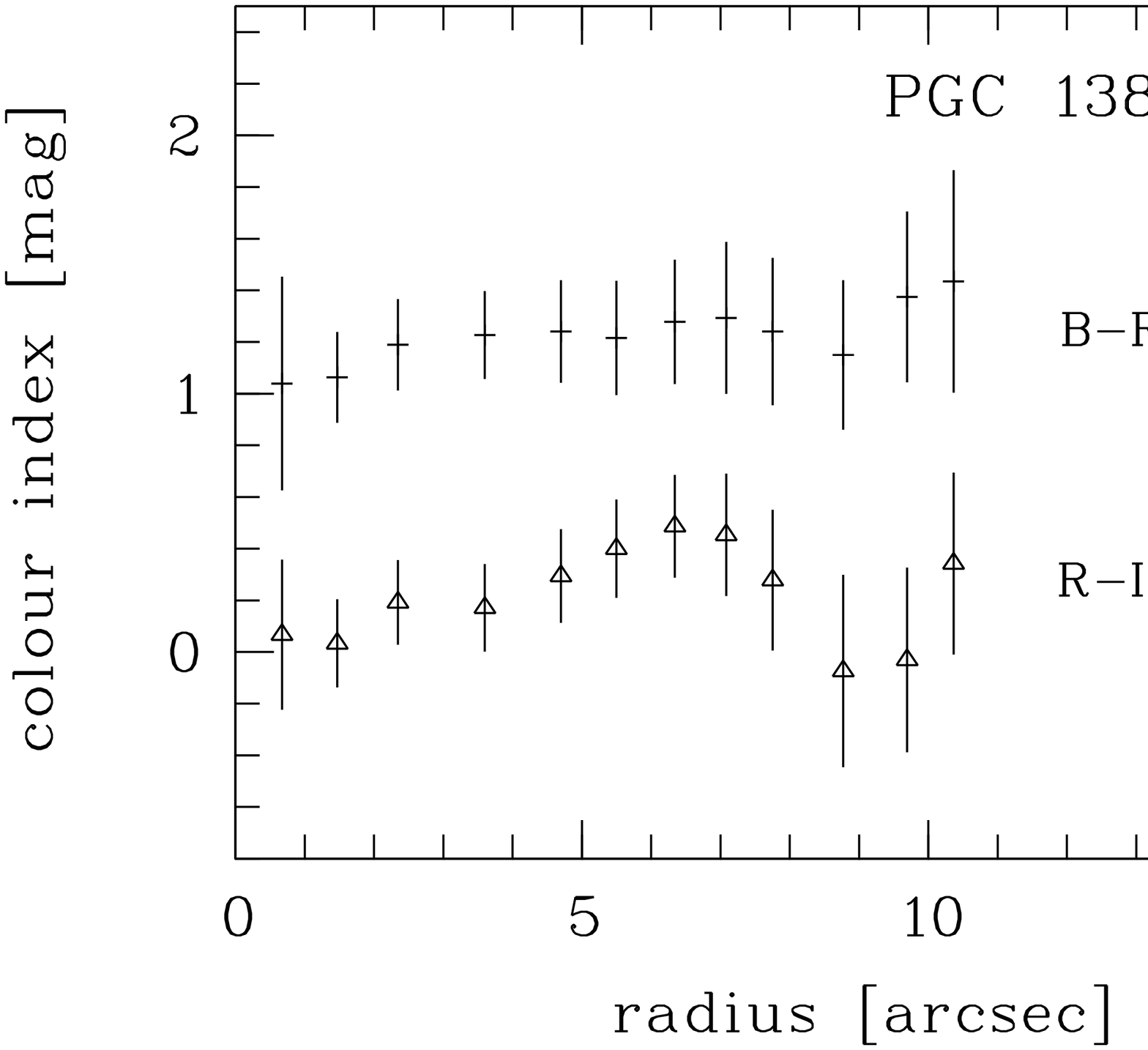}
\caption{Light distribution in PGC 138291. {\bf Left}:
the DPOSS red image 
(the field size is $\sim 2' \times 2'$, the north is top and the east is to the left);
 {\bf centre}:
radial surface brightness profiles in $B, R$, and $I$ with exponential disk models in
$B$ and
$R$ (continuous lines); {\bf right}: the colour-index profiles.
}
\label{LEDA.fig}
\end{figure*}

This galaxy was probably first discussed by vM83 
who referred to it as an \lq\lq edge-on\rq\rq ~galaxy.
Because its poor visibility only the position, velocity and \ion{H}{i}~flux are
reported in the literature. We derived its 
integral magnitudes and light distribution characteristics, using our
calibrated 
DPOSS frames. The most severe problem was proper subtraction
of the bright stellar halo. Fortunately, the stellar halo appears
nearly axisymmetric relative to the N-S stellar spike (but clearly
non-symmetric relative to the E-W spike). Therefore, we flipped the
image around the N-S spike, removed all other stellar images except
the bright star itself, and finally subtracted the flipped and cleaned
image from the original one.  
The derived $SB$ profiles
(Fig.~\ref{LEDA.fig}) become noise-dominated at relatively high
surface brightnesses. 
Nevertheless, we consider the central portion of the nearly exponential $SB$
profiles within $\leq 25.0$ mag~arcsec$^{-2}$, and $r \leq$ 11\arcsec  
as reliable. The galaxy is very flat with an axis ratio
of $b/a$ = 0.18, and it has generally a smooth regular shape. 
Neither the $SB$ profiles nor the optical images
prove the presence of a bulge while a tiny bulge may still remain hidden by
a dominating disk component, seen edge-on. 
The mean
colours ($B-R \simeq 1.0, R-I \simeq 0.2$) could have been explained by a   
dominating population of very young (a few Gyrs) and metal-poor ($\la$ 1/5Z$_{\sun}$) 
stars and probably low ISM/dust content  
in the LSB disk of this galaxy 
(Bruzual \& Charlot \cite{bruzual03}). 
Approximate dimensions of the galaxy were measured on the DPOSS red 
frame: $D_{25} \times d_{25} \simeq 1.1\arcmin \times$ 0.15\arcmin~that
yields a linear diameter of about 12.5 kpc.
The \ion{H}{i} contours in this \lq\lq edge-on\rq\rq galaxy as given by 
vM83 (his Fig. 29) appear regular.
%
\begin{figure*}
\includegraphics[height=51mm]{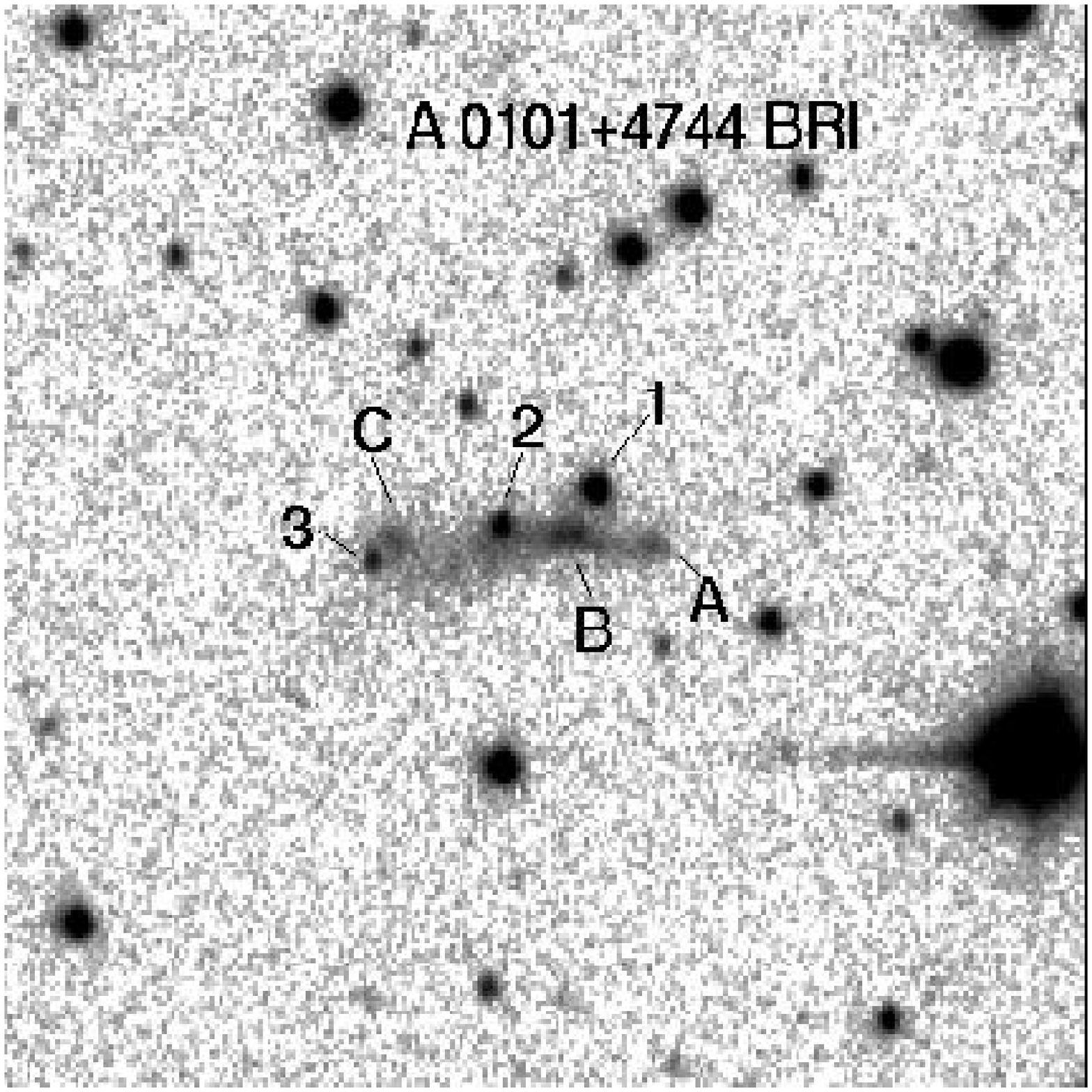}
\includegraphics[height=51mm,width=45mm]{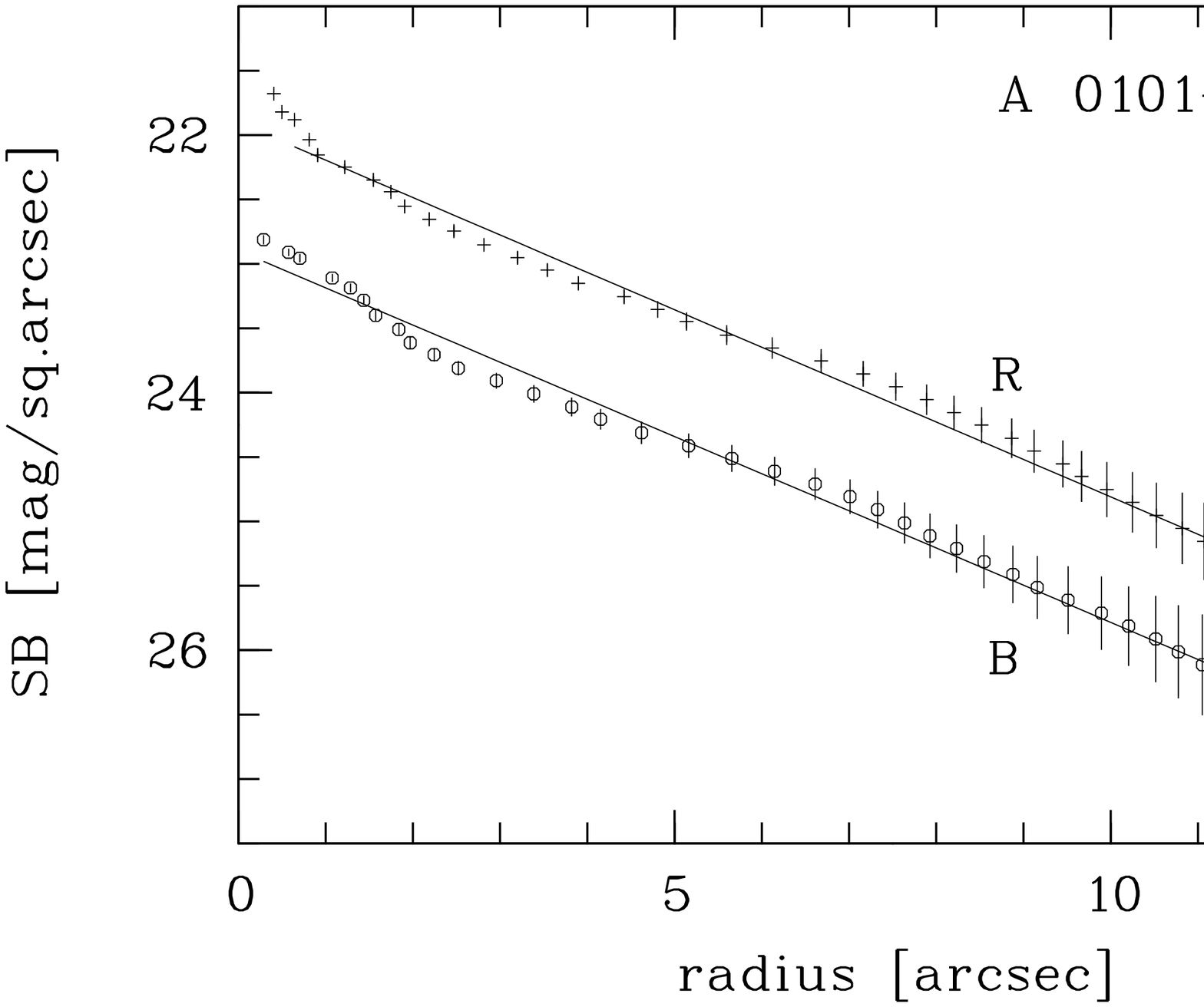}
\hspace{18mm}
\includegraphics[height=51mm,width=45mm]{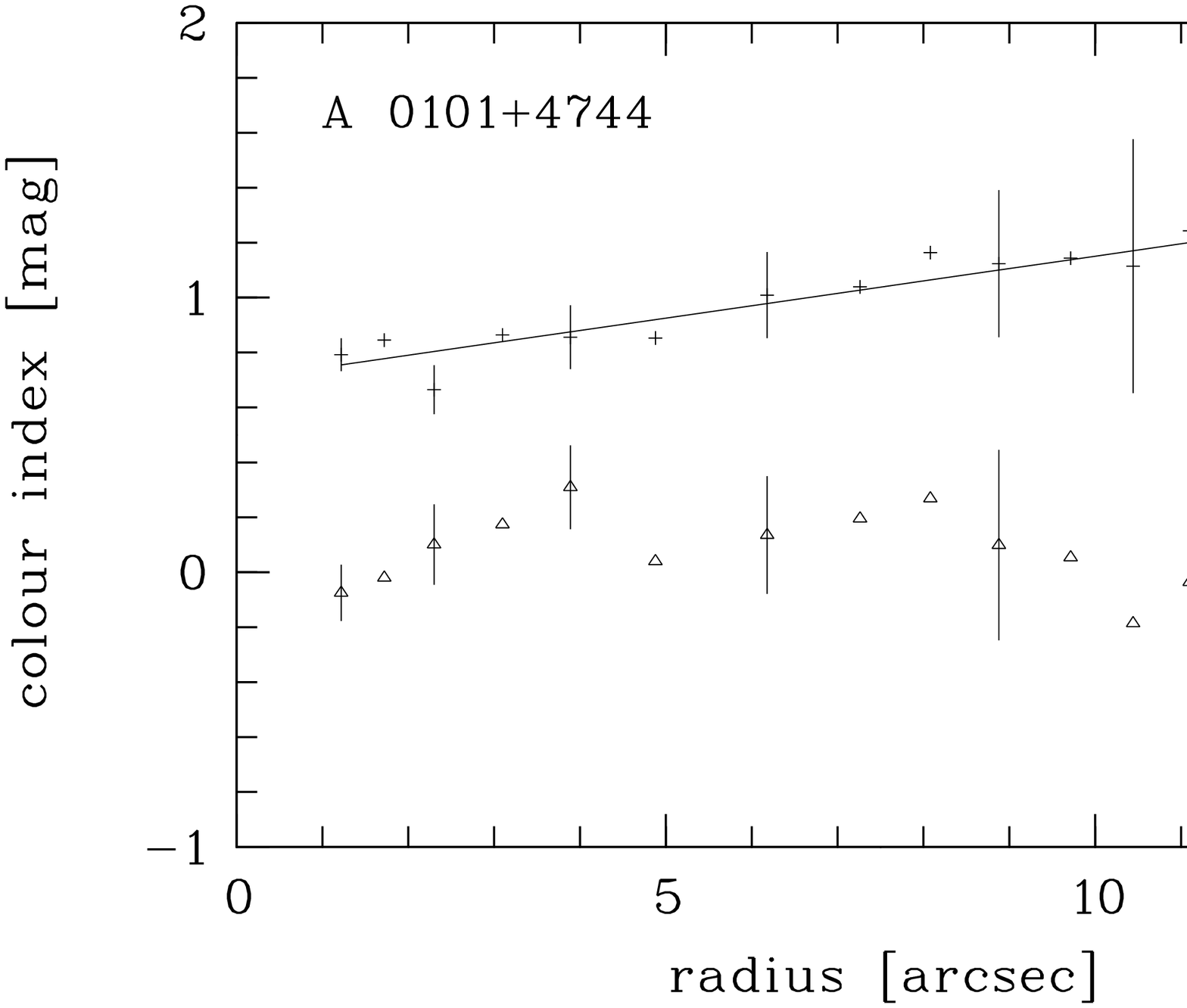}
\\
\caption{Light distribution in an anonymous galaxy A 0101+4744.
            {\bf From left to right: 1)} the $BRI$ composite image.
                        Unresolved stellar knots
                       are labelled with 1, 2 and 3; resolved 
                       probable star-forming regions in the galaxy are
                       labelled with A, B and C.
                       The field size is $2'\times 2'$;
                       the north is top and the east is to the left.
            {\bf 2)}  radial surface brightness ($SB$) profiles in
                       $B$ and $R$. The lines represent the fits of
                       the exponential disk model.
            {\bf 3)} colour-index profiles $B-R$ and $R-I$.
                       Typical errors are shown by bars.}
\label{IC65F1Dw}
\end{figure*}
\subsection {A 0101+4744} 
\noindent
This is a faint LSB anonymous galaxy located $\sim 5'$ (56 kpc in projection)  
to the NE of the IC 65. This galaxy was probably first discussed
by vM83 who describes it as \lq\lq ...  a barely
significant \ion{H}{i} detection coinciding with faint optical feature ...\rq\rq. 
This galaxy has an irregular head-tail 
 shape with a major light concentration
in a luminous and blue knot B (Fig.~\ref{IC65F1Dw}), which is located in the
western part of the galaxy, and a faint diffuse curved tail on the
opposite side.
The very blue colours of the knot B ($B-R = 0.6 \pm$ 0.15 and $R-I = 0.1 \pm$ 0.2) 
may correspond to a recently formed ($\leq$ 100 Myr) stellar
supercluster (Vennik \& Hopp \cite{vennik07}). 
Two further diffuse knots are distributed in the western (knot A) and eastern
(knot C) periphery. Those knots are redder ($B-R \simeq$ 1.0 for the knot
A, and $B-R \simeq$ 1.2 for the knot C), when compared to the knot B.
Three Galactic
stars (labelled with 1, 2, and 3 in Fig.~\ref{IC65F1Dw}) 
could have been distinguished through
their stellar PSF, 
and much redder colours  ($B-R = 2.0 \pm$ 0.1 and $R-I = 0.7 \pm$ 0.05), when
compared to the colours of the underlying LSB disk of the galaxy.
  
The light distribution in this LSB galaxy is nearly exponential.  
The $B-R$ colour index profile (starting from the 
knot B) shows a marginal radial colour (i.e. stellar population)
gradient in the underlying disk and/or in bright knots. 
Because of the concordant \ion{H}{i} radial velocity 
and the LSB morphology this galaxy could be considered as a certain dwarf irregular
member of the group with an absolute blue magnitude of $M_B$ =
-15.4, a major axis length of about $D_{25} \simeq$ 5 kpc 
and the exponential scale length of $h_B \simeq$ 0.7 kpc. 
\subsection {A 0100+4756} 
\begin{figure*}
\includegraphics[width=47mm]{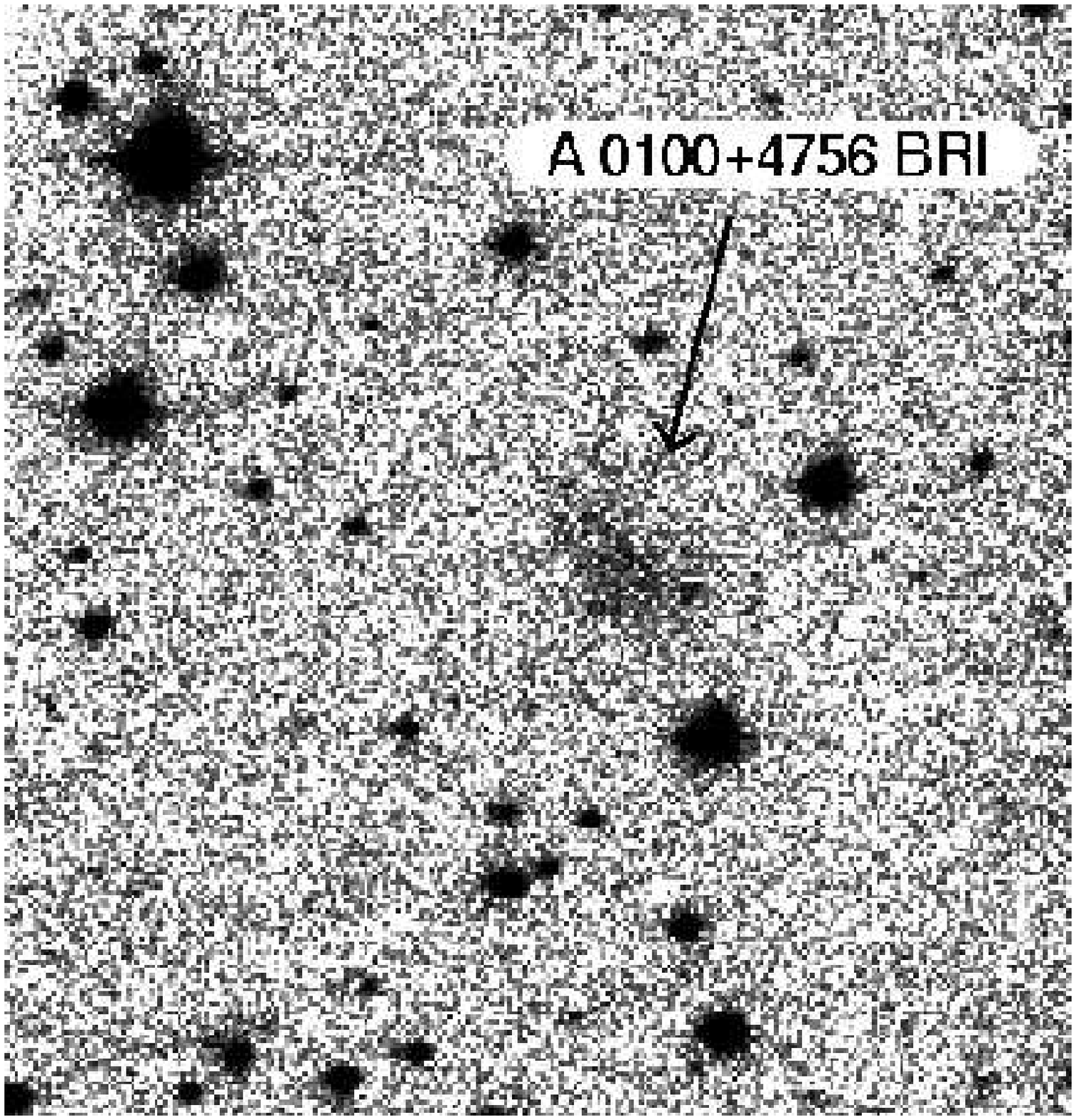}
\hspace{3mm}
\includegraphics[height=49mm,width=45mm]{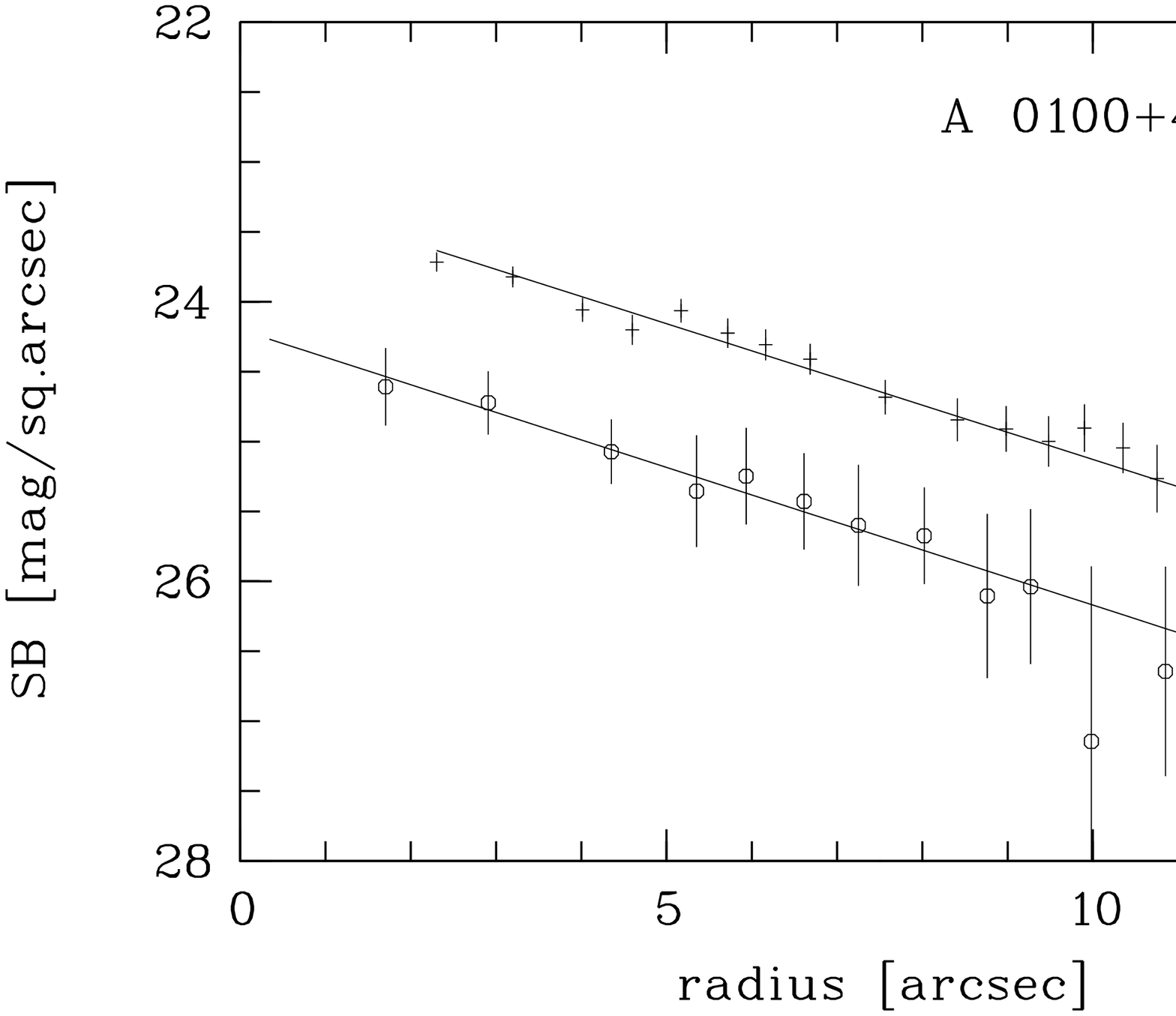}
\hspace{18mm}
\includegraphics[height=49mm,width=45mm]{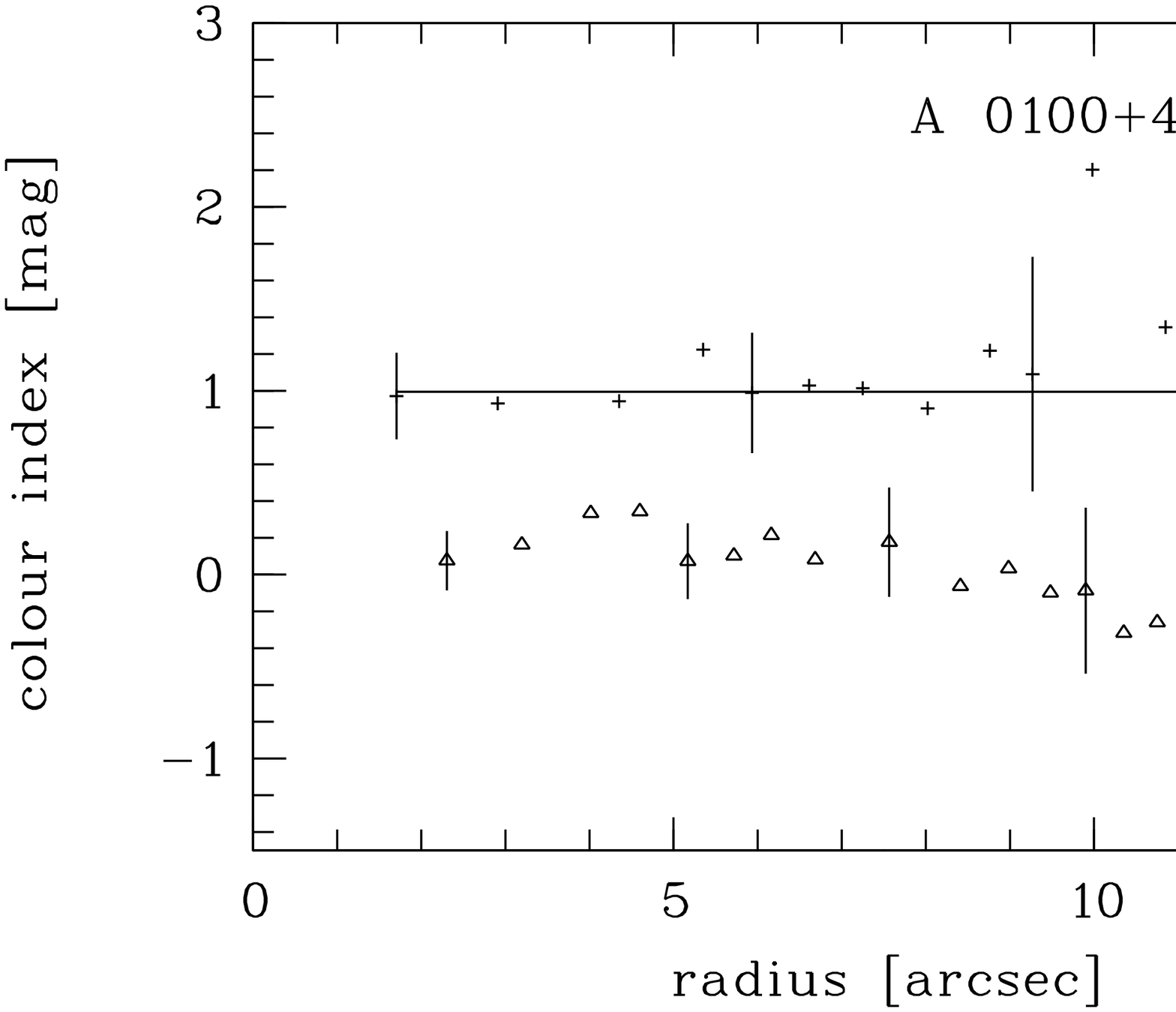}
\\
\caption{Light distribution in an anonymous galaxy
         A 0100+4756. Coded as in Fig.~\ref{IC65F1Dw}
}
\label{F2Dw}
\end{figure*}

This is a very faint LSB object located $\sim 5'$ (56 kpc in projection) to the SW of the UGC 622. 
The image (Fig.~\ref{F2Dw}) has an amorphous light distribution with a
maximal blue $SB$ of 23.9 $\pm$0.3 mag~arcsec$^{-2}$ and without any
luminous knots.  
The derived $SB$ profiles are nearly exponential 
and the $B-R$ and $R-I$ colour-index profiles are essentially flat, 
showing that nearly homogeneous stellar populations are distributed throughout the 
LSB  disk. 
The de-reddened mean colours $B-R = 0.76 \pm$ 0.2 and $R-I = 0.15 \pm$ 0.2  
are nearly consistent with a dominating population of metal-poor ($\sim$ 1/20Z$_{\sun}$) 
and young 
stars (Bruzual \& Charlot \cite{bruzual03}).

The distance of this object is unknown. 
Its LSB dwarfish morphology, nearly exponential SB profile and, particularly, 
its blue colours very similar to those
of the confirmed member A 0101+4744, and other two studied dwarf galaxies, as well as its
location close to UGC 622 encourage us
to consider it as a possible new dwarf member of the IC 65 group of galaxies.
If it is true, then its absolute magnitude is 
$M_{\rm B}$ = -14.4 (but $M^{\rm exp}_{\rm B} \simeq$ -15.0), the
diameter is $D \sim$ 3.6 kpc, and the scale length $h \simeq$ 1.0 kpc.
However, we should be aware that, according to Karachentsev et al. (\cite{karachentsev00}),  
an isolated Galactic cirrus with very low $SB$ can easily be confused with an irregular galaxy.  
\begin{figure*}
\includegraphics[height=46.5mm]{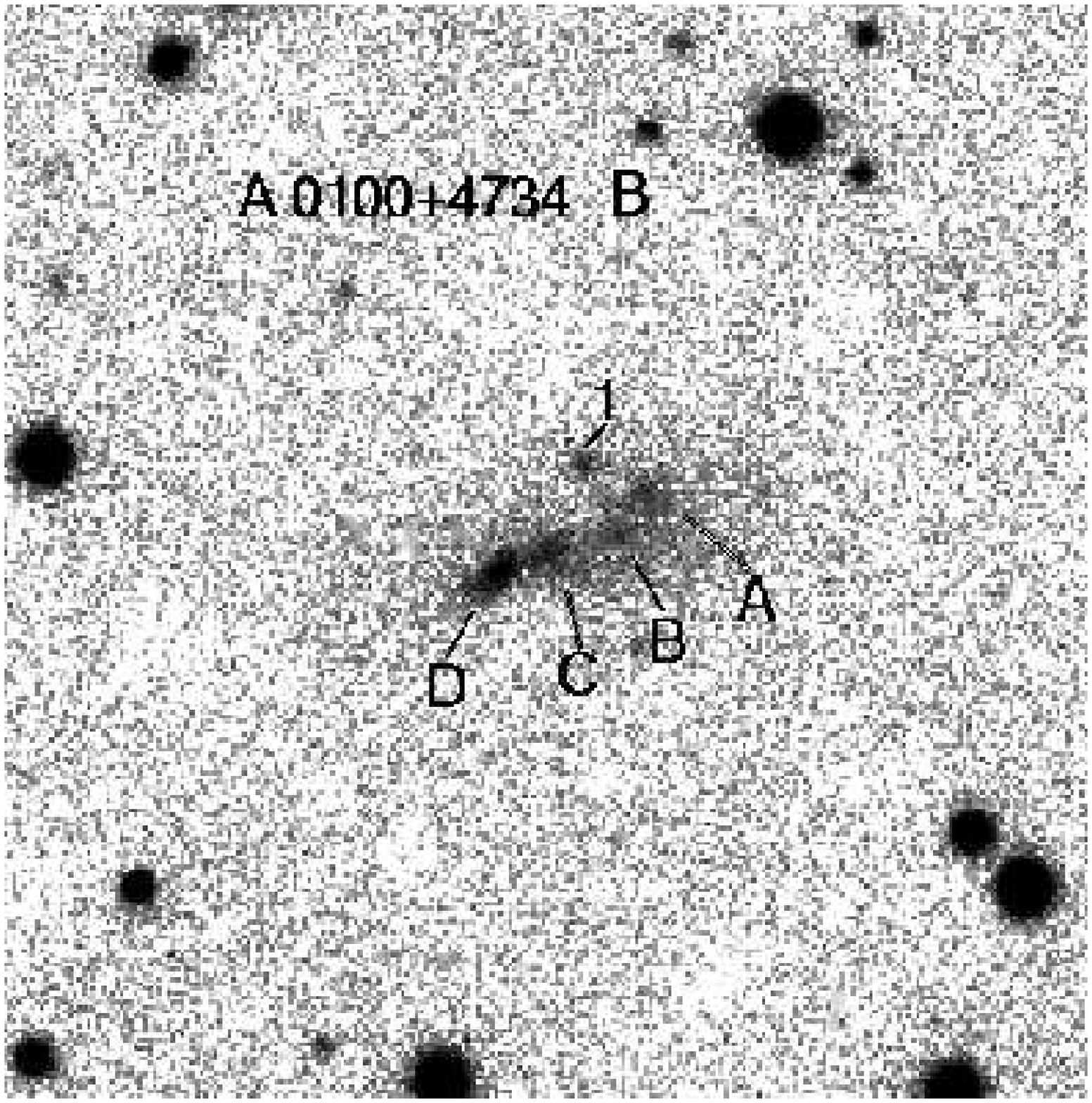}
\hspace{2mm} 
\includegraphics[height=46.5mm,width=45mm]{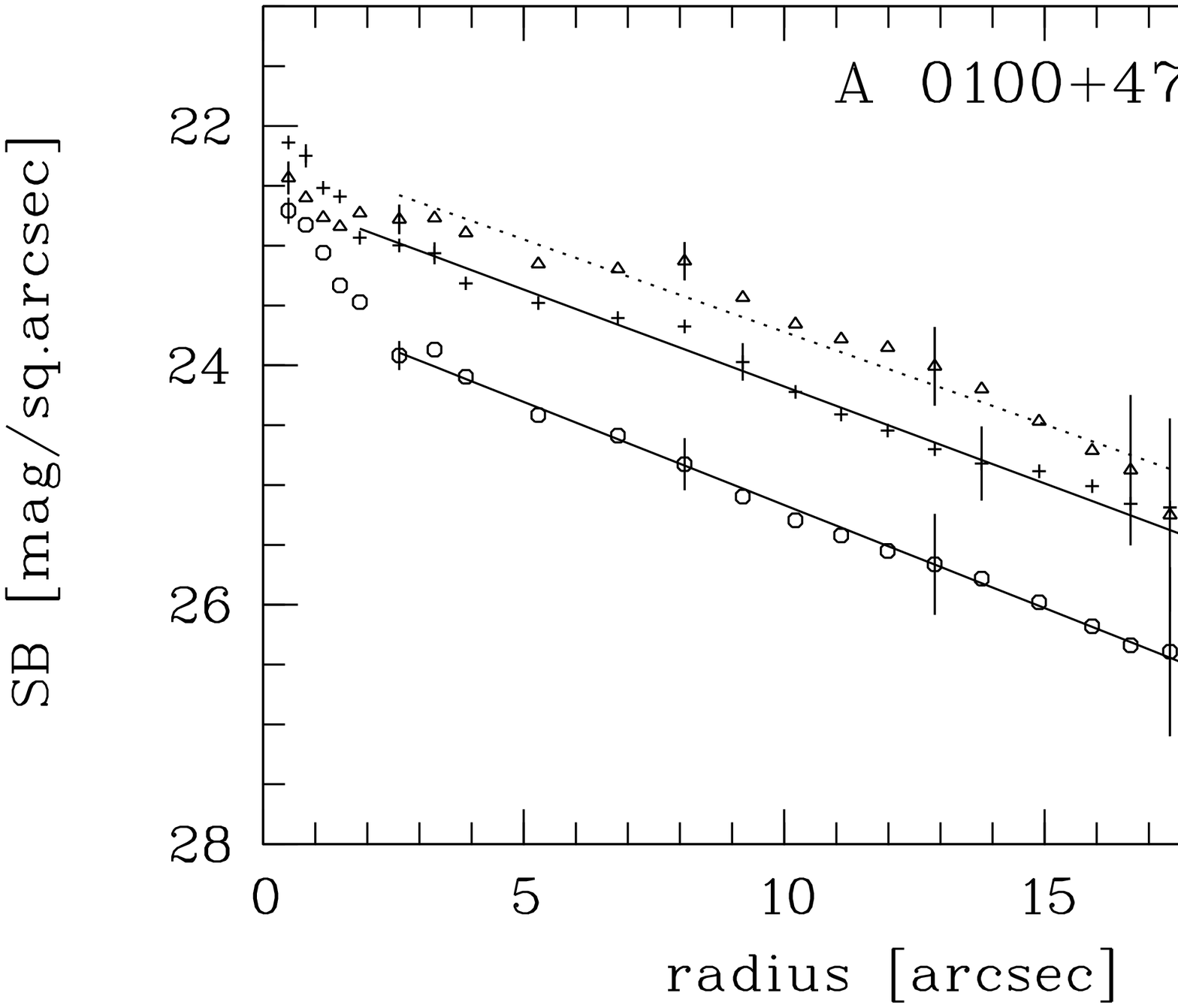}
\hspace{18mm} 
\includegraphics[height=46.5mm,width=45mm]{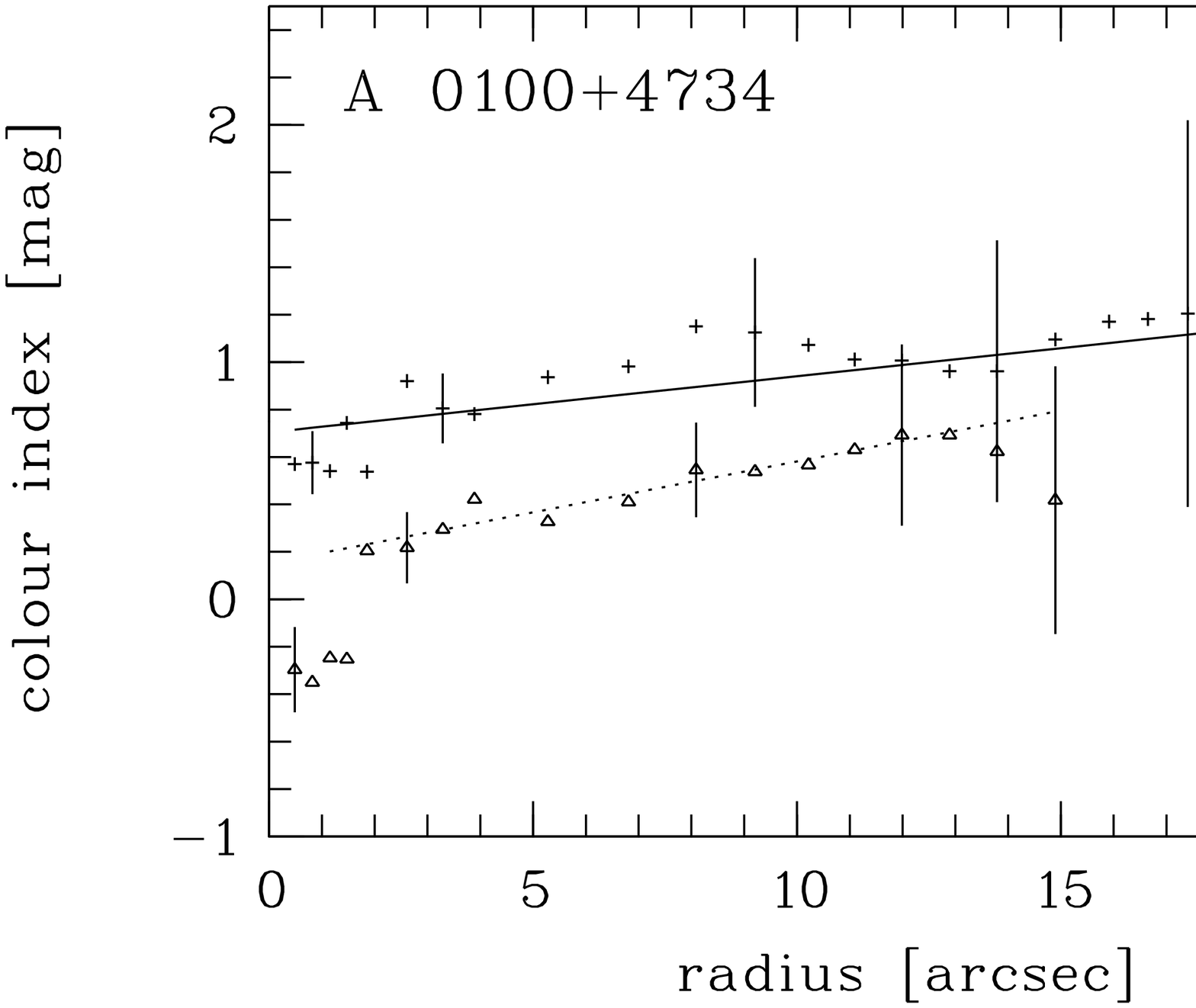}
\\
\caption{Light distribution in an anonymous galaxy A 0100+4734. Coded as
 in Fig.~\ref{IC65F1Dw} 
}
\label{F5Dw}
\end{figure*}

\subsection {A 0100+4734}
\noindent This is an anonymous irregular galaxy located
$\sim 10.3'$ {\bf (115 kpc in projection)} 
to the SW of the principal galaxy IC 65. It contains a
number of bright resolved knots labelled A, B, C, and D  
(Fig.~\ref{F5Dw}) with the brightest knot D located at the
eastern periphery and, therefore, giving the galaxy a head-tail appearance.  
The luminous blue knots which are probably centres of active
star formation, are embedded into the diffuse LSB underlying component of the  
size of $D \times d \simeq 49'' \times 19''$.
The radial $SB$ profiles
are well approximated by a simple exponential disk model. 
The brightest knot 
D is also the bluest one with colours of
$B-R = 0.4 \pm0.1$ and $R-I = -0.15 \pm0.2$, probably indicating very young ages 
of the stars in the range of 10 - 100 Myrs in it (Vennik \& Hopp \cite{vennik07}). 
The individual knots show a
range of colours with the fainter knots getting redder along the
distance from the brigthest/bluest knot D. 
These colour
variations could probably be interpreted as an age effect, and they are
generally used to support the self-propagating star formation mode
hypothesis (e.g. Gerola et al. \cite{gerola80}). 
The colours of the underlying disk are similar to those of the A 0101+4744. \\
An attempt to measure the \ion{H}{i} flux of this actively  
star-forming galaxy is described in Appendix B.  
Recent spectroscopic observations with the Hobby-Eberly Telescope have shown that this particular dIrr galaxy 
may be located in front of the IC 65 group (Hopp et al. \cite{hopp07}). 
\subsection {A 0101+4752}
\begin{figure*}
\includegraphics[height=46.5mm]{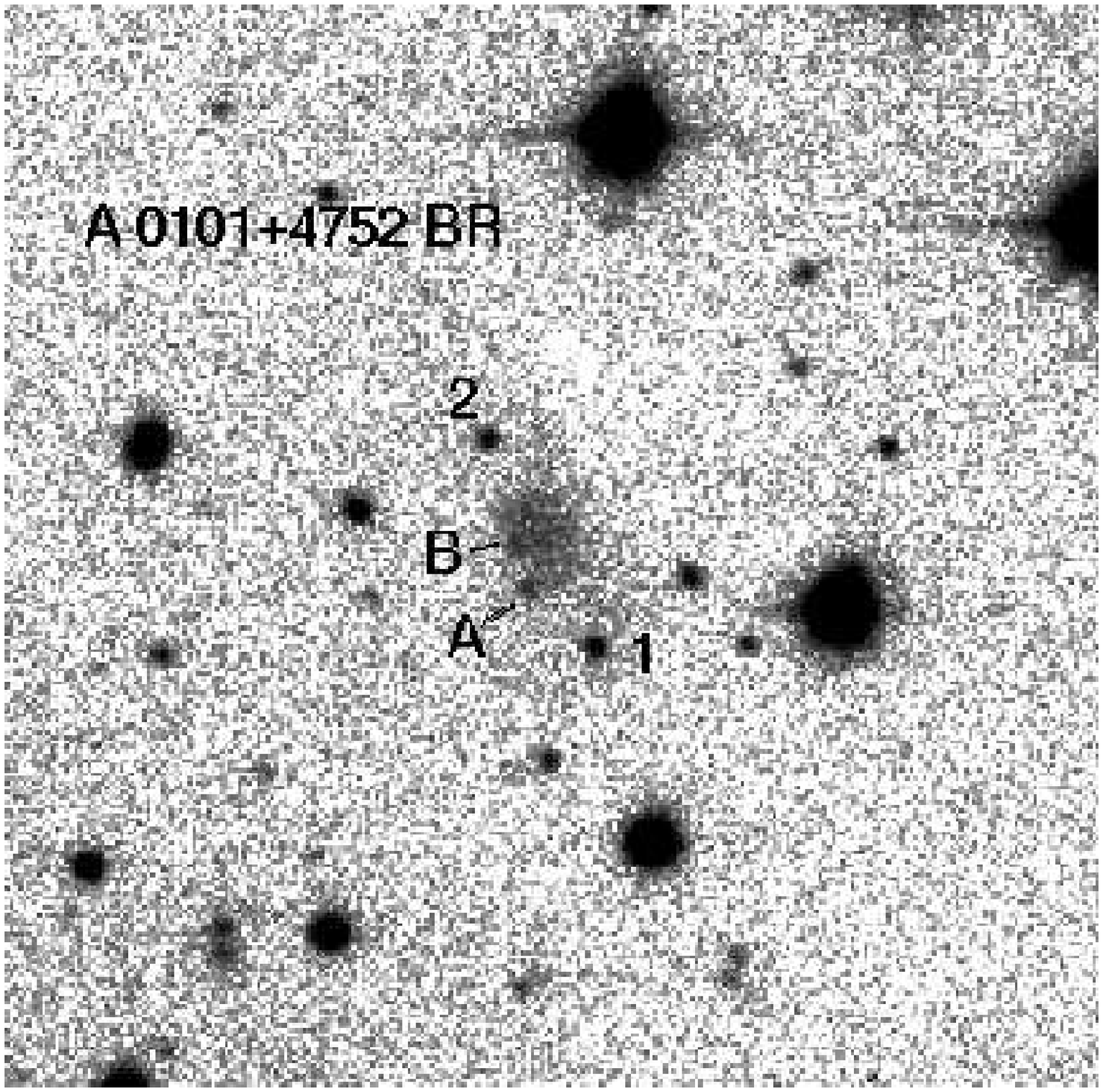}
\hspace{2mm}
\includegraphics[height=46.5mm,width=44mm]{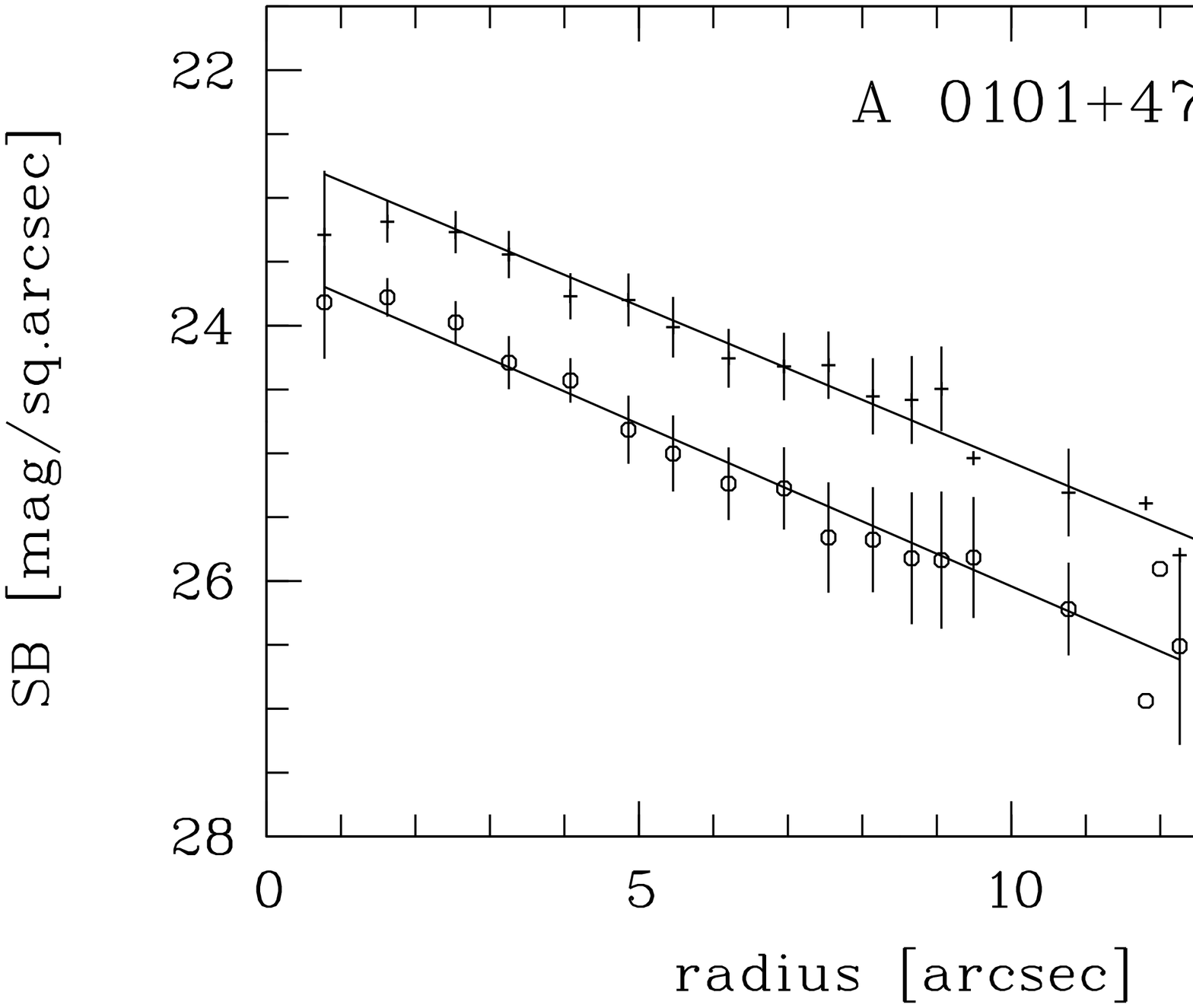}
\hspace{19mm}
\includegraphics[height=46.5mm,width=45mm]{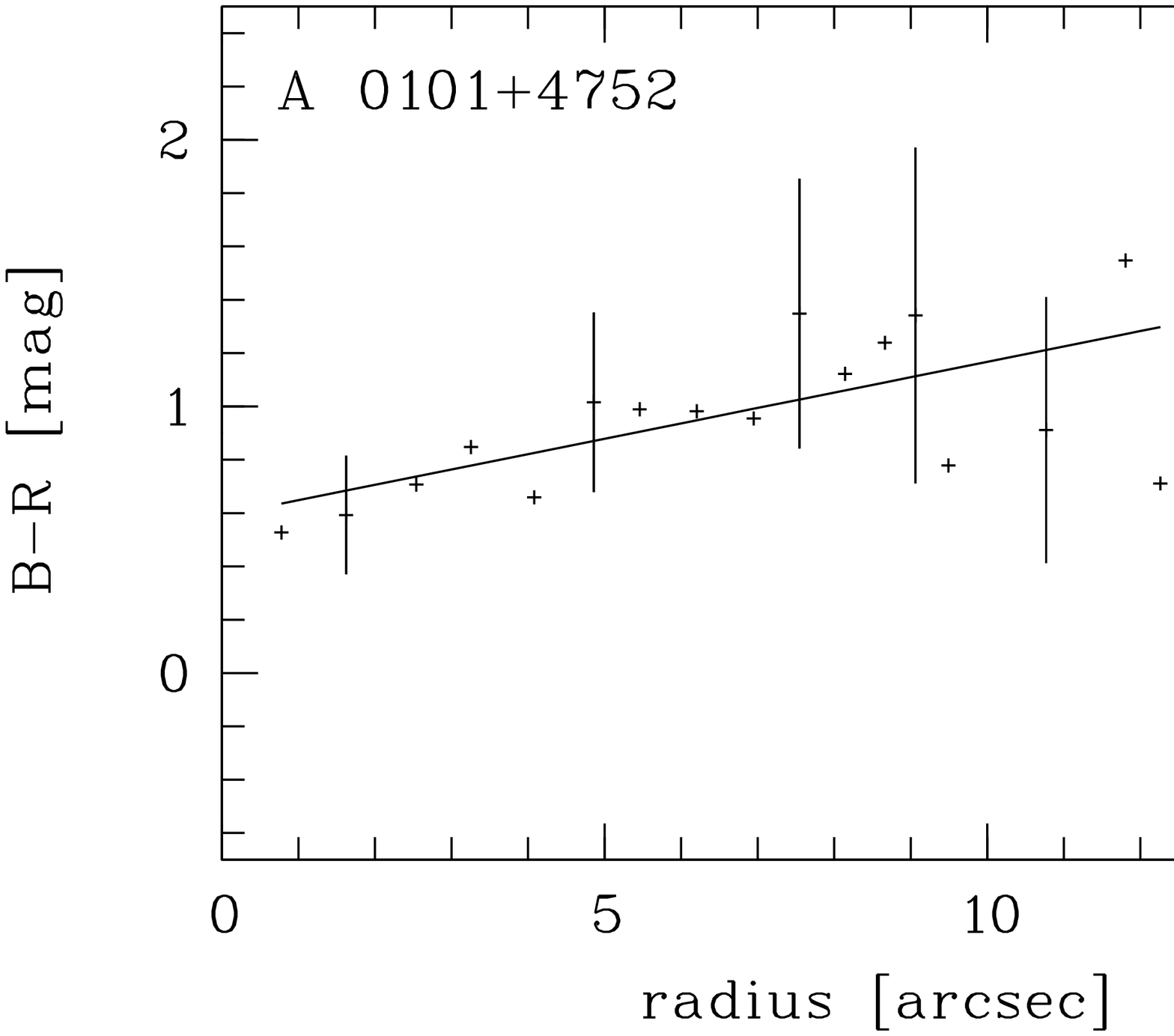}
\\
\caption{Light distribution in an anonymous galaxy A 0101+4752.
Coded as in Fig.~\ref{IC65F1Dw}
}
\protect\label{F6Dw}
\end{figure*}

This is a faint irregular galaxy located 14$'$ to the NE of IC~65,  
with a smooth LSB underlying light distribution and several
faint resolved knots, labelled with A and B in Fig~\ref{F6Dw}, embedded in it.  
Faint arc-like features are evident on the northern and southern 
periphery of this galaxy, both on our CCD frames and on deeper DPOSS images. 
The bright non-stellar knot B is extremely blue 
($B-R \simeq$ 0.24) when compared to the colours of another, but a semi-stellar knot A
($B-R \simeq$ 1.2) or to the colours of two 22$^{\rm nd} B$-magnitude Galactic
stars (labelled with 1 and 2) of $B-R \simeq 2.0 \pm$ 0.15.
The blue colour of the knot B is typical of a very young and short-lived starburst, 
which is not unusual in a dIrr galaxy.  
The azimuthally averaged 
$SB$ profiles (Fig.~\ref{F6Dw}) are nearly exponential. The
$B-R$ colour profile is very noisy and the reddening towards the
periphery may appear unreliable. 
We have checked our CCD
photometry by using the calibrated
DPOSS blue and red images. As a result, we obtained slightly more
extended $SB$ profiles, 
and the independently derived colour index profile confirms
the reddening tendency towards the periphery.\\
The redshift of this LSB galaxy are not known yet. So
we can only speculate about its relation to the IC 65 group of
galaxies on the morphological and colour grounds. 
Assuming that A 0101+4752 is located at the distance of the IC 65
group of galaxies, we derive the follwing global characteristics: $M_B = -14.9$, 
linear diameter of $\sim$ 4.5 kpc, scale length $\sim$ 0.8 kpc,
and rather blue integral colours of $B-R \simeq$ 0.65, and $R-I \simeq$ 0.20, 
which are also similar to the
colours of other new dwarf galaxy candidates of the 
group.  Nevertheless, A~0101+4752 may also be associated with the
bright early type galaxy MCG +08-03-006,  
which is only 2.1$'$ away, and certainly located in the background of the IC 65 group 
(Hopp et al. \cite{hopp07}).  
\section {Discussion}
\subsection {Dynamical properties of the group} 
\begin{table}[]
\caption{\normalsize Gaseous and dynamical masses and mass-to-light 
          ratios of \ion{H}{i} detected galaxies of the group} 
\protect\label{Dynamic.tab}
\begin{center}
\begin{small}
\begin{tabular}{lccccc}
\hline\hline 
 Galaxy     & $L_B^{g,i}$ & ${\cal M}$(\ion{H}{i}) & $\frac{{\cal M}(\ion{H}{i})}{L_B}$ & 
${\cal M}_{\mathrm dyn}$ & $\frac{{\cal M}_{\mathrm dyn}}{L_B}$ \\
 & [$10^{10} L_\odot$] & [$10^{10} {\cal M}_\odot$] & & [$10^{10} {\cal M}_\odot$] \\ 
\hline
 (1)        &     (2)      &    (3)      &  (4)     &   (5)       &  (6)   \\
\hline
\\
 UGC 608    & 0.46         & 0.44        & 0.97     &    2.82    &  6.1 \\
 PGC 138291 & 0.16         & 0.14        & 0.91     &    1.00    &  6.2 \\
 UGC 622    & 0.98         & 0.315       & 0.32     &    4.44    &  4.5 \\
 IC 65      & 4.16         & 1.425       & 0.31     &    17.51   &  4.2 \\
 A0101+4744 &  0.023:      &             &          &    0.38:    &  16: \\
\hline
\end{tabular}
\end{small}
\end{center}
\end{table}
Combining the new photometry with the \ion{H}{i} fluxes and
rotational velocities given in vM83 we
calculate the gaseous content and dynamical masses of bright group
members and of the dwarf companion galaxy A~0101+4744. The results are
listed in Table~\ref{Dynamic.tab}, where the data are arranged as
follows: (1) name of the galaxy; (2) the $B$-band
luminosity corrected for the Galactic and internal absorption; 
(3) total \ion{H}{i} mass, calculated as ${\cal M}$(\ion{H}{i})/${\cal M}_{\odot}
= G^{-1}~D^2\int S dV$, where $D$ = 38.5 Mpc is the
distance of the group and $\int S dV$ is the total \ion{H}{i} emission in
(Jy~km~s$^{-1}$); (4) the \ion{H}{i} mass to blue luminosity ratio; 
(5) dynamical mass within the outermost 
rotational velocity measurement, calculated as ${\cal M}_{\rm dyn} (<R) =
G^{-1}~R~V_{\rm rot}^2(R)$; (6) the dynamical mass to blue luminosity
ratio.
 
The principal galaxy of the group, IC~65, considerably dominates 
other group members, emitting nearly 3/4 of the group total blue
light and containing $\sim$ 2/3 of the dynamical mass of the secure members 
of the group.
UGC~608 and PGC~138291 are rich in neutral
hydrogen, as is typical of late type spirals. IC~65 and UGC~622 have a
smaller gaseous fraction in agreement with the statistics 
of Sbc and Sc galaxies (Roberts \& Haynes \cite{roberts94}).  
The individual mass-to-light
ratios of the four bright group members fit into the range of
typical ratios of the given morphological types (Roberts \& Haynes
\cite{roberts94}). For the dwarf irregular galaxy A~0101+4744 in the
field of IC~65, vM83 measured the velocity
width $\Delta V$ = 130 km s$^{-1}$. If we adopt a crude relation
$V_{\mathrm rot}^{\mathrm max} \simeq \Delta V/2$ = 65 km s$^{-1}$ and the  (optical) 
radius of $\sim$ 3.9 kpc we obtain an estimate of its dynamical mass 
${\cal M}_{\mathrm dyn} \simeq 3.8 \times 10^9 {\cal M}_{\odot}$ and 
${\cal M}_{\mathrm dyn}/{L_B} \simeq$ 16 in solar units.
Since a typical isolated dIrr galaxy has a stellar and gaseous ${\cal M}/L_B \simeq$ 0.8 - 1 
in solar units (van Zee 2001, Roberts \& Haynes 1994), then this particular
star-forming dwarf galaxy appears to be severly dominated by dark matter.

Using the available exact \ion{H}{i} radial velocity data we attempt to
estimate the total mass and the fraction of dark matter 
distributed in this group. The virial mass of the group can be
calculated according to Karachentsev (\cite{karachentsev70}) as
follows:
\begin{equation}
 {\cal M}_\mathrm {vir} = 3 \pi~G^{-1} n(n-1)^{-1}~R_{\rm H}~\sigma_V^2,   \protect\label{Mvir}
\end{equation} 
where $n$ is the number of galaxies in the group, $\sigma_V$ is the
dispersion of radial velocities, and the mean harmonic radius $R_H$ is
calculated as
\begin{equation}
R_{\rm H} = D~\sin \left[\frac{n(n-1)}{2\sum_i \sum_{j > i}\theta_{ij}^{-1}}\right], \protect\label{Rharm}
\end{equation}
where $D$ is the distance of the group and $\theta_{ij}$ is the angular 
separation between the galaxies
$i$ and $j$.  We obtain the following dynamical characteristics of the
group: $\sigma_V = 77 \pm$ 13 km~s$^{-1}$, $R_{\rm H} = 135 \pm$ 24 kpc, the mean linear
projected radius $<R_{\mathrm p}> = 192 \pm$ 34 kpc, and the virial mass ${\cal M}_\mathrm {vir}
= (2.16 \pm 0.97) \times 10^{12} {\cal M}_{\odot}$.  Using the total blue luminosity of
the group $L_{\mathrm T} = 5.78\times10^{10}~L_{\odot}$, as determined in this
work, we obtain the dynamical mass-to-blue light ratio $({\cal M}/L)_B
\simeq 37 \pm 17 ({\cal M}/L)_{\odot}$. The rms errors are estimated using the jackknife method 
(e.g. Efron \cite{efron81}).   
The determination of individual virial masses of small groups is uncertain because of the projection 
effects, temporary departures from overall equilibrium and incomplete redshift data.  
Traditionally, spherical symmetry and isotropic velocities are assumed, however, the groups may
be actually flat aggregates (e.g. Haud \cite{haud90}), 
or prolate spheroids (Plionis \cite{plionis04}). 
Various numerical simulations of the dynamical evolution of poor galaxy systems 
(e.g. Aarseth \& Saslaw \cite{AS72}; Barnes \cite{barnes85}; Mamon \cite{mamon93}, \cite{mamon07}) 
have demonstrated 
that as net effect of the given uncertainties the true mass is severely underestimated by 
application of the virial theorem. 

Since the principal galaxy IC 65 is essentially
dominating the group dynamics, we can consider the other 
group members as the companions moving in arbitrarily oriented
Keplerian orbits around the principal galaxy, and apply an alternative mass 
estimator, the projected mass method, which is found to be generally more 
reliable than the virial theorem (Bahcall \& Tremaine \cite{BT81}). 
Assuming isotropic velocity distribution,  
which is proper when some relaxation has taken place in the past, 
we can compute the projected mass estimator as:
\begin{equation}
{\cal M}_\mathrm {proj} = \frac{16}{\pi Gn} \sum_{i=1}^{n} \Delta V_i^2 \times R_i = (2.89 \pm 1.6)\times 10^{12} {\cal M}_\odot.  \protect\label{Mproj}
\end{equation}
The standard deviation of the projected mass in Eq. (9) accounts for statistical uncertainties 
1.263~$n^{-1/2}$. An additional {\bf effect} related to the uncertainties in the eccentricity distribution 
can be up to a factor of 1.5 (Bahcall \& Tremaine \cite{BT81}). In effect, we can conclude a 
reasonably good agreement between these two dynamical mass estimates. 

The dynamical state of the
group is characterized by its crossing time. We used two different crossing time definitions 
to check for possible variances. First, the traditional virial crossing time, as defined 
by Huchra \& Geller (\cite{HG82}):
\begin{equation}
t_c^\mathrm {vir} = \frac{3}{5^{3/2}~H_0}\times \frac{\pi~R_{\rm H}}{\sigma_V} = 1.4 \pm 0.4~Gyr.
\end{equation}  \protect\label{Tvir} 
Another estimator, the linear crossing time $t_c^{\rm lin}$ (Gott \& Turner, \cite{gott77}), 
which should  be more stable in the presence of various irregularities as close pairs, central  
condensations etc: 
\begin{equation}
t_c^\mathrm {lin} = \frac{2}{\pi~H_0}\times \frac{<V><\sin \theta_{ij}>}{<V_i-V_j>} = 1.3 \pm 0.4~Gyr. \protect\label{tcros} 
\end{equation}
Both estimators yield concordant results, and the short crossing time shows that the group as a 
whole could have been dynamically relaxed. 
For comparison, the estimated crossing times in nearby poor groups
range from 1.8 to 5.9 Gyr with a median around 2.3 Gyr (Grebel \cite{grebel07}).
However, occurrence of an amount of diffuse IGM would increase 
the predicted relaxation time computed for {\bf a} point-mass configuration and, consequently, short 
$t_c$ does not necessarily indicate significant dynamical evolution of small bound groups. 
Therefore, interpretation of apparently relaxed small 
groups is not unique and even groups with short crossing times ($t_c \sim 0.1 H_0^{-1}$)
may still be in a pre-virialized collapse phase (Barnes \cite{barnes85}, 
Nolthenius \& White \cite{nw87}).
Mamon (\cite {mamon93}, \cite {mamon07}) modeled the cosmo-dynamical evolution of isolated 
galaxy systems and found that the crossing time suffers from degeneracies between  the expansion 
and initial collapse phases and also between the late collapse and rebound phases. 
He suggested that these degeneracies could have been partly solved  
by means of combining crossing times with the virial to true mass ratio 
(${\cal M}_\mathrm {VT}/{\cal M}_\mathrm {true}$), or virial mass to the optical light ratio 
(${\cal M}_\mathrm {VT}/L_\mathrm {opt}$), when adopting an universal mass-to-light ratio. 
The location of a group in the plane defined by the dimensionless crossing time ($t_\mathrm {c}~H_0$) 
versus the virial mass to light ratio (${\cal M}_\mathrm {VT}/L_\mathrm {B}$) 
(e.g. Mamon \cite {mamon07}, Fig.~1) could help us to judge about the evolutionary status of a 
particular system. In this plot the IC 65 group is located well below the theoretical 
\lq fundamental track\rq of the group evolution, (which
has been determined assuming an universal ${\cal M}_\mathrm {true}/L_\mathrm{B} = 440~h$,  
where $h = H_0/100$~km~s$^{-1}$~Mpc$^{-1}$), 
in the region of collapsing groups, and within the area
 occupied by other low-multiplicity 
groups. In poor groups of galaxies the statistical noise in the virial mass-to-light ratio  
and crossing time estimates increases but the errors are obviously not sufficient in explaining 
the large number of groups well below the \lq fundamental track\rq~ unless the groups either have  
intrinsically lower ${\cal M}_\mathrm {VT}/L$ (e.g. because many group members are undergoing star-burst), or are 
caused by the bias near turnaround (Mamon \cite{mamon07}), or are caused by mass segregation between 
galaxies and dark matter (Barnes \cite{barnes85}).    
\subsection{Substructure and galaxy interactions in subgroups}  

Numerical simulations of small galaxy systems (e.g. Barnes \cite{barnes85}) 
have demonstrated that they tend to form 
subgroups during their dynamical evolution which ultimately merge to a single final remnant 
(an E or cD galaxy). 
Observationally it has been found (e.g. Karachentsev \cite{kara96}, \cite{kara05}) 
that nearby groups of galaxies tend to consist of two subgroups, which are 
respectively concentrated around two massive galaxies 
(e.g. MW+M31, CenA+M83, IC342+Maffei).   
By analogy with nearby groups it is tempting to divide also the LGG 16, 
despite of very poor statistics, into two compact 
subgroups with a projected separation of $\sim$ 220 kpc.
(For comparison, only 5 - 7 brightest/largest members of the Local Group could have
been detected at the distance of the LGG 16.)
The southern subgroup consists of the
principal galaxy IC~65 ($<V_{\sun}> = 2614 \pm$ 8 km~s$^{-1}$), 
and at least of one confirmed dIrr companion A~
0101+4744 (the NE companion). 
The group membership of another {\bf nearby} dIrr galaxy A~0100+4734 is not yet clear. 
The northern subgroup consists of three galaxies with comparable masses - UGC~
622, UGC~608, and PGC~138291 ($<V_{\sun}>_3 = 2689 \pm$ 66km~s$^{-1}$) - 
and the probable dwarf galaxy A~
0100+4756. The fourth, newly found dwarf galaxy candidate A~0101+4752
is located between the two subgroups and could be considered as a
\lq\lq free-floating\rq\rq~ dwarf member of the group. 

Little is known about the depth of this group along the line of sight. We can speculate that 
the northern subgroup could possibly be located behind of the principal galaxy IC 65,
as suggested by the radial velocity difference, 
or even could be a chance projection along the line of sight or is approaching the principal galaxy.
We can apply a crude check using the Tully-Fisher (hereafter TF) relation combining the published
rotation velocities and newly determined optical and NIR magnitudes. 
We compare our data with the TF relation of 31 galaxies in the UMa cluster of galaxies
(Verheijen \cite{verheijen01}) and of 17 galaxies in the Eridanus group (Omar et al. \cite{omar06}). 
Our data for three galaxies (IC 65, UGC 622 and UGC 608), 
with representitives from both subgroups, fit nicely the relation defined by both
comparison samples in the $K$ band, with a very small dispersion of $\sim$ 0.15 mag, which
is less than the dispersion among comparison data themselves. The TF relation for our five 
galaxies (the former three and PGC 138291, A 0101+4744) in the $B$ band has 
again no systematic trend between the northern subgroup and the principal galaxy, but shows
a larger dispersion of $\sim$ 0.78 mag comparable to the dispersion among the comparison
data. A larger dispersion in optical bands is (partly) due to uncertainties in the Galactic
and internal absorption in these bands. 
%
\begin{table*}[t]
\caption{Interaction probabilities of galaxies in subgroups. For each galaxy ({\it Col.~1}),
located in a particular subgroup ({\it Col.~2}) and having the gaseous ({\it Col.~3}) and optical
({\it Col.~4}) diameter ($D_\mathrm {gal}$) the main perturber ({\it Col.~5}), which is located
at the distance $d$ ({\it Col.~6}) from the galaxy is determined according to the
perturbation parameter ($P_\mathrm {gg}$) calculated both for the gaseous ({\it Col.~7}) and
optical ({\it Col.~8}) disks. }
\protect\label{Interactions.tab}
\begin{center}
\begin{tabular}{lccccccc}
\hline\hline
Galaxy & subgr. & \multicolumn{2}{c}{$D_\mathrm {gal}$} & main & $d$ & \multicolumn{2}{c}{ $P_\mathrm {gg}$}\\
       &      & \ion{H}{i} & opt & perturber & & \ion{H}{i} & opt \\
\hline
 (1)        &     (2)      &    (3)      &  (4)     &   (5)       &  (6) & (7) & (8)  \\
\hline
\\
IC 65 & 1 & 6\farcm9 & 4\farcm4 & A 0101+4744 & 5\farcm2 & 0.0065 & 0.0017 \\
A 0101+4744    & 1 & 1.8 & 0.8: & IC 65 & 5.2 & 0.246 & 0.0216 \\
A 0100+4734     & 1 &     & 0.8: &IC 65 & 10.3 & & 0.0027 \\\\
UGC 622 & 2 & 3.2 & 1.4 & PGC 138291 & 5.2 & 0.0066 & 0.0006 \\
PGC 138291 & 2 & 2.1 & 1.3: & UGC 622 & 5.2 & 0.0365 & 0.0087  \\
UGC 608 & 2 & 3.0 & 2.1 & UGC 622 & 14.2 & 0.0019 & 0.0006 \\
A 0100+4756  & 2 && 0.5: & UGC 622 & 5.0 && 0.0015 \\\\
A 0101+4752  & - && 0.5: & IC 65 & 14.8 && 0.00005 \\
\hline
\end{tabular}
\end{center}
\end{table*}

Both dIrr galaxies around the IC 65  
contain a chain of blue knots,  
bringing to mind the self-propagating star formation scenario.
The eastern part of the LSB underlying
stellar halo of the confirmed NE companion appears  curved
towards the primary IC~65. Both effects could be triggered by mutual interaction with the nearby 
principal galaxy. 
In small groups, where galaxy encounters occur at low relative velocities, i.e. they may have 
sufficiently long duration, and when the galaxy radii are comparable to the average separation 
between galaxies, tidal interactions can be efficient in disturbing the morphology of the disk, 
and/or in enhancing the 
star-forming activity (SFA)  
both in the circumnuclear region and throughout the disk. 
The tidal force (or asymmetric gravitational potential) per unit mass of the primary produced by 
 a companion is proportional to ${\cal M}_\mathrm {comp} \times d^{-3}$, where ${\cal M}_\mathrm {comp}$ is the 
mass of the companion and $d$ is its distance from the centre of the primary (Dahari \cite{dahari84}).
The simulations of Byrd \& Valtonen (\cite{byrd90}) have shown that tidal interactions are 
efficient in enchancing the SF activity, when the perturbation parameter
\begin{equation}
P_\mathrm {gg} = ({\cal M}_\mathrm {comp}/{\cal M}_\mathrm {gal}) \times (2d/D_\mathrm {gal})^{-3}       \protect\label{Pgg} 
\end{equation}
is in the critical range $P_\mathrm {gg} \geq 0.006 - 0.1$. The given range depends on the  
disk-to-halo mass ratio of the galaxy. 
We have determined the main perturber of every single group member and the 
maximum values of the perturbation parameter are listed in Table~\ref{Interactions.tab} 
(only available in electronic form).  
Parameter $P_\mathrm {gg}$ has been calculated both for the largest extent of \ion{H}{i} of bright galaxies, as 
given in vM83, and for the optical (blue) diameters of all 
confirmed and potential group members.
The obtained perturbation estimates and the image analysis of the galaxies in the southern subgroup 
lead to the following conclusions:  
1) The NE companion is expected to be  
perturbed by the primary since the perturbation parameter is in the critical range 
($P_\mathrm {gg} \simeq$ 0.246/0.022, for the \ion{H}{i}/stellar components). 
The expected perturbation is confirmed both in its disturbed 
morphology (the LSB halo is curved towards IC 65), and in enhanced SFA (multiple \ion{H}{ii} knots).
2) The warped NW-part of the gaseous disk of the (otherwise regular) primary galaxy IC 65 could have 
been invoked during the possible close passage with the NE companion in the past, as indicated 
by the marginally significant perturbation parameter ($P_\mathrm {gg} \simeq$ 0.0065) 
for the gaseous disk of the IC 65.

In the northern compact subgroup the three luminous galaxies possess an almost undisturbed regular stellar  
morphology, which agrees with non-critical values of the perturbation parameter as given   
for their optical disks in Table \ref{Interactions.tab}.
The situation is different for their more extended gaseous disks.  
VM83 noted that 
each of these three galaxies shows 
asymmetric features in their \ion{H}{i} distribution, especially the UGC~622.  
This is in general agreement with perturbation estimates in Table \ref{Interactions.tab}.
The outer \ion{H}{i} isophotes of UGC~622 are clearly disturbed and its gaseous disk
appears not to be aligned with the stellar disk, rather
being aligned with the direction to the neighbouring \lq\lq edge-on\rq\rq~ galaxy.
Poor resolution of the available  \ion{H}{i} data for the \lq\lq edge-on\rq\rq~ galaxy itself 
do not permit to study its gaseous morphology in more detail.

Huchtmeier \& Richter (\cite{hr82}) noted the occurrence of a large proportion of interacting 
galaxies among those with large \ion{H}{i}-extensions.  
Concerning the interaction time-scales Boselli \& Gavazzi (\cite{boselli06}) noted that the asymmetric 
\ion{H}{i} distribution in a gaseous disk can last for a few disk revolution times 
($t_\mathrm {rev} = \pi D_\mathrm {gal}/V_\mathrm {rot}$), 
after which the differential rotation should re-distribute the gas uniformly over the disk.
Since the $t_\mathrm {rev} \simeq 1 \times 10^9$ yr for the IC 65, and $\sim 2.5 \times 10^8$ yr 
for the UGC 622, thus observing an \ion{H}{i} asymmetry, 
and possibly the disturbed stellar disk in the SE part of the IC 65 too, 
implies that the disturbing 
mechanism has been acting for a relatively short time.
The time during which two galaxies stay close enough to get tidally perturbed - the 
tidal encounter time - can be estimated as $t_\mathrm {enc} \simeq max(R_\mathrm {1.gal},R_\mathrm {2.gal},d)/\Delta V$, 
where $d$ is their separation at closest approach (Binney \& Tremaine \cite{BT87}). In both 
subgroups $t_\mathrm {enc} = (5 \pm 1) \times 10^8$ yr. Further, the frequency of encounters 
in subgroups could be estimated as an inverse of the relaxation time,  
$t_\mathrm {relax} = 0.1 \times (R_\mathrm {vir}/\sigma_V) \times (n_\mathrm {gal}/\ln n_\mathrm {gal})$. 
For both dense subgroups the relaxation time is in the range $t_\mathrm {relax} \simeq (3 - 10) \times 10^8$ 
yr, i.e. close encounters should be frequent enough to significantly disturb  the outer 
gaseous and possibly also the stellar morphology, and, over a longer period to strip the majority of 
the atomic gas. 
Rasmussen et al. (\cite{rasmussen06}) argue that the atomic gas could have been stripped  over the course 
of a few Gyr by the intragroup medium and this should have a significant impact on the galaxy 
populations of compact groups. 
However, as found in the previous subsection, the gaseous content in regular galaxies of both 
subgroups is nearly normal, and the group is populated only with late type spiral and irregular 
galaxies with weakly developed tidal tracers. 
\emph{This strongly argues for the dynamical youth of the 
whole IC 65 group of galaxies.  Probably, the group is still assembling and the galaxies in 
subgroups have had few encounters.}

There is a great number of sophisticated numerical modeling experiments of the cosmo-dynamical 
evolution of galaxy systems of different multiplicity, performed during the last decades (see earlier 
review e.g. in Barnes \& Hernquist \cite{bh92}; Mamon \cite{mamon93}, \cite{mamon07}; Moore et al. 
\cite{moore98}). Particularly, Moore et al. (\cite{moore98}) proposed a new mechanism of galaxy 
harassment, in which the frequent high-speed encounters with massive galaxies cause impulsive 
gravitational shocks that severely remodel the fragile disks of late-type spiral and irregular galaxies. 
The cumulative effect of such encounters can ultimately change a disk galaxy into a spheroidal galaxy. 
Galaxy harassment is slightly more effective at removing mass than tides alone and it will occur 
anytime that galaxy encounters occur at speeds that are much larger than the galaxy's circular 
velocities (Lake \& Moore \cite{lake99}). Harassment should be effective in rich (virialized) clusters. 
The relative velocities can be too slow in poor groups, leading to merging rather than harassment 
(Lake \& Moore \cite{lake99}). However, some dynamical models 
of poor groups like the Local Group 
with nearly 30 \lq perturbing lumps\rq~ have yielded that  
the timescale for harassment of a dIrr galaxy like GR8 into something resembling 
present-day dwarf spheroidal 
satellites of the Milky Way (Dra and UMi) is roughly 3 Gyr (Moore et al. \cite{moore98}), 
or a similar transformation could happen in $\sim$ 7 Gyr when modeling 2 - 3 close ($< 40$ kpc)  
tidal encounters of a LSB satellite with its primary (Mayer et al. (\cite{mayer00}).
At the present stage we are not able to identify the possible dwarf spheroidal \lq remnants\rq~ 
around the IC 65, 
owing to unsufficient resolution of the available imaging data. The observed star-forming activity 
in dIrr satellite(s) around the IC 65 can possibly be triggered by tidal compression of gas at 
pericenter passage and then the SF could possibly be truncated after the tidal stripping have removed  
the remaining gas (Mayer et al. (\cite{mayer00}). 

\subsection{Star-forming properties of the brightest galaxies in subgroups}
\begin{table*}[t]
\caption{Star-forming properties of the brightest galaxies in subgroups.
{\it Col.~1}: galaxy name;
 {\it Cols. 2-4}: the IRAS
flux densities ($S_{60}$ and $S_{100}$) and the 1.4GHz radio flux density ($S_{1.4\mathrm{GHz}}$);
{\it Cols. 5-7}: the FIR ($L_{\mathrm{FIR}}$) and the radio ($L_{1.4\mathrm{GHz}}$) luminosities
and the mean current SFR ($SFR_{\mathrm{current}}=1/2(SFR_{\mathrm{FIR}}+SFR_{\mathrm{1.4GHz}}$),
calculated as in Bell (\cite{bell03}); {\it Col. 8}: the stellar mass (${\cal M}^*$); {\it Col.~9}:
the mean past SFR ($<SFR>_{\mathrm{past}}$); {\it Col.~10}: the Scalo parameter ($b$);
{\it Col.~11}:
the gas depletion rate ($\tau_{\mathrm{gas}}$).
}
\label{SF.tab}
\centering
\begin{tabular}{lcccccccccc}
\hline\hline
 Galaxy  & $S_{60}$ & $S_{100}$ & $S_{1.4\mathrm{GHz}}$ & $L_{\mathrm{FIR}}$ & $L_{1.4\mathrm{GHz}}$ & $SFR_{\mathrm{current}}$ &
${\cal M}^*$ & $<SFR>_{\mathrm{past}}$ &
 $b$ & $\tau_{\mathrm{gas}}$ \\
 & [Jy] & [Jy] & [mJy] & [$10^{10}$ L$_\odot$] & [$10^{21}$ W~Hz$^{-1}$] & [${\cal M}_\odot$ yr$^{-1}$] &
[$10^{10} {\cal M}_\odot$] & [${\cal M}_\odot$ yr$^{-1}$] & & [Gyr] \\
\hline
 (1)      &    (2)    &  (3)    &  (4)   &   (5)   &  (6)   &   (7)  &  (8)  & (9) & (10) &
(11)  \\
\hline
\\
IC 65  & 2.37 & 6.58 & 18.8 & 1.06$\pm$0.10 & 3.35$\pm$0.37 & 1.9$\pm$0.3 & 4.2$\pm$1.3 & 3.5$\pm$1.1 & 0.54$\pm$0.25 & 7.5 \\
UGC 622 & 1.61 & 5.16 & 11.4 & 0.78$\pm$0.11 & 2.03$\pm$0.25 & 1.4$\pm$0.2 & 2.0$\pm$0.3 & 1.7$\pm$0.3 & 0.72$\pm$0.20 & 2.3 \\
\hline
\end{tabular}
\end{table*}

We attempted to estimate the current SF rates in the brightest galaxies of 
both subgroups, in the IC 65 and UGC 622, using the homogenized 
IRAS flux densities ($S_{60}$ and $S_{100}$), and the 1.4 GHz radio flux density 
($S_{1.4\mathrm{GHz}}$) as given in the NED.
Both the far-infrared (FIR) and radio emission are expected to be enhanced in the instance of tidal
interactions (Roberts \& Haynes \cite{roberts94}) and, hence 
may serve as a clue to their interaction history.
%
There are well-known complications with interpreting the 
FIR spectral data in terms of star-formation (SF). 
The FIR spectra contain both a \lq\lq warm\rq\rq~ component associated with dust around young
star-forming regions ($\lambda \sim 60 \mu$m),
and a cooler IR cirrus component ($\lambda \geq 100 \mu$m) associated with more extended
dust heated by the interstellar radiation field of older disk stars. In late type
star-forming galaxies the FIR luminosity has been found to correlate with other tracers of
SF (as UV-continuum or H$\alpha$ luminosity) and, consequently, dust heating from young
stars is expected to dominate the 40 - 120 {$\mu$}m emission (Kennicutt \cite{kennicutt98}).

We have estimated the FIR star-formation rate ($SFR_{\mathrm{FIR}}$) and the radio star-formation rate
($SFR_{\mathrm{1.4GHz}}$) using the relations determined by Bell (\cite{bell03}), which account 
for old stellar populations, 
and should be proper for late type spiral galaxies. The mean value ($SFR_{\mathrm{current}}$), and 
the deviations from the mean are listed in Table \ref{SF.tab} 
(only available in electronic form). 
Both luminous galaxies show relatively low current $SFR$s in the range of 1 - 2 ${\cal M}_\odot$ yr$^{-1}$,
and hence, there is no clear sign of SF enchancement by possible recent interactions. A useful parameter
for characterizing the SF history is the ratio of the current SFR to the past mean SFR  
the Scalo parameter $b = SFR_{\mathrm{current}}/<SFR>_{\mathrm{past}}$. 
To determine this parameter, we need to know the stellar mass of galaxies.
Combining the $B$ and $K$ band luminosities and $B-R$ colours  
determined in the present work, with stellar population models of Bell \& de Jong (\cite{BdJ01}), 
we have calculated the maximum disk stellar mass-to-light 
ratios in $B$ and $K$ bands. The stellar mass (${\cal M}^*$) obtained as a mean from $B$ and $K$ band data,  
and the past mean $SFR$ over 12 Gyr ($<SFR>_{\mathrm{past}}$), are listed in Table \ref{SF.tab}.
We conclude that the UGC 622 appears to have had a nearly constant SFR over its whole SF history  
(with $b \simeq 0.72$), and is now slowly running off its fuel with gaseous supplies enough to 
maintain the present SF level during the next $\tau_{\mathrm{gas}} = {\cal M}$(\ion{H}{i})/$SFR_{\mathrm{current}} \sim$ 2.6 Gyrs. 
The principal galaxy IC 65 appears to have had a 
more active star-forming past ($b \simeq 0.54$). This may either have included several 
short star-bursting activities, which possibly could have been triggered
through accretion of its gas-rich dwarf satellites, or the given value of the Scalo parameter  
simply results from exponentially declining SFR, as typical for unperturbed spirals. The available gas 
reserve is enough to maintain the current SFR during at least  a half of another Hubble time 
($\tau_{\mathrm{gas}} \simeq$ 7.5 Gyrs). 
Both IC 65 and UGC 622 appear to be quiescently star-forming galaxies. 
%
\section{Summary}
The main contributions of this paper are as follows: 
\begin{enumerate}
 \item We have selected  
four LSB dwarf companion candidates of the IC 65 group of galaxies on deep DPOSS frames according to
their surface brightnesses, colours and morphology. 
\item The $B, R$ and $I$ band surface photometry  
is presented for the first time for all certain and probable members of the group.
The bright group members were studied in the NIR $J, H$ and $K$ bands, too.  
An image gallery and the deduced $SB$ and colour
profiles are shown, permitting the detailed morphological analysis of
the galaxies studied. Their relevant physical and model characteristics are determined.
\item 
Dynamical masses and star-forming characteristics of the bright group members are estimated using the 
new optical photometry and the available NIR, FIR, \ion{H}{i} and radio data. The probable evolutionary status 
of the group is discussed. 
\end{enumerate}

An analysis of the available photometric and kinematic data of individual
galaxies with emphasis to study the possible mutual interactions between the group members 
leads to the following results: 

\noindent
$\bullet$ The available \ion{H}{i} imaging data show that all bright members and
at least one dwarf companion have a nearly normal gaseous fraction with \ion{H}{i} mass to blue
luminosity ratios in the range of 0.3 -- 1.0, consistent with their morphological type.  
The outer \ion{H}{i} isophotes 
of the IC 65, and especially of the UGC 622 appear disturbed, in agreement with 
perturbation estimates.\\  
$\bullet$ The optical morphology of the bright galaxies generally appears to be regular, 
with barely significant disturbances in isophotes of the outer stellar disk of the IC 65 and UGC 622.\\
$\bullet$ All bright group members (except PGC~138291, which we could
not study in such detail) consist of many blue star-forming knots and
plumes, especially UGC~608. A comparison of the surface
photometry with stellar population models of Bruzual \& Charlot (\cite{bruzual03}) indicates that these
blue knots must have formed recently. 
The available data do not allow to establish whether they formed simultaneously, e.g. in star-bursts 
possibly triggered by interactions.\\
$\bullet$ Two dIrr galaxies around the IC 65 both contain a number of \ion{H}{ii} regions, 
which show  
a range of stellar ages and provide an evidence of propagating star-formation. 
One of these galaxies - A~0101+4744 is a confirmed member of the group; the second one - A~0100+4734 
appears to be located in front of the group.\\ 
$\bullet$ The brightest galaxies in both subgroups can fuel their current star-forming rates 
of $\sim$ 1 - 2 ${\cal M}_\odot$ yr$^{-1}$ for about the next 3 - 7 Gyr.

The IC~65 group of
galaxies obviously belongs to the class of less evolved groups. It is
composed of late type spiral and irregular galaxies arranged in two
subgroups. No massive early type galaxies are present. No 
hot gas has been detected in it by the ROSAT survey. 
Some morphological irregularities and signs of enhanced SF in its members could be 
indicative of recent/ongoing  mutual interactions.
Yet, the individual
group members have retained much of their initial gas component. 
A few available velocities point to 
a short crossing time of only $\sim$ 0.1~$H_0^{-1}$. However, this hardly means that 
the group has already reached a stable (virialized) configuration. 
The evidence, discussed above, lets us conclude that the IC~65 group of galaxies is a
dynamically young system at a still relatively early stage of its
collapse.
\begin{acknowledgements}
We want to thank an anonymous referee whose comments and suggestions were precious 
for improving the paper. 
We would like to
thank Dr. G.M. Richter for using his software package and Dr. Walter
Huchtmeier for supplying the \ion{H}{i} data for two of our objects. JV
thanks Dr. Claus M\"ollenhoff for useful discussions.
The research of JV has been supported by the Estonian
Science Foundation grants 4702, 6106, and by the German Academic Exchange Service
(DAAD) fellowship A/01/02044. JV acknowledges the hospitality of the
Universit\"atssternwarte M\"unchen during his visit. 
This research
has made use of the NASA/IPAC Extragalactic Database (NED) which is
operated by the Jet Propulsion Laboratory, California Institute of
Technology, under contract with the National Aeronautics and Space
Administration. The Digitized Sky Surveys were produced at the Space
Telescope Science Institute under U.S. Government grant NAGW-2166. We
have made use of the Lyon-Meudon Extragalactic Database (LEDA), and
the Two Micron All-Sky Survey (2MASS) at IPAC. UH would like to thank the
Calar Alto staff for support during the imaging observations.
\end{acknowledgements}

\appendix  
\section{Surface photometry and error estimates}

\subsection {Image processing}

Noise is a fundamental problem of image processing in astronomy because of the 
specific interest in the faint signals. Smoothing stands for removal of the noise and 
filtering is the usual tool. Stationary filtering with an impulse response (or point-spread function) 
which is invariant over the image is unsuitable since the high resolution objects in the 
image would be degraded and smeared. 
We need to apply a space-variable (adaptive) filter which recognizes the 
local signal resolution and adapts its own impulse response to this resolution. As a result the 
adaptive filter smoothes extensively the background, less extensively the galaxian outskirts and 
not at all the highest resolution features. We used the Potsdam adaptive filtering facility 
which is described in Lorenz et al. (\cite{lorenz93}) and implemented in the ESO MIDAS 
environment (task FILTER/ADAPTIV).
The filter is based on the H-transform (Richter \cite{richter78}).  
The local signal-to-noise (S/N) ratio as a function of decreasing resolution is evaluated by means of comparing 
the mean gradients and curvatures over different scale lengths 
(obtained from the H-coefficients of different order)  
to the corresponding expectation values of the noise. The order for which this 
S/N ratio exceeds a given (threshold for significance)  parameter 
indicates the local resolution scale length of 
the signal, and determines the size of the impulse response of the filter at this point. 
Before applying the adaptive filter, a mask frame was built, where bright stars and galaxies 
were masked out. The mask frame is essential for a proper determination of the noise statistics  
made by filter and used to discriminate between a local signal and noise at each scale length.
We varied the maximum filter
size between 7$\times$7 pixels and 15$\times$15 pixels depending on
the quality of the frames. The filter strength, defined by the minimum
S/N ratio for the detection of a local signal, was
generally between 2.0 $\div$ 2.5 of the {\it rms} noise level of the sky
background at each scale length.  More details concerning the filtering
procedures are given in Vennik et al. (\cite{vennik96}).  After
filtering, the sky background was fitted by a tilted plane, created
from a two-dimensional polynomial, using a least-squares method
FIT/FLAT\_SKY.  Then, the galaxy image was cleaned from disturbing
objects, e.g. foreground stars, using the interactive polygon
editor of the Potsdam Image Processing Software (PIPS) package,
running within MIDAS environment.

In addition to the adaptive smoothing, we performed an adaptive Laplacian
filtering, too.  The Laplacian filter computes the second derivative
(i.e. curvature) of the $SB$ distribution which does not retain any
photometric information. However, it is useful to disentangle the
inner morphology as multiple nuclei, \ion{H}{ii} regions, spiral arms, and
bridges, usually hidden by the large luminosity gradients of
the central regions of galaxies.
\subsection{Surface photometry and profile extraction}
The structure of the galaxy is reflected to a large extent in the surface brightness
($SB$) distribution where the different structure components
produce characteristic light distributions within the overall $SB$
profile. Therefore, among the basic results of the
present study are the $SB$ profiles as well as a set of isophotal and
integral parameters derived from these profiles. A common approach
to obtaining the $SB$ profile of a regular galaxy is fitting a sequence
of ellipses to the galaxy isophotes at predefined intensity levels and
computing the mean $SB$ between successive ellipses. This simple
technique is inadequate when dealing with objects having complex
morphologies. For irregular galaxies, we applied an
alternative strategy of calculating equivalent light profiles 
not depending on any a priori assumptions of the galaxy morphology.
This approach is integrated in the PIPS package.

For this approach, we slice the smoothed image of a galaxy into $k$
areas where each area $n$ is enclosed by two isophotes $I_n$ and
$I_{n+1}$ ($n$ runs from 0 to $k$). The level of these isophotes is
generally spaced by 0.1 mag. 
Total intensity of all pixels between successive isophotes  $I_n$ and
$I_{n+1}$ 
divided by the area enclosed between the two
isophotes ($\Delta A = A_{n+1}-A_n$) yields the surface brightness for the equivalent light
profile at the  equivalent radius $r_{\rm {eq},n} = [(A_n + A_{n+1} )/2\pi]^{1/2}$. 
A slight disadvantage of this approach results from the noise,
especially at low galaxy light levels near the mean sky
background. The routine integrates into the equivalent profile the
random signal fluctuations that are a few sigmas above the mean sky level.
As an effect, the derived equivalent radius will be
artificially increased when approaching the sky level. Fortunately, this
effect can be largely reduced by adaptive smoothing of the signal in
the outskirts of the galaxy.

Finally, we determined the $SB$ profiles twice for every galaxy.
First, we calculated equivalent light profiles using the PIPS
software. The resulting equivalent light profile yields the
isophotal radii, and the mean $SB$. It further allows to determine
the best-fitting parameters of the particular density distribution
model. In addition, the light growth curve of the galaxy was
calculated and the isophotal magnitudes as well as the
effective radius and effective $SB$ were determined on its basis.
The total magnitude was estimated by asymptotic extrapolation
of the radial growth curve.

Secondly, we applied the ellipse fitting algorithm FIT/ELL3
within in the SURFPHOT package of MIDAS to
obtain the shape and orientation of the isophotes
of the regular galaxies. As a result, we generated a set of radial
profiles in each particular passband: surface brightness ($SB$),
minor-to-major axis ratio ($b/a$), position angle ($P.A.$),
and displacement of the centre of fitted ellipses ($x-x_0, y-y_0$).
Here, ($x_0,y_0$) are the coordinates of the centre of the innermost ellipse.

Finally, we obtained colour index profiles. Here, we forced the
surface photometry in the second (or further) band(s) to follow
the ellipticity fit of the first band, generally the $R$ band as the deepest one.

\subsection {The photometry errors}

The magnitude errors consist of internal and external components.
The internal errors of the instrumental (non-calibrated) magnitudes and
those of the particular $SB$ profile are dominated by the error of the adopted
sky background value, and, in addition, includes, the random count
error  determined from the photon statistics in the aperture measurements.
Following Vader \& Chaboyer (\cite{vader94}) we calculated
the internal errors in intensity as

\begin{equation}
\Delta I = \sqrt{N_{\rm tot}+(\delta~n_{\rm sky}~A)^2},
\end{equation}
\noindent
where $N_{\rm tot}$ is the total number of counts, as measured on the
adaptive filtered frame within aperture $A$. The aperture $A$ is
defined as the area between successive isophotes (in pixels),  $n_{\rm
sky}$ is the mean sky counts per pixel, and $\delta$ is the fractional
error in the mean sky value. We measured the sky background
fluctuations on the filtered CCD frames within small apertures in the
vicinity of each galaxy. The variations of the sky background were
typically about 0.45\% (in $B$), 0.35\% (in $R$), and 0.3\% (in $I$).
Adopting the typical sky brightnesses of 22.0 ($B$), 21.2 ($R$), and 19.7 ($I$)
mag~arcsec$^{-2}$ (Table~\ref{Obs}) the mean uncertainty introduced
by the inaccuracy of the sky background determination is in the range of 0.10 - 0.06 mag.
This uncertainty is
primarily caused by the background variations across the frame due to 
some problems with flat-fielding, particularly on the $B$ frames.
The equivalent $SB$ profiles were measured typically up to the level 27 to 28 mag~arcsec$^{-2}$ 
in the $B$ band, up to 26 mag~arcsec$^{-2}$ in the $R$ band, and up to $\sim$ 25 mag~arcsec$^{-2}$ 
in the $I$ band.
The photometric accuracy of the 2MASS images is estimated to be better than
$\sim$ 0.1 mag (Jarret et al. \cite{jarret00}).
The azimuthally averaged $SB$ sensitivities (3 $\sigma$), obtained
from the coadded NIR images, are $\sim 21.0,
\sim 20.5$, and $\sim$ 20.0 mag~arcsec$^{-2}$ in the $J, H$, and $K$ bands, respectively
(Omar \& Dwarakanath (\cite{omar06}).
Typical internal
uncertainties are shown with error bars on the $SB$ and colour profiles
in Section 4.

\begin{figure}
\hspace{5mm}
\resizebox{0.33\textwidth}{!}{\includegraphics{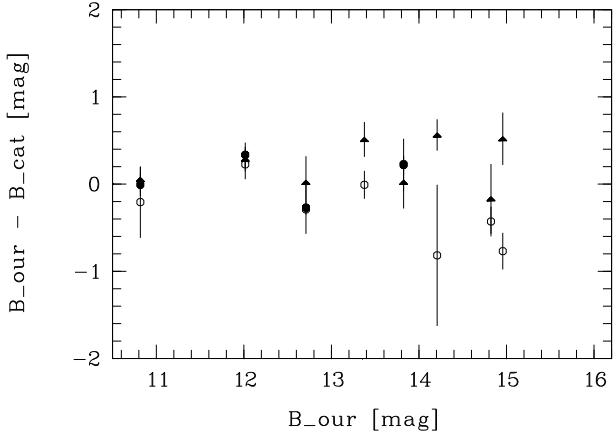}}
\caption{Comparison of our measured total $B$ magnitudes (B\_our)
         with those taken from 
the RC3 ($B_{\rm T}$ - filled circles, $m_{\rm B}$ - filled triangles), 
and from the LEDA (open circles). 
         The quoted errors of catalogued magnitudes are indicated
         with error-bars.}
\label{Compare.fig}
\end{figure}

To estimate the true (external) magnitude calibration errors
independently, we compared our total magnitude measurements with
published data for the brighter galaxies, as quoted in the RC3 (de
Vaucouleurs et al. \cite{rc3}), and/or in the LEDA database (Paturel
et al. \cite{paturel95}). The IC 65 group of galaxies contains only
three bright galaxies (IC 65, UGC 608 and UGC 622) with reliable
magnitudes in the above catalogues. To establish a comparison with a
better statistical data base, we include here the galaxies NGC~2591,
NGC~2655, NGC~2715, UGC~4701, and UGC~4714.  They belong to the
NGC~2655 group of galaxies and were observed during the same observing
session. The data were handled exactly the same way as described for
the IC~65 group members.  All other results concerning this particular
group will be presented in the forthcoming paper of this series.
The results of comparisons are shown in Figure~\ref{Compare.fig}, and
summarized in Table~\ref{Compare.tab}, where columns
contain the name of the catalogue and the magnitude system (1),
number of common galaxies (2), their mean quoted magnitude error (3),
the mean difference between our measurements and catalogued data
($\Delta m = m_{\rm our}-m_{\rm cat}$) with its 1$\sigma$ standard deviation
(4); and the mean uncertainty of our measurements, calculated as
follows: $\sigma_{\rm our}^2 = \sigma^2-\sigma_{\rm cat}^2$ (5).

The small number of available comparison sets as well as the
relatively large intrinsic errors of the literature data complicate
the estimate of our own systematic errors. We believe that the
RC3 magnitudes are still the most reliable ones and give them a higher
weight for our conclusions. We find no significant indication for
systematic magnitude zeropoint differences and there appear no trends
with the apparent magnitude.
Our error estimates in the $B$ band range from $\pm$ 0.15
mag for bright galaxies to $\pm$ 0.20 mag for fainter ($B_{\rm T} \geq$ 15) galaxies. A
similar error range should be proper for our $R$ and $I$  band magnitudes,
too.

\begin{table}
\caption {Comparison of magnitudes
from the present work with those taken from literature}
\label{Compare.tab}
\centering
\begin{tabular}{cccccc}
\hline\hline
Cat./ Mag. & n & $<\sigma_{\rm cat}>$ & $<\Delta m> \pm \sigma$ & $\sigma_{\rm our}$ \\
\hline
  (1) & (2) & (3) & (4) & (5)  \\

\hline
\\
RC3/$B_{\rm T}$  & 4  &  0.23      & +0.07   $\pm$0.27      & 0.135 \\
RC3/$mB$         & 8  &  0.26      & +0.23  $\pm$0.28      & 0.11  \\
LEDA/$B_{\rm T}$ & 8  &  0.35     & -0.26  $\pm$0.40      & 0.21 \\
LEDA/$I_{\rm T}$ & 4  &  0.08     & -0.20  $\pm$0.26     & 0.25 \\
\hline
\end{tabular}
\end{table}

\subsection{Profile fitting}

The $SB$ profiles depend on the morphological type and bear some
information about the possible environmental influences on the
dynamical evolution of galaxies. A preliminary inspection of the
$SB$ profiles reveals that those of the LSB dwarf galaxies show only
minor deviations from the pure exponential. These small deviations mostly
appear in their inner parts of the galaxies and can be attributed
to the occurence of star-forming knots.
%

As first step we have fitted the equivalent light profiles of the dwarf galaxies 
with a Sersic (\cite{sersic68}) power law 

\begin{equation}
\mu (r) = \mu_0 + 1.086 (\frac{r}{h})^{1/n},  \protect\label{Sersic.mu}
\end{equation}
\noindent
and determined the best-fitting value of the parameter $n$.
The radial range of the light profiles over which the best fit has been computed 
goes from outside the region dominated by seeing (typically 2 - 3 arcseconds) out to where 
the uncertainties in the sky subtraction render the light profiles unreliable.  
We found the best-fitting values in the range $n$ = 0.7~-~1.2, and concluded  
that the underlying stellar component (i.e. when excluding the luminous blue knots) 
of all four LSB dwarf galaxies could be reasonably 
well fitted with a pure exponential model ($n = 1$). In the second attempt we 
fitted the (outer) linear part of each profile with the exponential disk model and 
determined the best-fitting model parametrs ($\mu_0, h$), which we aim to compare 
(for our full sample of groups) with results of earlier studies of grouped and field 
LSB dwarf galaxies (e.g. Barazza et al. \cite{barazza01}, Parodi et al. \cite{parodi02}, 
Vennik et al. \cite{vennik96}). 
The total light emitted by the exponential disk can be computed as 

\begin{equation}
m^\mathrm {exp} = \mu_0^\mathrm {exp} - 5~\log(h) - 1.995. \protect\label{mexpo}
\end{equation}
The goodness of the pure exponential fit can then be expressed as the
difference $\Delta m$ between $m^{\rm exp}$ and the actual measured total
magnitude $m_{\rm T}$: $\Delta m = m^{\rm exp} - m_{\rm T}$.

Regular late-type spiral galaxies IC~65 and UGC~608 both show a reliable 
light excess above the exponential disk
model. This extra light was modelled with a Sersic (\cite{sersic68}) power law
%
which has been found  proper when describing the light distribution of
spiral bulges (Andredakis et al. \cite{andredakis95}). 
The parameters ($\mu_0, h$)
of the best-fitting exponential model and the magnitude difference
$\Delta m$ for the individual galaxies are listed in
Table~\ref{Model.tab}.
The results of fitting are discussed individually for each galaxy in Section 4.

\section {Radio-observations of the A 0100+4734}

\begin{figure}
\resizebox{\hsize}{!}{\includegraphics{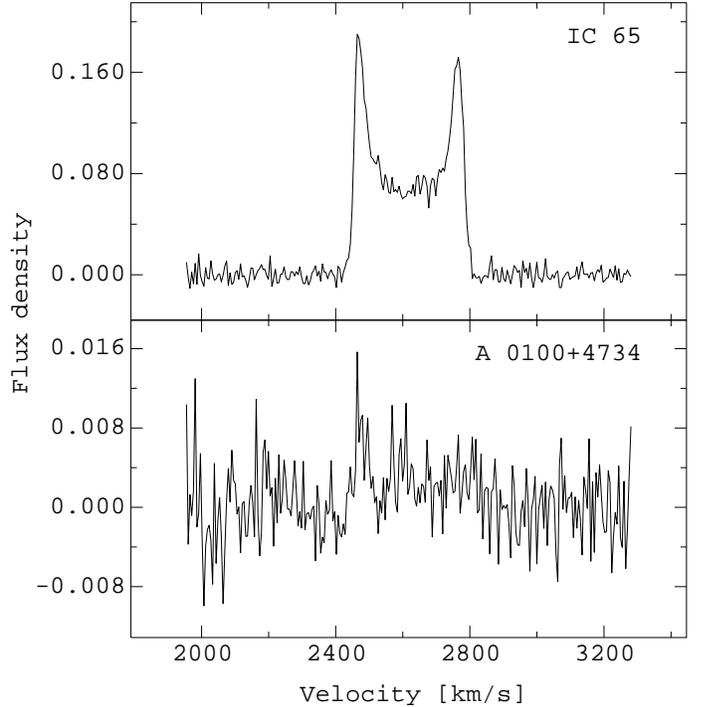}}
\caption{The \ion{H}{i} spectrum of IC 65
  ({\bf top}) and of its dwarf companion A 0100+4734
  ({\bf bottom}). Flux densities along the y-axis are in relative units.
}
\protect\label{F5Dw.radio}
\end{figure}

We attempted to measure the \ion{H}{i} flux of this actively
star-forming late type galaxy.
Pilot observations have been made by W.
Huchtmeier with the 100-m radio telescope at Effelsberg (HPBW = 9.$'$3)
in July 2002.
The principal galaxy IC 65 shows a broad
($\Delta V_{20}$ = 343 km s$^{-1}$)
double-horned global \ion{H}{i} profile (Fig.~\ref{F5Dw.radio}),
as typical of a classical rotation curve with an extended
flat part in outer regions.
The blueshifted horn shows a slightly higher amplitude, in accord with the
result obtained earlier by vM83 (his Fig.~26).
The dwarf galaxy A 0100+4734 was
searched for emission in the same frequency band 6.25 MHz (1250
km s$^{-1}$) centered on the velocity of IC 65.
The obtained spectra are compared to each other in Fig.~\ref{F5Dw.radio}.
A marginal signal 
registered by pointing the telescope in the direction of that dwarf
is certainly corrupted by signal of the parent galaxy, due to the close proximity of these two galaxies.
New \ion{H}{i} observations with better angular resolution and in broader redshift interval are needed
in order to 
determine the \ion{H}{i} properties of the A~0100+4734.

\end{document}